\documentclass[lettersize,twoside,journal]{IEEEtran}
\usepackage{graphicx}
\usepackage{amssymb}
\usepackage{amsmath,cases}
\usepackage{cite}
\usepackage{xcolor}
\usepackage{subfigure}
\usepackage{booktabs}
\usepackage{balance}
\usepackage{acro}
\usepackage{stfloats}

\usepackage{array,booktabs,tabularx}
\usepackage{multirow}
\usepackage[figuresleft]{rotating}

\newcolumntype{C}{>{\centering\arraybackslash}X}
\newcolumntype{L}{>{\raggedright\arraybackslash}X}
\newcolumntype{R}{>{\raggedleft\arraybackslash}X}

\usepackage[mathscr]{euscript}
\usepackage{makecell}

\DeclareAcronym{1D}{
short = 1D,
long = one-dimensional
}

\DeclareAcronym{2D}{
short = 2D,
long = two-dimensional
}

\DeclareAcronym{3D}{
short = 3D,
long = three-dimensional
}

\DeclareAcronym{5G}{
short = 5G,
long = fifth generation
}

\DeclareAcronym{6G}{
short = 6G,
long = sixth-generation
}

\DeclareAcronym{AI}{
short = AI,
long = artificial intelligence
}

\DeclareAcronym{ANN}{
short = ANN,
long = artificial neural network
}

\DeclareAcronym{AR}{
short = AR,
long = augmented reality
}

\DeclareAcronym{AO}{
short = AO,
long = absorptive object
}

\DeclareAcronym{AWGN}{
short = AWGN,
long = additive white Gaussian noise
}

\DeclareAcronym{BB84}{
short = BB84,
long = Bennett and Brassard 1984
}

\DeclareAcronym{BEP}{
short = BEP,
long = bit error probability
}

\DeclareAcronym{BLIP}{
short = BLIP,
long = bootstrapping language-image pretraining
}

\DeclareAcronym{BPSK}{
short = BPSK,
long = binary phase-shift keying
}

\DeclareAcronym{BS}{
short = BS,
long = base station
}

\DeclareAcronym{BSM}{
short = BSM,
long = Bell-state measurement
}

\DeclareAcronym{CKA}{
short = CKA,
long = conference key agreement
}

\DeclareAcronym{CNN}{
short = CNN,
long = convolutional neural network
}

\DeclareAcronym{CNOT}{
short = CNOT,
long = controlled-NOT
}

\DeclareAcronym{CPS}{
short = CPS,
long = cyber-physical system
}

\DeclareAcronym{CPTP}{
short = CPTP,
long = completely positive trace-preserving
}

\DeclareAcronym{CPU}{
short = CPU,
long = central processing unit
}

\DeclareAcronym{CSI}{
short = CSI,
long = channel state information
}

\DeclareAcronym{CV}{
short = CV,
long = continuous-variable 
}

\DeclareAcronym{DNN}{
short = DNN,
long = deep neural network
}

\DeclareAcronym{DL}{
short = DL,
long = deep learning
}

\DeclareAcronym{DQL}{
short = DQL,
long = deep Q-learning
}

\DeclareAcronym{DRL}{
short = DRL,
alt = deep RL,
long = deep reinforcement learning
}

\DeclareAcronym{DT}{
short = DT,
long = digital twin
}

\DeclareAcronym{DV}{
short = DV,
long = discrete-variable 
}

\DeclareAcronym{FD}{
short = FD,
long = full-duplex
}

\DeclareAcronym{GAN}{
short = GAN,
long = generative adversarial network
}

\DeclareAcronym{GEO}{
short = GEO,
long = geostationary Earth orbit
}

\DeclareAcronym{GHZ}{
short = GHZ,
long = Greenberger--Horne--Zeilinger
}

\DeclareAcronym{GPS}{
short = GPS,
long = global positioning system
}

\DeclareAcronym{GPT}{
short = GPT,
long = generative pretrained transformer
}

\DeclareAcronym{GPU}{
short = GPU,
long = graphics processing unit 
}

\DeclareAcronym{GUS}{
short = GUS,
long = geometrically uniform symmetry
}

\DeclareAcronym{HQC}{
short = HQC,
long = hybrid quantum-classical
}

\DeclareAcronym{IIoT}{
short = IIoT,
alt = industrial IoT,
long = industrial Internet of Things
}

\DeclareAcronym{IoIV}{
short = IoIV,
long = Internet of intelligent Vehicles
}

\DeclareAcronym{IoT}{
short = IoT,
long = Internet of Things
}

\DeclareAcronym{ISAC}{
short = ISAC,
long = integrated sensing and communication
}

\DeclareAcronym{KPI}{
short = KPI,
long = key performance indicator
}

\DeclareAcronym{LEO}{
short = LEO,
long = low Earth orbit
}

\DeclareAcronym{LLaMA}{
short = LLaMA,
alt = large language model Meta AI,
long = LLM Meta AI
}

\DeclareAcronym{LLM}{
short = LLM,
long = large language model
}

\DeclareAcronym{LPIPS}{
short = LPIPS,
long = learned perceptual image patch similarity
}

\DeclareAcronym{LOCC}{
short = LOCC,
long = local operations and classical communication
}

\DeclareAcronym{LSTM}{
short = LSTM,
long = long short-term memory
}

\DeclareAcronym{MEO}{
short = MEO,
long = medium Earth orbit
}

\DeclareAcronym{MIMO}{
short = MIMO,
long = multiple-input multiple-output
}

\DeclareAcronym{mmWave}{
short = mmWave,
long = millimeter-wave
}

\DeclareAcronym{ML}{
short = ML,
long = machine learning
}

\DeclareAcronym{MLP}{
short = MLP,
long = multi-layer perceptron
}

\DeclareAcronym{MPA}{
short = MPA,
long = mean pixel accuracy
}

\DeclareAcronym{MR}{
short = MR,
long = mixed reality
}

\DeclareAcronym{MSE}{
short = MSE,
long = mean squared error
}

\DeclareAcronym{NISQ}{
short = NISQ,
long = noisy intermediate-scale quantum
}

\DeclareAcronym{NMSE}{
short = NMSE,
alt = normalized MSE,
long = normalized mean squared error
}

\DeclareAcronym{NLP}{
short = NLP,
long = natural language processing
}

\DeclareAcronym{NTN}{
short = NTN,
long = non-terrestrial network
}

\DeclareAcronym{PISQ}{
short = PISQ,
long = perfect intermediate-scale quantum
}

\DeclareAcronym{POVM}{
short = POVM,
long = positive operator-valued measure
}

\DeclareAcronym{PQC}{
short = PQC,
long = parametrized quantum circuit
}

\DeclareAcronym{QAA}{
short = QAA,
long = quantum anonymous authentication
}

\DeclareAcronym{QAB}{
short = QAB,
long = quantum anonymous broadcast
}

\DeclareAcronym{QAC}{
short = QAC,
long = quantum anonymous communication
}

\DeclareAcronym{QACKA}{
short = QA-CKA,
alt = quantum anonymous CKA,
long = quantum anonymous conference key agreement
}

\DeclareAcronym{QACD}{
short = QACD,
long = quantum anonymous collision detection
}

\DeclareAcronym{QAE}{
short = QAE,
long = quantum anonymous entanglement
}

\DeclareAcronym{QAIA}{
short = QAIA,
long = quantum anonymous identity authentication
}

\DeclareAcronym{QAIR}{
short = QAIR,
long = quantum anonymous information retrieval
}

\DeclareAcronym{QAM}{
short = QAM,
long = quadrature amplitude modulation
}

\DeclareAcronym{QAN}{
short = QAN,
long = quantum anonymous network
}

\DeclareAcronym{qAN}{
short = QAN,
long = quantum anonymous notification
}

\DeclareAcronym{QANO}{
short = QANO,
long = quantum anonymous notification
}

\DeclareAcronym{QAOA}{
short = QAOA,
long = quantum approximate optimization algorithm
}

\DeclareAcronym{QAP}{
short = QAP,
long = quantum anonymous publication
}

\DeclareAcronym{QAR}{
short = QAR,
long = quantum anonymous ranking
}

\DeclareAcronym{QAS}{
short = QAS,
long = quantum anonymous sensing
}

\DeclareAcronym{QAT}{
short = QAT,
long = quantum anonymous teleportation
}

\DeclareAcronym{QAV}{
short = QAV,
long = quantum anonymous voting
}

\DeclareAcronym{QCC}{
short = QCC,
long = quantum covert communication
}

\DeclareAcronym{QCRB}{
short = QCRB,
long = quantum Cram\'er--Rao bound
}

\DeclareAcronym{QCS}{
short = QCS,
long = quantum clock synchronization
}

\DeclareAcronym{QEC}{
short = QEC,
long = quantum error correction
}

\DeclareAcronym{QED}{
short = QED,
long = quantum entanglement distribution
}

\DeclareAcronym{QFI}{
short = QFI,
long = quantum Fisher information
}

\DeclareAcronym{QFIM}{
short = QFIM,
alt = QFI matrix,
long = quantum Fisher information matrix
}

\DeclareAcronym{QFT}{
short = QFT,
long = quantum Fourier transform
}

\DeclareAcronym{QIoT}{
short = QIoT,
alt = quantum IoT,
long = quantum Internet of Things
}

\DeclareAcronym{QKD}{
short = QKD,
long = quantum key distribution
}

\DeclareAcronym{QLSTM}{
short = QLSTM,
alt = quantum LSTM,
long = quantum long short-term memory
}

\DeclareAcronym{QML}{
short = QML,
alt = quantum ML,
long = quantum machine learning
}

\DeclareAcronym{QNN}{
short = QNN,
long = quantum neural network
}

\DeclareAcronym{QOC}{
short = QOC,
long = quantum optimal control
}

\DeclareAcronym{QPU}{
short = QPU,
long = quantum processing unit
}

\DeclareAcronym{QSC}{
short = QSC,
alt = quantum SC,
long = quantum semantic communication
}

\DeclareAcronym{QSN}{
short = QSN,
long = quantum sensing network
}

\DeclareAcronym{QUBO}{
short = QUBO,
long = quadratic unconstrained binary optimization
}

\DeclareAcronym{ReLU}{
short = ReLU,
long = rectified linear unit
}

\DeclareAcronym{ResNet}{
short = ResNet,
long = residual network
}

\DeclareAcronym{RGB}{
short = RGB,
long = red-green-blue
}

\DeclareAcronym{RL}{
short = RL,
long = reinforcement learning
}

\DeclareAcronym{RMSE}{
short = RMSE,
alt = root MSE,
long = root mean squared error
}

\DeclareAcronym{RNN}{
short = RNN,
long = recurrent neural network
}

\DeclareAcronym{RSA}{
short = RSA,
long = Rivest--Shamir--Adleman
}

\DeclareAcronym{SAM}{
short = SAM,
long = segment anything model
}

\DeclareAcronym{SC}{
short = SC,
long = semantic communication
}

\DeclareAcronym{SEP}{
short = SEP,
long = symbol error probability
}

\DeclareAcronym{SNR}{
short = SNR,
long = signal-to-noise ratio
}

\DeclareAcronym{SRM}{
short = SRM,
long = square-root measurement
}

\DeclareAcronym{SQL}{
short = SQL,
long = standard quantum limit
}

\DeclareAcronym{SVM}{
short = SVM,
long = support vector machine
}

\DeclareAcronym{SwinT}{
short = SwinT,
long = shifted window transformer
}

\DeclareAcronym{THz}{
short = THz,
long = terahertz
}

\DeclareAcronym{TN}{
short = TN,
long = terrestrial network
}

\DeclareAcronym{UAV}{
short = UAV,
long = unmanned aerial vehicle
}

\DeclareAcronym{U-Net}{
short = U-Net,
long = U-shaped network
}

\DeclareAcronym{VAE}{
short = VAE,
long = variational autoencoder
}

\DeclareAcronym{V2I}{
short = V2I,
long = vehicle-to-infrastructure
}

\DeclareAcronym{V2X}{
short = V2X,
long = vehicle-to-everything
}

\DeclareAcronym{VFM}{
short = VFM,
long = vision foundation model
}

\DeclareAcronym{ViT}{
short = ViT,
long = vision transformer
}

\DeclareAcronym{VQA}{
short = VQA,
long = variational quantum algorithm
}

\DeclareAcronym{VQC}{
short = VQC,
long = variational quantum circuit
}

\DeclareAcronym{VQS}{
short = VQS,
long = variational quantum sensing
}

\DeclareAcronym{VR}{
short = VR,
long = virtual reality
}

\DeclareAcronym{YOLO}{
short = YOLO,
long = you only look once
}

\DeclareAcronym{EB-QCS}{
short = EB-QCS,
alt = entanglement-based QCS,
long = entanglement-based quantum clock synchronization
}

\DeclareAcronym{FE-QCS}{
short = FE-QCS,
alt = frequency-entangled QCS,
long = frequency-entangled quantum clock synchronization
}

\DeclareAcronym{FEP}{
short = FEP,
long = frequency-entangled pulse
}

\DeclareAcronym{FFT}{
short = FFT,
long = fast Fourier transform
}

\DeclareAcronym{GNSS}{
short = GNSS,
long = global navigation satellite system
}

\DeclareAcronym{HOM}{
short = HOM,
long = Hong-Ou-Mandel
}

\DeclareAcronym{HOM-QCS}{
short = HOM-QCS,
alt = Hong-Ou-Mandel QCS,
long = Hong-Ou-Mandel quantum clock synchronization
}

\DeclareAcronym{LO}{
short = LO,
long = local oscillator
}

\DeclareAcronym{MDI}{
short = MDI,
long = measurement-device-independent
}

\DeclareAcronym{NMR}{
short = NMR,
long = nuclear magnetic resonance
}

\DeclareAcronym{qan}{
short = QAN,
long = quantum access network
}

\DeclareAcronym{QBER}{
short = QBER,
long = quantum bit error rate
}

\DeclareAcronym{QND}{
short = QND,
long = quantum non-demolition
}

\DeclareAcronym{QSTT}{
short = QSTT,
long = quantum secure time transfer
}

\DeclareAcronym{RMS}{
short = RMS,
long = root mean square
}

\DeclareAcronym{SCS}{
short = SCS,
long = superposition of coherent state
}

\DeclareAcronym{SPD}{
short = SPD,
long = single-photon detector
}

\DeclareAcronym{SPDC}{
short = SPDC,
long = spontaneous parametric down-conversion
}

\DeclareAcronym{TCB-QCS}{
short = TCB-QCS,
alt = temporal correlation-based QCS,
long =  temporal correlation-based quantum clock synchronization
}

\DeclareAcronym{TMSV}{
short = TMSV,
long = two-mode squeezed vacuum
}

\DeclareAcronym{ToA}{
short = ToA,
long = time-of-arrival
}

\DeclareAcronym{ToA-QCS}{
short = ToA-QCS,
alt = time-of-arrival QCS,
long = time-of-arrival quantum clock synchronization
}

\DeclareAcronym{TQH}{
short = TQH,
long = ticking-qubit handshake
}

\DeclareAcronym{TQH-QCS}{
short = TQH-QCS,
alt = ticking-qubit handshake QCS,
long =  ticking-qubit handshake quantum clock synchronization
}

\usepackage{xcolor}
\usepackage{braket}
\usepackage{bbm}
\usepackage{pifont}

\DeclareMathAlphabet{\mathpzc}{OT1}{pzc}{m}{it}

\DeclareMathAlphabet{\mathitsf}{OML}{cmbr}{m}{it}
\DeclareMathAlphabet{\mathsf}{OT1}{cmbr}{m}{n}

\definecolor{darkgreen}{RGB}{0,128,0}

\AtBeginDocument{}

\newcommand{\M}[1]{\boldsymbol{#1}}

\newcommand{\imU}{\imath}

\newcommand{\PX}{\M{\sigma}_\mathrm{x}}
\newcommand{\PY}{\M{\sigma}_\mathrm{y}}
\newcommand{\PZ}{\M{\sigma}_\mathrm{z}}

\begin{document}
\bstctlcite{IEEEexample:BSTcontrol}

\title{ Quantum Clock Synchronization Networks:\\ A Survey}

\author{
Uman~Khalid,
Muhammad~Shohibul~Ulum,
Mujirin, 
Giuseppe~Thadeu~Freitas~de~Abreu,~\IEEEmembership{Senior~Member,~IEEE},
Emil Bj\"ornson,~\IEEEmembership{Fellow,~IEEE},
and
Hyundong~Shin,~\IEEEmembership{Fellow,~IEEE}


\thanks{
U.~Khalid,
M.~S.~Ulum,
Mujirin,
and H.~Shin (corresponding author)
are with the Department of Electronics and Information Convergence Engineering,
Kyung Hee University,
1732 Deogyeong-daero, Giheung-gu,
Yongin-si, Gyeonggi-do 17104, Republic of Korea
(e-mail: hshin@khu.ac.kr). 
U.~Khalid and M.~S.~Ulum contributed equally to this paper.
}
\thanks{
G.~T.~F.~de~Abreu is with the School of Computer Science and Engineering, Constructor University, Campus Ring 1, 28759, Bremen, Germany 
(e-mail: gabreu@constructor.university).
}
\thanks{
E.~Bj\"ornson is with the Division of Communication Systems,
KTH Royal Institute of Technology, 
100 44 Stockholm, Sweden 
(e-mail: emilbjo@kth.se).
}
}
\markboth{
   }{ 
Khalid and Ulum \textit{\MakeLowercase{et al.}}:
Quantum Clock Synchronization {Networks}: A Survey
}

\maketitle  

\begin{abstract}
\Ac{QCS} aims to establish a shared temporal reference between distant nodes by exploiting uniquely quantum phenomena such as entanglement, single-photon interference, and quantum correlations. In contrast to classical synchronization and time-transfer techniques, which are limited by signal propagation delays, atmospheric disturbances, and oscillator drift, \ac{QCS} protocols offer the potential to surpass classical precision bounds and enhance resilience against adversarial manipulations. As precise and secure time synchronization underpins distributed quantum networks, navigation systems, and emerging quantum Internet infrastructures, understanding \ac{QCS} principles, capabilities, and implementation challenges has become increasingly important.
This survey provides a unified and critical overview of the rapidly growing \ac{QCS} research landscape, highlighting fundamentals, protocol types, enabling resources, performance constraints, security considerations, and practical implementations of \ac{QCS}. We first introduce the theoretical underpinnings of \ac{QCS}, including entanglement-assisted time transfer, \acl{HOM} interference-based synchronization, and quantum slow-clock transport. We then categorize the main \ac{QCS} protocols, ranging from ticking-qubit and entanglement-based schemes to time-of-arrival correlation methods, conveyor-belt synchronization, and quantum-enhanced two-way time transfer. This organization clarifies the relationships between protocol families and their achievable precision advantages over classical methods.
Key quantum resources such as \acl{SPDC}-based entangled photon pairs, \acl{GHZ} and W multipartite states, squeezed and frequency-entangled light, quantum frequency combs, and quantum memories are reviewed in the context of scalability and robustness. Practical limitations arising from decoherence, channel loss, dispersion, detector timing jitter, and relativistic effects are examined alongside emerging noise mitigation strategies. Security aspects---including eavesdropping on time correlations, intercept-resend attacks, and adversarial delay manipulation---are evaluated with respect to trusted and untrusted network architectures. Recent proof-of-concept demonstrations in fiber networks, free-space optical links, chip-integrated photonic systems, and satellite-ground platforms are summarized, followed by discussions of open challenges toward precise and secure global quantum-synchronized networks.
\end{abstract}



\section*{Acronyms}
\noindent
\begin{tabular}{ll}
\Acs{6G} & 	\Acl{6G}\\
\Acs{BSM} & 	\Acl{BSM}\\
\Acs{CNOT} & 	\Acl{CNOT}\\
\Acs{EB-QCS} & 	\Aca{EB-QCS}\\
\Acs{FFT} & 	\Acl{FFT}\\
\Acs{GHZ} & 	\Acl{GHZ}\\
\Acs{GNSS} & 	\Acl{GNSS}\\
\Acs{HOM} & 	\Acl{HOM}\\
\Acs{HOM-QCS} & 	\Aca{HOM-QCS}\\
\Acs{HQC} & 	\Acl{HQC}\\
\Acs{LEO} & 	\Acl{LEO}\\
\Acs{LO} & 	\Acl{LO}\\
\Acs{MDI} & 	\Acl{MDI}\\
\Acs{NMR} & 	\Acl{NMR}\\
\Acs{QBER} & 	\Acl{QBER}\\
\Acs{QCRB} & 	\Acl{QCRB}\\
\Acs{QCS} & 	\Acl{QCS}\\
\Acs{QKD} & 	\Acl{QKD}\\
\Acs{QND} & 	\Acl{QND}\\
\Acs{QSTT} & 	\Acl{QSTT}\\
\Acs{RMS} & 	\Acl{RMS}\\
\Acs{SCS} & 	\Acl{SCS}\\
\Acs{SNR} & 	\Acl{SNR}\\
\Acs{SPD} & 	\Acl{SPD}\\
\Acs{SPDC} & 	\Acl{SPDC}\\
\Acs{SQL} & 	\Acl{SQL}\\
\Acs{TMSV} & 	\Acl{TMSV}\\
\Acs{ToA} & 	\Acl{ToA}\\
\Acs{ToA-QCS} & 	\Aca{ToA-QCS}\\
\Acs{TQH} & 	\Acl{TQH}\\
\Acs{TQH-QCS} & 	\Aca{TQH-QCS}\\
\end{tabular}

\acresetall		

\color{black}

\section{Introduction}
\label{sec:1}

\IEEEPARstart{Q}{uantum}
clock synchronization (\acs{QCS}) has emerged as a fundamentally new paradigm for establishing high-precision temporal references across distributed systems by exploiting uniquely quantum resources that surpass the intrinsic limits of classical synchronization \cite{GLM:01:N}.
Advancements in atomic clocks have pushed timekeeping precision to unprecedented levels, establishing them as the foundation of modern timing systems. This progress has been largely driven by sophisticated optical clock designs based on trapped ions and neutral atoms, which enable the most accurate temporal standards \cite{LBP:15:RMP}. Central to this progress is the quantum-mechanical characterization of stability and coherence---the Allan variance serves as a fundamental metric for clock noise, wherein quantum extensions provide ultimate bounds for entangled atomic ensembles \cite{Wis:03:PS, CLD:16:NJP}. Furthermore, lasers can also act as practical sources of clock coherence rather than mere quantum channels, anchoring time standards to agreed conventions instead of absolute phase references. Beyond individual clock units, recent theoretical advances reveal exponential improvements in clock stability as the number of atoms and atomic ensembles increases, driving the evolution of quantum metrology for synchronizing distributed time standards \cite{RL:13:AQP}.

Classical clock synchronization has undergone significant evolution, leveraging a range of approaches to reducing errors and improving stability across large-scale distributed networks \cite{Ein:1905:Anna, BS:13:PRL, MDH:15:PRA, Pit:17:Pres}. Early works emphasized the theoretical aspects, such as Einstein’s exploration of the electrodynamics of moving bodies, which set a basis for understanding synchrony over relativistic frames \cite{Rie:17:NPh,TCZB:15:SR, CSDGN:24:APLP}. Subsequent developments have advanced into practical systems utilizing atomic clocks, where noise and instability, especially in optical lattice clocks, are meticulously studied to ensure high accuracy in timekeeping. Modern synchronization techniques incorporate the simultaneous remote transfer of both timing and optical frequency across fiber networks, achieving remarkable precision over continental distances \cite{DSGS:16:PRX,AHCRA:24:ICSCA, LKP:13:APB,PRFFB:11:ARS, LGQ:16:NC,FNAK:17:QIM,YVHS:22:PRD}. Optical networks and satellite links now allow for remote atomic clock synchronization at the scale of geodesy and navigation systems, providing essential infrastructure for metrology and global positioning. Furthermore, the deployment of space-based optical clock networks and the application of quantum-limited optical time transfer in contemporary research underscore the role of photonic technologies in achieving uninterrupted, high-fidelity clock comparisons over space and ground-based platforms \cite{RLR:15:AQP}.

As demonstrated by field experiments, classical methods, though increasingly supplemented by quantum principles, remain foundational to global time dissemination, with ongoing efforts directed at minimizing environmental noise, technical errors, ensuring stability, and expanding scalability for future scientific and industrial demands. Building on classical foundations, recent research has shifted toward quantum-enhanced synchronization and the exploration of quantum phase dynamics in atomic systems. It is worthwhile to distinguish between quantum-enhanced and fully quantum synchronization. Quantum-enhanced synchronization refers to protocols in which classical synchronization procedures are improved through the use of quantum resources, such as nonclassical states or quantum measurement techniques, while the overall framework remains largely classical. In contrast, fully quantum synchronization relies fundamentally on quantum correlations---particularly entanglement---to establish timing relationships between distant parties, thereby enabling performance and security advantages that cannot be achieved with classical resources alone.
Numerous studies investigate the use of teleportation protocols for transferring quantum information relevant to synchronization, emphasizing that only speakable quantum information can be transmitted effectively.
Advances include the conceptualization of quantum stopwatches for optimal storage of temporal information and methods integrating quantum intelligence to enhance synchronization precision beyond classical limits \cite{CGMP:12:PRA, YCH:18:PRS, CSRS:19:AQT, MZZ:20:QIP}. Investigations into continuous variable systems and quantum oscillators reveal criteria and mechanisms for maintaining phase coherence remotely, even under noisy environmental conditions, such as non-Markovian contexts \cite{YKY:21:SST, XLZL:23:PRA}.





Accurate clock synchronization underpins a wide range of modern infrastructures, including navigation and communication systems, financial networks, and distributed scientific experiments \cite{EPGFBP:10:IEEE_J_UFFC, VNCB:19:IEEE_M_AES, AHCRA:24:ICSCA, CHKWR:10:PRL, N:15:NC, HSLTP:16:PRL, S:17:NPh, T:16:NPh, CHRW:10:Science, DL:13:ActaFut, L:17:JGeod, G:18:NP, B:15:GJI, K:16:PRD, D:17:PRL, DP:14:NP, AHVT:15:PRD, W:16:Nat_Astron, HGABW:16:PRL, R:17:NC}. Classical methods such as network time protocols \cite{M:88:RFC_1059, M:89:RFC_1119, M:92:RFC_1305} and precision time protocol mechanisms provide baseline synchronization across packet-switched computer networks, while \acp{GNSS} \cite{GGWZSGS:19:RS, CED:19:AAP} and two-way satellite time-and-frequency transfer extend timing distribution to global scales. Nevertheless, these classical techniques are approaching their fundamental performance limits, typically achieving only microsecond-to-nanosecond precision---insufficient for emerging applications demanding sub-picosecond accuracy, enhanced robustness, and resilience to interference \cite{B:25:SSRN, HLTLHCYLC:25:EPJQT,  MCGL:18:N, CAMP:17:N, CDESSBNS:23:N, NGBF:24:WCNC}.
Meanwhile, advances in optical and atomic clock science now reach fractional uncertainties at the $10^{-18}$ level, creating a growing mismatch between the performance of local clocks and the capabilities of existing synchronization channels. This performance gap has motivated a new paradigm such as \ac{QCS}, which leverages quantum resources---including entanglement, correlated photons, and quantum frequency combs---to surpass classical constraints on accuracy and security \cite{GM:25:QST, GAGLBM:24:APLP}. Experimental progress confirms that quantum correlations can directly enhance synchronization fidelity, with single-atom systems demonstrating entanglement-enabled phase locking and coherent control \cite{LAMK:20:PRL, GMVD:24:PRA}. Complementary advances in optomechanical and oscillator-network platforms further enable simultaneous control of phase synchronization and entanglement, providing operational pathways for connecting theoretical quantum synchronization criteria with experimental implementations \cite{MGG:13:SR, MFD:13:PRL}.

Beyond timing applications, recent developments in quantum positioning systems and networked quantum communication underscore the broader relevance of \ac{QCS}. Such analyses and network simulations illustrate the feasibility of quantum-enhanced positioning---particularly schemes employing entangled photon pairs---which offer secure and resilient solutions for location-aware tasks across terrestrial and non-terrestrial environments \cite{NGNFBF:23:EWC, DCS:21:IJMS, F:19:IOPCS}. Parallel efforts demonstrate the transition from theoretical models to real-world deployments through quantum-limited optical time transfer for geophysical networks and integration of \ac{QKD} within telecom bands \cite{HL:15:3PGCIC, CPP:24:CN, CDESSBNS:23:N}. Related milestones include high-rate generation and fiber-based transmission of entangled photon states in conventional telecom infrastructures, marking significant progress toward scalable quantum network architectures \cite{WEH:18:S, LXXQHL:23:COL}. These developments highlight an accelerating shift from classical timing frameworks to quantum-enabled synchronization, communication, and positioning technologies, reflecting their potential to deliver unprecedented synchronization fidelity, enhanced security, and improved resilience beyond what classical systems can achieve.



Recent progress in \ac{QCS} has been demonstrated primarily in fiber-based experiments employing correlated photon pairs. For example, correlated photons generated in a periodically poled lithium niobate waveguide have enabled synchronization over a 20-kilometer (km) fiber link with sub-picosecond stability using a common reference clock \cite{GFDSYWGZ:24:IEEE_CONF_OGC}. Temporal photon correlations have also been exploited to achieve synchronization jitter below 68\;picoseconds (ps) without external timing references, demonstrating feasibility under realistic high-loss conditions \cite{STSKCCDRRS:23:PRAppl}. Furthermore, two-way fiber-optic \ac{QCS} has been extended to 50\;km, reaching femtosecond-scale stability and accuracy, thereby highlighting the practicality of quantum-enhanced synchronization for intracity optical links \cite{HQXX:22:IEEE_J_JLT}.

International metrology initiatives, such as those led by the Consultative Committee for Time and Frequency, are charting roadmaps toward redefining the SI second using optical clock standards, which already surpass cesium-based realizations in precision. In parallel, recent research highlights the role of \ac{QCS} in emerging infrastructures. For instance, \ac{QCS} has been proposed as a complement to classical synchronization protocols in future \ac{6G} wireless systems and packet-based networks \cite{B:25:SSRN}. It has also been explored for satellite-based positioning and navigation, offering scalability and improved reliability for emerging \acp{GNSS} and non-terrestrial \ac{6G} components \cite{NRHIBMPWBDCF:24:ComNet}. Further work envisions satellite constellations equipped with quantum resources operating as a global master clock, where mutual synchronization through \ac{QCS} protocols enables sub-nanosecond timing distribution worldwide \cite{DATAH:25:PRAppl}. In these architectures, synchronized qubits act as carriers of highly precise timing signals with stabilities approaching $10^{15}$ oscillations per second \cite{NLPNF:23:IEEE_CONF_AERO}. Simulations have demonstrated that a constellation of approximately 50 \ac{LEO} satellites can realize such quantum-enhanced timing, distributing global time with sub-nanosecond precision \cite{DATAH:25:PRAppl}.

\begin{figure*}[t!]	
\centering
\includegraphics[width=0.82\textwidth]{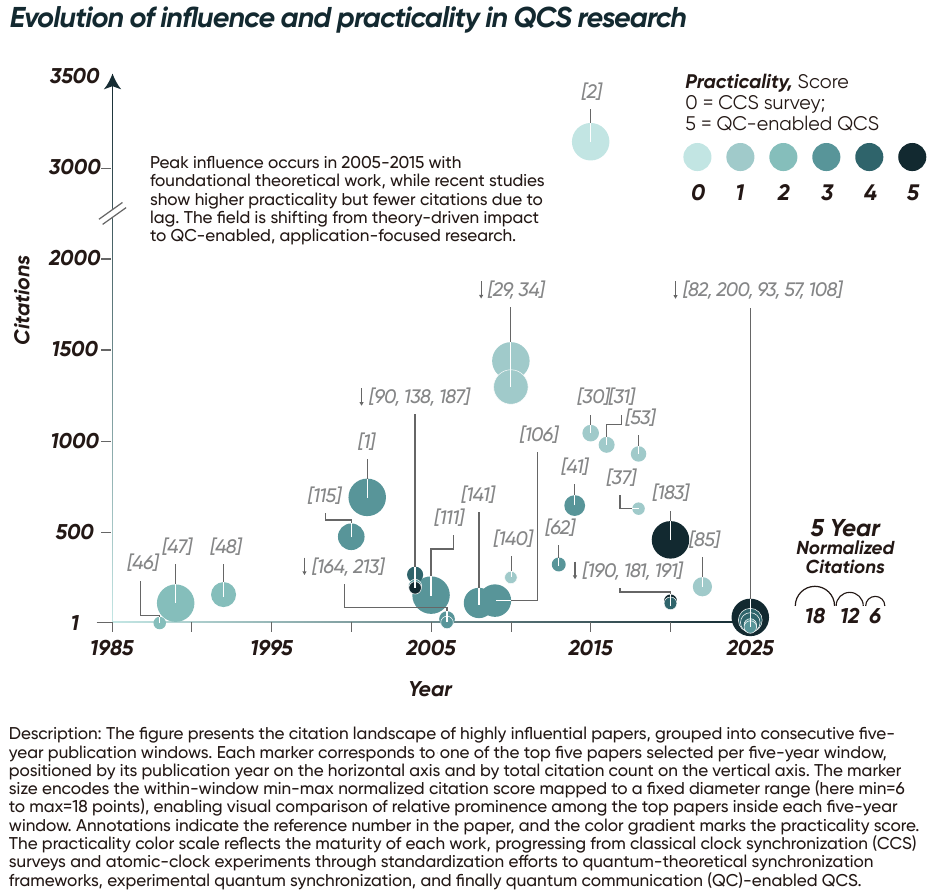}
\caption{ 
Evolution of QCS research, jointly illustrating scientific influence and deployment practicality from classical atomic clocks to QCS networks.
}
\label{fig:1}
\end{figure*}

Collectively, these developments underscore that \ac{QCS} is rapidly moving from theory to practical system-level deployment, positioning it as a core enabling technology for future global time standards, precision timekeeping, and secure synchronization within emerging quantum Internet infrastructures (see Fig.~\ref{fig:1}). This paper provides a comprehensive survey of the field of \ac{QCS}, spanning its fundamental principles, protocols, key resources, noise sources, security considerations, applications, and implementations. 
The main contributions of this survey are summarized as follows.
\begin{itemize}
\item We provide a unified overview of the fundamental principles and models of \ac{QCS}, including entanglement-assisted synchronization, \ac{HOM} interference, quantum Einstein synchronization, and slow clock transport. This helps clarify how uniquely quantum phenomena can be exploited to improve synchronization precision beyond classical limits.
\item We systematically classify and compare major \ac{QCS} protocols, including qubit-based handshakes, entanglement-assisted schemes, multiparty synchronization approaches, and \ac{HQC} techniques. This structured comparison clarifies their operational characteristics, performance trade-offs, and applicability of different synchronization protocols in quantum networks.
\item We review the key quantum resources and practical constraints that influence \ac{QCS} performance, including entangled photon sources, frequency combs, single photons, and quantum memories, together with noise sources such as dispersion, decoherence, and environmental disturbances. This analysis provides insight into practical implementation challenges and highlights potential strategies to mitigate synchronization errors.
\item We examine security aspects of \ac{QCS} protocols, including adversarial delay manipulation, intercept-based attacks, and vulnerabilities in distributed synchronization settings. This discussion highlights mechanisms required to ensure secure and reliable timing in quantum communication infrastructures.
\item We summarize emerging applications and experimental progress in \ac{QCS}, including use cases in \ac{QKD}, quantum teleportation, distributed sensing, relativistic geodesy, and the quantum Internet, along with demonstrations across fiber, satellite, and chip-integrated platforms. This overview highlights current capabilities and identifies open challenges, thereby outlining future research directions toward scalable quantum-synchronized networks.
\end{itemize}

The rest of the paper is organized as follows. Section~\ref{sec:2} introduces the fundamental concepts and theoretical background of \ac{QCS}. Section~\ref{sec:3} reviews representative synchronization \ac{QCS} protocols. Section~\ref{sec:4} discusses key enabling quantum resources, while Section~\ref{sec:5} analyzes practical noise sources and performance limitations. Section~\ref{sec:6} examines security considerations in \ac{QCS}. Section~\ref{sec:7} outlines major application domains, and Section~\ref{sec:8} reviews experimental implementations across different platforms. Section~\ref{sec:9} discusses open challenges and future research directions. Finally, Section~\ref{sec:10} concludes the survey.



\section{\ac{QCS} Fundamentals} \label{sec:2}

This section outlines the fundamental principles of \ac{QCS}, including the underlying quantum systems, entanglement-based mechanisms, the \ac{HOM} effect, and core models that define quantum and relativistic limits on achievable timing precision.

\subsection{Quantum Systems for Synchronization}
\begin{figure}[t!]	
\centering
\includegraphics{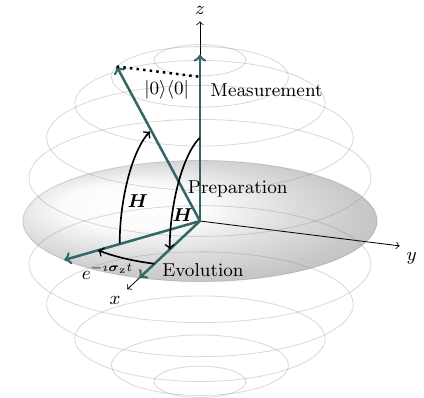}
\caption{Bloch-sphere representation of the \ac{QCS} protocol using a two-level quantum system, illustrating the stages of state preparation, evolution, and measurement.  The quantum system is initialized via the Hadamard gate $\M{H} = \left(\PX + \PZ\right)/\sqrt{2}$ to prepare the  superposition state $\ket{+}=\left(\ket{0}+\ket{1}\right)/\sqrt{2}$, where $\PX=\ket{0}\!\bra{1}+\ket{1}\!\bra{0}$ and  $\PZ=\ket{0}\!\bra{0}-\ket{1}\!\bra{1}$ are the Pauli-$\mathrm{x}$ and Pauli-$\mathrm{z}$ operators, respectively. During evolution, the system accumulates a phase that encodes the time information. The measurement is performed in the Hadamard basis, which is equivalent to applying a Hadamard gate followed by a computational-basis measurement.
}
\label{fig:2}
\end{figure}

Quantum systems serve as the foundational physical resources for implementing \ac{QCS} protocols, enabling the encoding, transmission, and extraction of temporal information at the quantum level. These systems are described by a state vector $\ket{\psi}$ in a complex Hilbert space $\mathscr{H}$, with their evolution governed by the Schr\"odinger equation and their measurement outcomes dictated by the probabilistic framework of quantum mechanics. Within this framework, \ac{QCS} protocols can be implemented in the simplest quantum mechanical system---namely, a two-level system---using a sequence of stages that includes preparation, evolution, and measurement, as illustrated in Fig.~\ref{fig:2}. The choice of quantum system fundamentally determines the capabilities and limitations of the synchronization protocol in terms of coherence time, control fidelity, scalability, and interaction with the environment.

The most commonly used systems for \ac{QCS} include photonic qubits and atomic ensembles, with photons being particularly preferred due to their intrinsic advantages \cite{RH:16:PRA,BGMP:16:PRA}. Photons exhibit long coherence times, can propagate over long distances with minimal decoherence, and are relatively immune to thermal fluctuations, making them ideal carriers of quantum information across distributed networks \cite{LLS:17:PRE}. Photonic qubits can be encoded in various degrees of freedom as follows.
 
\begin{itemize}

\item Polarization encoding uses horizontal and vertical polarization states as the basis.

\item Time-bin encoding represents qubits as early and late arrival times of a single photon. 

\item Frequency-bin encoding uses photons with different center frequencies to define the logical basis.

\item Spatial-mode encoding exploits different propagation modes or path degrees of freedom, such as those utilized in dual-rail systems.

\end{itemize} 

The manipulation of photonic qubits relies on linear optical components such as half-wave or quarter-wave plates, polarizing beam splitters, phase shifters, and Mach-Zehnder interferometers. The measurements are performed using \acp{SPD}, e.g., avalanche photodiodes or superconducting nanowire \acp{SPD}, and time-tagging electronics \cite{WRKRKH:20:RSI, THAK:22:arXiv}.

Quantum communication systems open new opportunities for synchronization by leveraging qubit states, coherence, and quantum correlations. In contrast to classical methods that depend on separate hardware-based timing references, qubit-based schemes can embed synchronization information directly into the transmitted quantum states. For example, in \ac{QKD} systems, subsets of the transmitted qubits have been used to achieve clock and pulse synchronization without the need for auxiliary channels, thereby reducing system complexity and cost \cite{SLCLDW:25:IEEE_J_COML}. Other approaches, such as the iQSync protocol, enable efficient clock-offset recovery on resource-constrained hardware platforms, demonstrating robustness even under high channel loss \cite{KWHF:25:PRAppl}. Experimental studies have further shown that quantum-based clock recovery methods can replace classical service channels in time-bin \ac{QKD} without degrading key generation performance \cite{ZRMRZGKOB:23:AVS_QSci}.
Experiments have further demonstrated robustness under emulated high-loss conditions, including atmospheric turbulence, indicating the practicality of such schemes for quantum communication networks \cite{STSKCCDRRS:23:PRAppl}. 
These results highlight how quantum optical resources can serve as the physical layer for emerging \ac{QCS}.

\subsection{Entanglement and Nonlocal Correlation}
Entanglement is a quintessential quantum resource that lies at the heart of many quantum information protocols by enabling nonlocal correlations between spatially separated systems. These correlations transcend the limitations of classical physics and serve as a basis for enhanced performance in various quantum tasks, particularly in the domain of precision metrology \cite{SKHD:16:PRL, CPAL:22:NP}. For instance, maximally entangled states, such as Bell and \ac{GHZ} states exhibit exceptional sensitivity to relative phase shifts. This heightened sensitivity enables measurement precision that surpasses the \ac{SQL}, achieving the Heisenberg limit \cite{TVHH:20:PRL}.
Phase shifts in entangled systems can arise from time delays, frequency detunings, or path-length differences, making them particularly valuable in applications involving timekeeping and synchronization. In \ac{QCS}, entangled states can be used to detect and correct for temporal misalignments between spatially separated clocks by translating relative phase information into timing offsets \cite{GLM:02:PRA, Sha:02:AQP, HS:12:CPB}.

Entanglement allows spatially separated parties to synchronize their time without relying on classical timing signals subject to stochastic delays. A key example is the use of entangled photon pairs generated through \ac{SPDC}, where the detection of one photon at a given location instantaneously projects its entangled counterpart---regardless of spatial separation---into a correlated quantum state due to the nonlocal nature of entanglement. When such pairs are subject to coincidence measurements, the joint statistics exhibit high-visibility interference fringes or \ac{HOM} dips. These features are highly sensitive to relative temporal delays and can be exploited to estimate time offsets between the detection events \cite{VSS:04:APL}.

Despite their advantages, achieving the Heisenberg limit often suffers from limitations due to assumptions of unbiased estimators and asymptotic conditions. Such challenges can be addressed by formulating a criterion that yields a binary decision on whether the accuracy lies within a specific threshold \cite{ZZM:13:PRA}.
Furthermore, entangled systems are inherently fragile and susceptible to decoherence, especially in noisy or lossy quantum channels. Loss and environmental noise degrade entanglement quality, thereby limiting the scalability of entanglement-based protocols. To address these challenges, techniques such as entanglement purification and quantum error correction are often employed to mitigate the decoherence effect and improve the fidelity of  entangled states \cite{OTD:18:npjQI}.

Photon correlations, whether from non-degenerate pairs generated in nonlinear crystals or from time-correlated emissions in quantum communication setups,  enable high-precision synchronization across fiber links. 
For example, frequency-bin entangled photons have been used to demonstrate nonlocal modulation cancellation, achieving sub-picosecond stability ($\sim$ 0.5\;ps) over 5.5\;km of fiber in a feedback-free design~\cite{CSLPMWL:25:OpticaQ}. 
Together, these results showcase the important role of both photon correlation and entanglement to provide powerful resources for enabling precise and scalable \ac{QCS}.

\subsection{\ac{HOM} Interference Effect}
\begin{figure}[t!]	
\centering
\includegraphics{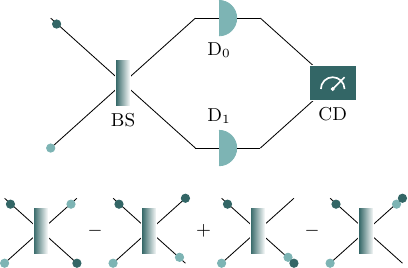}
\caption{\ac{HOM} interferometer. Two photons impinge on a $50:50$ beam splitter (BS) from separate input ports. Due to two-photon interference, both photons bunch and exit through the same output port if they are indistinguishable, leading to a suppression of coincident detections at detectors $\mathrm{D}_0$ and $\mathrm{D}_1$. The coincidence detector (CD) registers a dip in the coincidence count when the photons are perfectly indistinguishable in time, manifesting the \ac{HOM} interference effect.
}
\label{fig:3}
\end{figure}

The \ac{HOM} effect is a two-photon interference phenomenon that occurs when two indistinguishable photons enter a $50:50$ beam splitter from different separate ports, as illustrated in Fig.~\ref{fig:3}. If the photons are identical in all degrees of freedom---temporal, spectral, polarization, and spatial mode---the quantum amplitude for both photons to exit through different output ports interferes destructively, resulting in both photons exiting the same port \cite{BSZFM:20:RPP}. This results in a dip in the coincidence detection rate, known as the \ac{HOM} dip, as a function of the relative time delay between the photons. In the case of no delay, the coincidence probability reaches its minimum, indicating perfect two-photon interference. As the delay increases and the photons become distinguishable, the interference fades, and the coincidence rate returns to the classical level \cite{GWFS:24:NPJQI}.

The \ac{HOM} effect serves as a critical tool in quantum optics experiments, including indistinguishability characterization, photon source diagnostics, and quantum gate calibration \cite{HWWWZ:25:SA, HLHPD:24:PRXQ, JAL:22:PRA}. 
Beyond its fundamental interest, the \ac{HOM} effect plays a pivotal role in quantum optics and quantum information processing \cite{WLLW:19:SR}. It is widely used as a benchmark for assessing photon indistinguishability---a key resource for quantum communication and photonic quantum computing. High-visibility \ac{HOM} interference is essential for implementing linear optical quantum gates, such as the \ac{CNOT} gate \cite{CAMLG:08:PRL}. The effect also serves as a diagnostic tool for characterizing single-photon sources such as quantum dots and \ac{SPDC} systems \cite{YXG:19:SR}.
Moreover, due to its extreme sensitivity to temporal mismatches and optical path-length variations, the \ac{HOM} effect has been employed in precision metrology and time-delay interferometry \cite{TSTIMY:17:OE}. In \ac{QCS} protocols, particularly those involving photonic qubits, \ac{HOM}-type interference can be used to align clocks by identifying the temporal overlap condition that yields maximal interference.

The \ac{HOM} interferometer provides a sensitive method for detecting relative arrival-time differences between photons, making it a valuable tool for quantum synchronization. Recent experiments have synchronized independent photon sources to an external reference with sub-picosecond precision using \ac{HOM} interference, demonstrating its suitability for distributed quantum networks \cite{LBBJKGSP:24:OE}. In addition, \ac{HOM}-based synchronization has been applied in measurement-device-independent \ac{QKD}, where it enables clock alignment between users without requiring extra hardware, thereby reducing system complexity \cite{HLLWWYHCGH:25:IEEE_J_JLT}. These advances highlight \ac{HOM} interference as a practical mechanism for achieving high-precision synchronization in quantum communication settings.

\subsection{\ac{QCS} Models}
\begin{figure}[t!]	
\centering
\includegraphics{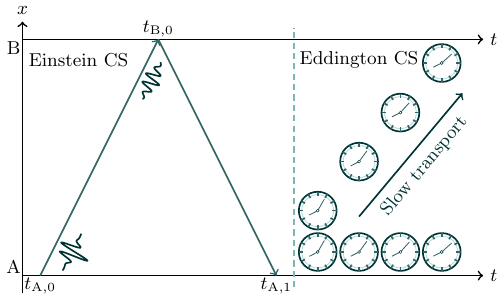}
\caption{Illustration of Einstein and Eddington synchronization methods. On the left, Einstein clock synchronization (CS) is depicted, where synchronization occurs via a two-way exchange of signals between spatially separated points $\mathrm{A}$ (say, Alice) and $\mathrm{B}$ (say, Bob). On the right, Eddington synchronization involves slow-clock transport, physically moving clocks along a defined trajectory at negligible speeds. The comparison highlights that, in the limit of infinitesimally slow transport, Eddington synchronization becomes operationally equivalent to Einstein’s signal-based synchronization.
}
\label{fig:4}
\end{figure}

\Ac{QCS} models aim to extend classical synchronization frameworks by leveraging quantum properties for time transfer or offset estimation. Two prominent models inspired by classical methods are the quantum analogs of Einstein synchronization and Eddington slow-clock transport, as illustrated in Fig.~\ref{fig:4}.

\subsubsection{Quantum Einstein Synchronization}

Quantum Einstein synchronization is the quantum adaptation of the classical Einstein synchronization protocol, where two parties exchange light pulses to determine the relative time offset. However, this method suffers from uncertainties due to variable transmission delays, particularly in the presence of atmospheric fluctuations, and its precision is fundamentally limited by the \ac{SQL} \cite{GLMW:01:PRL}. In the quantum version, entangled photon pairs or other quantum resources are used to perform synchronization exploiting quantum phenomena, such as interference and correlation to enhance the synchronization precision. By measuring the arrival times and classically correlating detection events, accounting for known channel delays, the relative clock offset can be computed \cite{GLM:02:PRA}. Notably, the entangled nature of the photon pairs allows the parties to estimate relative timing directly from quantum correlations \cite{HLK:09:NJP}.
While entanglement can surpass the \ac{SQL} and even reach the Heisenberg limit, entangled states are notoriously fragile---i.e., highly susceptible to decoherence and photon loss---and distributing entanglement across a channel requires complex and resource-intensive procedures \cite{GLM:01:N}.

Theoretical studies have also examined how classical synchronization notions extend into the quantum regime. Using Ramsey theory, synchronization procedures such as those of Einstein and Eddington can be represented as graph-coloring problems on lattices of clocks, where the transitivity of synchronization determines the resulting Ramsey structure. Unlike Einstein synchronization, which is transitive, both relativistic and quantum synchronization schemes lack transitivity, reflecting fundamental departures from classical assumptions \cite{B:25:Foundations}. More recently, operational models introduce privileged-frame frameworks that reconcile relativistic navigation with quantum requirements for global coherence. By applying a magnitude-matching selection rule to world-line data, these models identify a simultaneity frame that is empirically testable, thereby collapsing synchronizing freedom and grounding quantum synchronization in observable criteria \cite{H:25:SSRN}.

\subsubsection{Quantum Eddington Slow-Clock Transport}

In contrast with the Einstein protocol, the Eddington slow-clock transport method avoids transmission delays altogether by synchronizing a portable clock and adiabatically transporting it, assuming negligible relativistic effects during transport. This enables direct comparison between the transported clock and the local clock to determine the time offset.
The quantum analog of the Eddington protocol uses ticking qubits---nondegenerate two-level quantum systems evolving under a known Hamiltonian---to encode timing information \cite{YZWZ:12:PEOM}. This ticking behavior acts as an internal clock that carries timing information as the qubit propagates through a quantum channel. In the simplest one-way protocol, Alice prepares a superposition state and sends it to Bob. This state accumulates timing information in the form of a relative phase, which Bob can estimate from interference measurements \cite{Chu:00:PRL}. Notably, by coherently exchanging a single qubit multiple times, we can simulate the effect of a faster-ticking clock or an entangled state, enabling Heisenberg-limited scaling of synchronization precision \cite{BB:05:PRA}.

Beyond Einstein-type synchronization, researchers have also investigated protocols that resemble clock transport at the quantum level. A recent study demonstrated a  relativistic syntonization method, in which highly stable hydrogen quantum clocks were physically relocated while accounting for gravitational potential differences and second-order Doppler effects. The experiment achieved a relative error of $4.6 \times 10^{-16}$, outperforming \ac{GNSS}-based comparisons by more than an order of magnitude, and has been proposed as a foundation for building networks of quantum clocks \cite{FSK:23:MeasTech}. Although less developed than photon-exchange approaches, such relocation-based methods highlight an important complementary route to quantum synchronization.

\subsection{Performance Metrics for \ac{QCS}}
An important aspect of \ac{QCS} protocols is the definition of quantitative performance metrics for evaluating synchronization accuracy, stability, and implementation efficiency \cite{GM:25:QST, GAGLBM:24:APLP}. These metrics provide a common framework for comparing different \ac{QCS} protocols and assessing their suitability for deployment in quantum networks. A primary metric for evaluating \ac{QCS} performance is the residual time offset between spatially separated clocks after the synchronization protocol has been executed. This offset represents the estimation error of the relative clock phase or time difference and is typically extracted from photon arrival-time statistics, coincidence measurements, or interference patterns. The synchronization accuracy is often characterized by the variance or standard deviation of the estimated time offset, which depends on factors such as detector timing resolution, photon indistinguishability, and channel noise.

Beyond instantaneous accuracy, timing and frequency stability are critical for maintaining synchronization over extended periods. These properties are commonly quantified using statistical measures such as Allan deviation or time deviation, which characterize the temporal fluctuations of the clock offset as a function of the averaging time \cite{MGG:13:SR}. These metrics capture the impact of noise processes, including photon timing jitter, detector uncertainty, environmental perturbations, and channel fluctuations that may accumulate during long-distance quantum communication. For interferometric or entanglement-based \ac{QCS} protocols, an additional important metric is the interference visibility, which quantifies the contrast of two-photon interference patterns such as those arising from \ac{HOM} interference. High interference visibility indicates strong temporal indistinguishability of photons generated at remote nodes, which is essential for accurate estimation of the synchronization delay. Any degradation in visibility due to dispersion, spectral mismatch, or timing jitter directly reduces the precision of the synchronization process \cite{HWWWZ:25:SA, HLHPD:24:PRXQ}. \Ac{QCS} protocols also can be assessed in terms of their robustness to physical channel impairments, including optical attenuation, detector dark counts, background noise, and dispersion effects \cite{STSKCCDRRS:23:PRAppl}. Performance under these conditions is typically analyzed by examining how the synchronization error scales with channel loss or noise parameters. Such analysis is particularly important for evaluating the feasibility of \ac{QCS} deployment in large-scale quantum networks, where long-distance links and heterogeneous hardware platforms introduce additional sources of error.

\subsection{Quantum Limits for Time Synchronization}

The ultimate precision of \ac{QCS} is governed by quantum estimation theory, with the \ac{QCRB} setting the minimum variance achievable in time-offset estimation. Optical frequency-comb studies have shown that temporal-mode strategies can reduce timing deviations by factors of 2 to 10 relative to conventional intensity-based methods, provided that sufficient prior information about system parameters is available \cite{GM:25:QST}. In parallel, quantum frequency combs that exploit squeezing and entanglement have been proposed as a means to surpass the \ac{SQL}, offering performance unattainable by purely classical approaches \cite{GAGLBM:24:APLP}. Together, these insights delineate the quantum limits of synchronization and guide the design of resource-efficient \ac{QCS} protocols.

\subsection{Relativistic Considerations in QCS}

Relativistic effects such as gravitational redshift, motion-induced time dilation, and frame rotation must be accounted for in any high-precision synchronization scheme, whether classical or quantum. Recent studies have explicitly incorporated these considerations into \ac{QCS} models. In satellite navigation contexts, \ac{QCS} protocols based on multi-qubit architectures have been shown to reduce timing-related positioning errors to below one meter, demonstrating both scalability and suitability for small-satellite platforms \cite{NRHIBMPWBDCF:24:ComNet}. 
Furthermore, Ramsey-graph analyses reveal that while Einstein synchronization is transitive, quantum and general-relativistic synchronization are not, requiring graph-theoretic methods to model networks of entangled clocks \cite{B:25:Foundations}. Together, these efforts emphasize that \ac{QCS} must be formulated as an inherently relativistic problem, particularly for applications in space-based communication, navigation, and global timekeeping.

\section{\ac{QCS} Protocols} \label{sec:3}

This section outlines the principal \ac{QCS} protocols by describing their operational principles, performance characteristics, and suitability for different synchronization scenarios. Table~\ref{tab:QCS-Protocols} outlines a comparison of major \ac{QCS} protocols, focusing on their underlying principles, performance benchmarks, and key limitations.

\subsection{Ticking-Qubit Handshake Clock Synchronization}

The \ac{TQH} \ac{QCS} (\acs{TQH-QCS}) protocol basically follows the Eddington slow-clock transport protocol but employs a qubit as the clock \cite{Chu:00:PRL}. The qubit uses its superposition state and keeps time in the form of the relative phase between the computational basis $\ket{0}$ and $\ket{1}$. This phase evolves as the qubit goes in transit, and by measuring the qubit to extract its phase information, we can determine the time difference $\Delta t$ between spatially separated clocks. Due to the probabilistic nature of quantum measurements, multiple qubits must be used to obtain an accurate estimate of the phase and thus of $\Delta t$.

Let Alice and Bob synchronize their clocks using the \acs{TQH-QCS} protocol, outlined as follows \cite{Chu:00:PRL}.

\begin{enumerate}

\item 
\emph{Alice Transmission:}
Alice sends the prepared qubit in the state $\ket{\psi}$ and the tick rate $\omega$ at $t_\mathrm{A}$ to Bob. 
The transmitted qubit evolves as governed by the Hamiltonian $\M{\mathcal{H}}$ during transit unitarily as  
\begin{align}
	\ket{\psi \left(t\right)}
	=
	e^{-\imU \M{\mathcal{H}} t}
	\ket{\psi}
\end{align}
where $\imU=\sqrt{-1}$. For example, consider the qubit in the form of a spin-$1/2$ particle governed by the Hamiltonian $\M{\mathcal{H}} = \omega \PZ$. Then, during transit, the qubit accumulates a phase due to the unitary evolution $e^{-\imU \omega t_{\mathrm{AB}} \PZ}$ where $t_{\mathrm{AB}}$ is the time the qubit spends in transit. Upon receiving the qubit from Alice at $t_\mathrm{B}$, Bob applies the operation $\PX e^{\imU \omega \left(t_\mathrm{B} - t_\mathrm{A}\right) \PZ}$. 

\item 
\emph{Bob Transmission:}
At time $t_{\mathrm{B}'}$, Bob sends the resulting state $\PX e^{\imU \omega \Delta t \PZ} \ket{\psi}$. Similarly, the qubit accumulates the phase in transit by $e^{-\imU \omega t_{\mathrm{B}' \mathrm{A}'} \PZ}$ where $t_{\mathrm{B}' \mathrm{A}'}$ is the transit time from Bob to Alice. Upon receiving the qubit at $t_{\mathrm{A}'}$, Alice applies the operation $\PX e^{\imU \omega \left(t_{\mathrm{A}'} - t_{\mathrm{B}'} \right) \PZ}$. Using the identity
\begin{align}
	\PX e^{\imU \omega \PZ}
	=
	e^{-\imU \omega \PZ} \PX,
\end{align}
we then arrive at the \ac{QCS} state $e^{2 \imU \omega \Delta t \PZ} \ket{\psi}$.

\end{enumerate}

Since $\Delta t$ is encoded via the unitary evolution generated by $\PZ$, the prepared qubit $\ket{\psi}$ cannot be the eigenstate of $\PZ$. Otherwise, $\Delta t$ is encoded in the global phase of the state $\ket{\psi}$, making it unobservable. Instead, Alice can prepare a superposition state $\ket{\psi} = \ket{+}$ 
and get the \ac{QCS} state as follows:
\begin{align}
e^{2 \imU \omega \Delta t \PZ} \ket{\psi}
=
\frac{1}{\sqrt{2}}
\left(
	e^{2 \imU \omega \Delta t} \ket{0}
	+
	e^{-2 \imU \omega \Delta t} \ket{1}
\right)
\label{eq:TE}
\end{align}
where $\ket{\pm}=\left(\ket{0} \pm\ket{1}\right)/\sqrt{2}$ are the Hadamard-basis states. Measuring this state directly still does not resolve the encoded $\Delta t$ as the probabilities of the measurement outcomes are determined by the squared norm of the probability amplitudes. 
Hence, Alice performs the Hadamard operation $\M{H}$ on the \ac{QCS} state \eqref{eq:TE} to interfere with the probability amplitudes, leading to the final \ac{QCS} state as $\cos \left(2\omega \Delta t\right) \ket{0} + \imU \sin \left(2 \omega \Delta t\right) \ket{1}$.

Upon measurements, Alice obtains the state $\ket{0}$ with probability $\cos^2\left(2\omega \Delta t\right)$. However, estimating this probability with $n$-bit precision requires $2^{2n}$ qubits, which is inefficient due to its exponential growth with respect to the number of digits.
This inefficiency can be tackled using quantum phase estimation to determine $n$ bits of $\Delta t$ requiring only $O\left(n\right)$ transmitted qubits---an exponential advantage \cite{CEMM:98:PRSAMPE}. This advantage comes at the cost of $m+1$ additional qubits for data processing and a non-trivial quantum operation that introduces intricate entanglement among these extra qubits. Furthermore, quantum phase estimation relies on the inverse \ac{QFT} that requires $O\left(2^n\right)$ elementary single- and two-qubit operations. If $2^m \omega \Delta t$ is an integer, we can get $\omega \Delta t$ with probability $1$. Otherwise, it can be estimated to $n$ bits of accuracy with a success probability of at least $1-\epsilon$ by using $m = n + \left\lceil\log_2\left(2 + 1/\left(2 \epsilon\right) \right)\right\rceil$ extra qubits where $\left\lceil x \right\rceil$ denotes the smallest integer greater than or equal to $x$. 
Despite this improvement, certain challenges remain. In particular, the tick rate $\omega$ must cover an exponentially large range while maintaining sufficient stability to prevent shifts smaller than half a wavelength.
A further linear improvement in the number of transmitted qubits can be achieved, as proven in \cite{HK:01:AQP} that the product of the range of $\omega$ and the number of transmitted qubits scales as $\Omega\left(2^n\right)$.

\begin{figure}[t!]	
\centering
\includegraphics{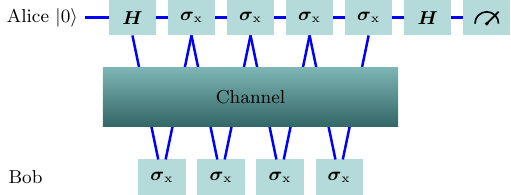}
\caption{
Illustration of the \acs{TQH-QCS} protocol. A single qubit is exchanged multiple times between Alice and Bob, with each party applying a sequence of $\PX$ operations during each pass through the channel. The accumulated phase encodes the time difference between their clocks, which Alice extracts through a final Hadamard rotation and subsequent measurement.
}
\label{fig:5}
\end{figure}

Instead of relying on entanglement to increase synchronization accuracy, it was proposed in \cite{BB:05:PRA} to leverage additional communication resources. In this protocol, Alice and Bob engage in bidirectional communication, repeatedly exchanging a single qubit to accumulate phase information, as shown in Fig.~\ref{fig:5}. This approach achieves a near-quadratic improvement in synchronization accuracy compared to the \ac{SQL}, as the synchronization uncertainty scales as $O\left(\log_2 N_\mathrm{e} /N_\mathrm{e}\right)$ where $N_\mathrm{e}$ is the number of qubit exchanges. Notably, this enhancement does not depend on entanglement but instead arises from the increased complexity of coherent communication.

\subsection{Entanglement-Based Clock Synchronization} 

The \aca{EB-QCS} (\ac{EB-QCS}) protocol requires Alice and Bob to share a prior entangled quantum state \cite{JADW:00:PRL}. This entangled state requires not to evolve in time. Alice and Bob acquire their synchronized clock by means of quantum measurements and classical communication without exchanging relative location and timing information. Consequently, this protocol is insensitive to the transmission medium.

The \ac{EB-QCS} protocol starts by sharing an entangled state between Alice and Bob in the form of a singlet state:
\begin{align}
\ket{\M{\psi}}
=
\frac{1}{\sqrt{2}}
\left(
	\ket{01}_\mathrm{AB} 
	-
	\ket{10}_\mathrm{AB}
\right).
\end{align}
Since both parties experience identical time evolution, the singlet state accumulates only an unobservable global phase. Hence, the singlet state effectively remains unchanged over time.
Alternatively, the singlet state can also be expressed as
\begin{align}
\ket{\M{\psi}}
=
\frac{1}{\sqrt{2}}
\left(
	\ket{+-}_\mathrm{AB}
	-
	\ket{-+}_\mathrm{AB}
\right).
\end{align}
From this representation, we can directly observe the post-measurement state in the Hadamard basis $\ket{\pm}$, which collapses to either $\ket{+-}_\mathrm{AB}$ or $\ket{-+}_\mathrm{AB}$ with equal probability. 
The post-measurement state evolves over time due to the operation $e^{-\imU \omega t \PZ}$. For the state $\ket{+}$, the time evolution follows the same form as in \eqref{eq:TE}, with $\Delta t$ replaced by $t/2$. For the state $\ket{-}$, an additional substitution $\ket{1} \rightarrow -\ket{1}$ is required. Followed by the Hadamard operation and subsequent measurement in the computational basis $\left\{\ket{0},\ket{1}\right\}$, Alice and Bob can resolve the time $t$ by inferring from the probability of obtaining the state $\ket{0}$. 
Since Alice’s measurement on the singlet state produces a probabilistic outcome, she announces her measurement results to Bob using classical communication. This allows Bob to determine the probability of obtaining the state $\ket{0}$ (after the Hadamard operation $\M{H}$) for each possible post-measurement state of the singlet state.

The \ac{EB-QCS} protocol faces challenges in sharing the prior entangled quantum state and tackling the scenario of relatively moving parties. The shared prior entanglement can be established by generating the entangled state locally and then distributing it between Alice and Bob before they separate. However, Alice and Bob can synchronize their clocks locally and carry them slowly, as in the Eddington \ac{QCS} protocol. 
This protocol also assumes that the spatial reference frames of the parties are synchronized, which does not necessarily hold in real scenarios. The misalignment of spatial reference frames between Alice and Bob introduces distortions in the measurement probabilities, leading to incorrect time inference. However, this issue can be addressed by modifying the protocol as proposed in \cite{LLT:03:CTP}, where time and space can be synchronized simultaneously using an ensemble of singlet states. This protocol relies on the fact that Pauli operators associated with each party can be related through their relative orientations, characterized by the polar and azimuthal angles $\theta$ and $\phi$, respectively.
Instead of measuring the singlet states solely in the Hadamard basis, Alice also performs measurements in the computational basis with equal probability. After announcing measurement results to Bob, he can post-select the collapsed states $\ket{-}_\mathrm{B}$ (or $\ket{0}_\mathrm{B}$) and calculate the expectation values of the Pauli operator $\PZ$ (or $\PY=\imU \PX \PZ$). The expected values are given by
\begin{align}
\braket{-|\PZ|-}_\mathrm{B} 
&=
\sin \theta  
\cos \omega t \\  
\braket{0|\PY|0}_\mathrm{B} 
&=
\sin \theta
\sin \phi.
\end{align}
Hence, calculating the expectation values of the operator $\PZ$ with respect to the collapsed states $\ket{-}_\mathrm{B}$ at two-time instances as well as the expectation value of the operator $\PY$ with respect to the collapsed state $\ket{0}_\mathrm{B}$ can determine the time $t$, polar angle $\theta$, and azimuth angle $\phi$.

Alternatively, Bob can first synchronize the space by calculating the expectation values of $\PZ$ and $\PX$ with respect to the collapsed states $\ket{1}_\mathrm{B}$ and $\ket{0}_\mathrm{B}$, respectively. These expectation values take a simpler form as follows: 
\begin{align}
\braket{1|\PZ|1}_\mathrm{B} 
&=
\cos \theta \\
\braket{0|\PX|0}_\mathrm{B} 
&=
\sin \theta 
\cos \phi.
\end{align}
Using this information, Bob can align his spatial orientation with Alice's using rotation operators. Once the space synchronization is achieved, he then proceeds with the original clock synchronization protocol.
Although this method addresses the issue of spatial synchronization, it still faces the fundamental challenges associated with singlet-based protocols, particularly the distribution of singlet states.

Entanglement provides a nonclassical resource for directly correlating time between distant parties. In \ac{EB-QCS}, the shared phase coherence between entangled photons enables clock alignment without transmitting dedicated synchronization signals. A metropolitan field trial has demonstrated this approach in a 50-km energy–time entangled \ac{QKD} link, where clocks were synchronized entirely through entangled photon detection, enabling secure and autonomous operation \cite{PSCLAMT:23:PRAppl}. 
Taken together, these results show that both entanglement and photon correlations serve as powerful resources for practical \ac{QCS} implementations.

\subsection{Multiparty \ac{EB-QCS}} 

The extension of \ac{EB-QCS} to a multiparty \ac{QCS} protocol was proposed in \cite{KP:02:PRA}, aiming to synchronize $\nu$ spatially separated clocks. 
This protocol utilizes a $\nu$-partite entangled state shared among participants, where each party holds a single qubit. The state is chosen to exhibit nonzero entanglement between each pair of qubits, which can be realized using the symmetric entangled W state:
\begin{align}
\ket{\M{W}_\nu}
=
\frac{1}{\sqrt{\nu}}
\sum_{k=0}^{\nu-1}
\ket{\M{k}_\nu}
\end{align}
where $\ket{\M{k}_\nu}$ denotes the $\nu$-qubit state such that only the $k$th qubit is in the state $\ket{1}$ and all others are in the state $\ket{0}$ (e.g., $\ket{\M{1}_3}=\ket{100}$).
Then, a designated party, say Alice, which serves as a master clock, measures
the W state in the Hadamard basis and broadcasts the measurement outcome using classical communication. This operation leaves all participants (say, Bobs) with a collapsed state that evolves over time. By subsequently measuring their respective qubits in the Hadamard basis, the other participants can infer how long their qubits have evolved and adjust their clocks to the master clock. Assuming the chosen party (Alice) obtains the state $\ket{+}_\mathrm{A}$ upon measurement, the probability for the other participants (Bobs) to measure $\ket{\pm}_\mathrm{B}$ is given by
\begin{align}
P_{\ket{\pm}}
=
\frac{1}{2}
\pm
\frac{1}{\nu}
\cos \omega t.
\label{WS}
\end{align}
As the number $\nu$ of participants increases, the accuracy of estimating the time $t$ based on the measurement probability decreases. This decline stems from the fact that the pairwise entanglement between qubits diminishes as $\nu$ increases.
Therefore, finding an alternative entangled state that does not exhibit this limitation can enhance the \ac{QCS} accuracy.

Maximally entangled \ac{GHZ} states have been proposed for multiparty clock synchronization in \cite{RH:12:PRA,RH:14:QCMC}. 
To ensure that distribution times do not affect the state evolution, the bit-flip operator $\PX$ is applied to half of the total qubits in the \ac{GHZ} state. The resulting \ac{GHZ}-type state for even $\nu$  is 
\begin{align}
\ket{\M{\psi}}
=
\frac{1}{\sqrt{2}}
\left(
	\ket{0}^{\otimes \nu/2} \ket{1}^{\otimes \nu/2}
	+
	\ket{1}^{\otimes \nu/2} \ket{0}^{\otimes \nu/2}
\right).
\end{align} 
where $\otimes$ denotes the tensor product.
By randomly distributing the flipped and unflipped qubits among the participants and collecting measurement outcomes from all parties, each participant can infer the difference between their local clock and the average time of all clocks. Consequently, each participant can adjust their clock to synchronize with the average time.
A main feature of this protocol is that its synchronization accuracy remains independent of the number of participants. Furthermore, this approach significantly reduces the number of required qubits. Specifically, it employs only half the number of qubits used in the pairwise multiparty protocol of \cite{JADW:00:PRL} and merely one-fourth of those required by another multiparty protocol achieving comparable accuracy in \cite{BE:11:PRA}.

These multiparty \ac{EB-QCS} protocols rely on a shared entangled state that remains invariant under natural time evolution---that is, it is an energy eigenstate of the evolution operator. The time evolution begins once one party measures its respective qubit, thereby imprinting time information into the state upon measurement. A more general formulation of such protocols is provided by operation-triggered \ac{QCS}, where time information is encoded into a shared entangled state following the execution of a prescribed operation by each party involved in the synchronization process \cite{YZF:15:PRA}. This protocol also depends on a shared entangled eigenstate to prevent time information from leaking into the system during the distribution stage.
Once the entangled state is distributed, each party performs a local unitary operation $\PX$ on its respective qubit. This operation encodes the local clock time of each party into the shared quantum state. The operator $\PX$ flips the qubit state, thereby triggering time-dependent evolution. The initial shared state is designed so that once all nodes have performed their respective operations, the collective state returns to an invariant form. This re-invariance is essential, as it allows the state to retain the encoded time information without further evolution.
Following the local operations, the resulting state is transmitted back to a central node, where global measurements---designed by quantum metrology techniques---are performed to extract the encoded local times. The global measurement outcomes are then used to reconstruct the time offsets of all participating parties. These offsets are subsequently announced back to the parties via classical communication, thereby completing the synchronization process.

Extending \ac{EB-QCS} protocols to multiparty settings enables simultaneous alignment of multiple clocks within a network. From a theoretical standpoint, recent work has established that genuine multipartite entanglement is necessary to achieve a quantum advantage in networked synchronization. While local correlations alone cannot reach the Heisenberg limit, a probabilistic protocol has been shown to attain this ultimate bound when successful, offering both enhanced precision and inherent privacy guarantees \cite{YYX:24:PRL}. Simulation studies have examined the use of \ac{GHZ} states to synchronize atomic clock qubits interconnected by optical fiber, demonstrating feasibility for large-scale infrastructures such as energy-science networks and highlighting the resource requirements in terms of qubit numbers and operation times \cite{YBGKR:24:IEEE_CONF_QCNC}. These findings position \ac{EB-QCS} as a promising pathway toward network-wide synchronization for future quantum Internet applications.

\subsection{Time-of-Arrival Clock Synchronization}

The \ac{ToA} \ac{QCS} (\acs{ToA-QCS}) protocol aims to enhance clock synchronization that utilizes the arrival time of the transmitted photons by leveraging the quantum properties of pulsed photons \cite{GLM:01:N}. 
In this protocol, the accuracy of \ac{ToA} measurement is determined by the number of photons per pulse, the number of pulses, and the bandwidth.
Using only classical resources, the achievable accuracy scales as $\sqrt{M}$ with $M$ pulses and as $\sqrt{N}$ with $N$ average photons. However, using $M$ frequency-entangled pulses with the same bandwidth, it is possible to improve the accuracy by a factor of $\sqrt{M}$ by measuring the correlations between \acp{ToA}. This enhancement arises because the entanglement reduces the uncertainty in the average arrival time across $M$ pulses. Furthermore, employing number-squeezed states for each pulse can improve the accuracy by an additional factor of $\sqrt{N}$ as compared to that achieved by the coherent states with the average photon number $N$. Thus, by combining frequency entanglement and squeezed states, the \ac{ToA-QCS} protocol can deliver a total enhancement of $\sqrt{MN}$ in synchronization accuracy.

Despite its promising performance advantages, this protocol faces practical challenges in implementation. Generating and maintaining the required entangled pulses, especially over long distances in noisy environments, poses significant difficulties. Notably, entangled states are highly sensitive to photon loss: if one or more photons are lost, the timing information cannot be reliably resolved from the remaining photons. For two pulses, a parametric down-converter can be utilized to generate the required entangled pulses. However, entangling more than two pulses, as required, is still practically challenging.


\begin{figure}[t!]	
\centering
\includegraphics{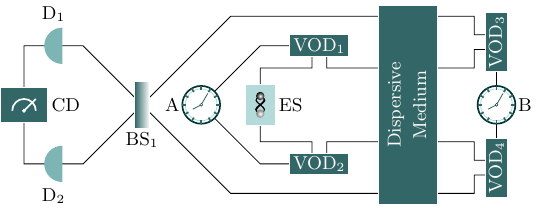}
\caption{
Illustration of the conveyor-belt \ac{QCS} protocol. The entanglement source (ES) generates entangled photons that are exchanged between Alice and Bob, with each photon passing through variable optical delay (VOD) modules and a dispersive channel. The photons are modulated according to each party’s local clock, then recombined at Alice’s station using the beam splitter (BS), and detected at detectors $\mathrm{D}_1$ and $\mathrm{D}_2$, where the relative clock offset is extracted from the coincidence statistics obtained by the coincidence detector (CD).
}
\label{fig:6}
\end{figure}

\subsection{Conveyor-Belt Clock Synchronization} 

The conveyor-belt \ac{QCS} circumvents the need for the arrival times of transmitted signals to synchronize distant clocks, making it robust against dispersion in the transmission channel \cite{GLM:04:PRA}. This protocol can be implemented using classical coherent-state pulses, with performance further enhanced by quantum resources such as frequency-entangled photons.
The conveyor-belt protocol requires two parties, Alice and Bob, to exchange signals over a medium where the transit times in both directions are equal (see Fig.~\ref{fig:6}). Each party operates nearly perfect clocks with negligible relative drift over the signal round-trip time. The implementation involves Alice and Bob modulating polarized-laser pulses according to their respective clock readings. The resulting interference pattern observed by Alice reveals the time difference between their clocks.

A key advantage of the conveyor-belt protocol is its immunity to dispersion, a common challenge in synchronization methods that rely on \ac{ToA} measurements. This robustness arises from its structure, which requires only amplitude measurements of the returned signals rather than their exact arrival times. Additionally, quantum resources can further enhance the protocol performance. In particular, frequency-entangled pulses enable quantum dispersion cancellation, allowing the protocol to remain resilient against a broad class of dispersive effects compared to purely classical implementations.
Beyond point-to-point schemes, recent work has explored quantum synchronization protocols in more complex network scenarios. Reviews of classical and quantum methods highlight how entanglement and correlated-photon techniques, including time-correlated entangled photons and time-bin encoding, could extend synchronization capabilities far beyond current classical limits, with potential precision gains of several orders of magnitude \cite{B:25:SSRN}. Complementary simulation studies have modeled networks of quantum clocks interconnected by optical fiber, using \ac{GHZ} states of atomic clock qubits to investigate the resource requirements for multi-node synchronization \cite{YBGKR:24:IEEE_CONF_QCNC}. While these approaches remain largely theoretical, they provide important insights into how \ac{QCS} can be scaled to distributed architectures.

\subsection{\ac{HOM} Clock Synchronization}

The \ac{HOM} \ac{QCS} (\acs{HOM-QCS}) protocol relies on the \ac{ToA} correlation of entangled photons to determine the clock offset between Alice and Bob \cite{BG:04:AIP}. Unlike traditional methods, this protocol does not require knowledge of geometric distances between the clocks and the presence of an optical medium, making it suitable for space-based applications.
The \ac{HOM-QCS} protocol assumes that the entangled photon source and the clocks held by Alice and Bob are all at rest. Furthermore, the \ac{HOM} interferometer is assumed to be spatially co-located with the entangled-photon source, and an adjustable refractive index exists between the source and Bob.

\begin{figure}[t!]	
\centering
\includegraphics{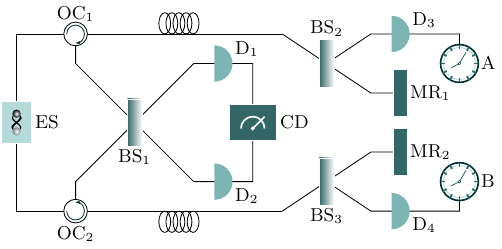}
\caption{
Illustration of the \acs{HOM-QCS} protocol. Entangled photons from the source propagate to Alice and Bob via optical circulators (OC) and delay lines. After reflection, the photons enter the \ac{HOM} interferometer to measure its coincidence rate. When the \ac{HOM} dip is minimized, the forward-and-backward transit times are balanced, allowing Alice and Bob to calibrate their clocks using the subsequent time-tagged photon arrivals at detectors $\mathrm{D}_3$ and $\mathrm{D}_4$.
}
\label{fig:7}
\end{figure}

The protocol starts by tuning the transit times of entangled photons traveling from the source to Alice and Bob. The emitted entangled photons are reflected back to the source by both parties, where the coincidence rate of these returned photons is then observed in the \ac{HOM} interferometer, as shown in Fig.~\ref{fig:7}. This procedure is repeated while Bob adjusts the refractive index in his optical path until the coincidence rate reaches a minimum. This minimum indicates that the \ac{HOM} interferometer is balanced, implying that the transit times from the source to Alice and Bob are identical.
Once the interferometer is balanced, a set of entangled photons are transmitted to Alice and Bob, who record the arrival times of their respective photon pairs as $t_{\mathrm{A},i}$ and $t_{\mathrm{B},i}$, $i=1,2,\ldots,N$, respectively.
Let $\Delta t_\mathrm{A}$ and $\Delta t_\mathrm{B}$ be the clock offsets of Alice and Bob relative to a global clock, respectively. The relationship between their photon arrival times is given by
\begin{align}
t_{\mathrm{B},i}-t_{\mathrm{A},i}
=
\Delta t_\mathrm{A}-\Delta t_\mathrm{B}.
\end{align}
Alice and Bob generate functions $f_\mathrm{A}\left(\tau\right)$ and $f_\mathrm{B}\left(\tau\right)$, respectively, as follows:
\begin{align}
f_\mathrm{A}\left(\tau\right)
&=
\frac{1}{\sqrt{N}}
\sum_{i=1}^N
\delta
\left(
	\tau - t_{\mathrm{A},i}
\right)\\
f_\mathrm{B}\left(\tau\right)
&=
\frac{1}{\sqrt{N}}
\sum_{i=1}^N
\delta
\left(
	\tau - t_{\mathrm{B},i}
\right)
\end{align}
where $\delta\left(x\right)$ denotes the Dirac delta function. Alice then sends her function $f_\mathrm{A}\left(\tau\right)$ to Bob using classical communication. Upon receiving it, Bob calculates the correlation function: 
\begin{align}
g\left(t\right)
&=
\int_{-\infty}^{+\infty}
f_\mathrm{A}\left(\tau\right)f_\mathrm{B}\left(\tau-t\right)
dt\\
&=
\frac{1}{N}
\sum_{i=1}^N
\sum_{j=1}^N
\delta
\left(
	t - t_{\mathrm{A},i} + t_{\mathrm{B},j}
\right).	
\end{align}
This correlation function attains its peak at $t_0 = t_{\mathrm{A},i} - t_{\mathrm{B},i} = \Delta t_\mathrm{B}-\Delta t_\mathrm{A}$. Finally, Bob synchronizes his clock with Alice's by adding $t_0$ to his local time $t_{\mathrm{B}}$.

\subsection{Time-Offset Clock Synchronization}

The time-offset \ac{QCS} protocol utilizes the strong temporal correlation between entangled photon pairs to enable clock synchronization among remote parties \cite{HLK:09:NJP}. This protocol uses the cross-correlation of detection time stamps to find both the time offset and frequency difference between two clocks. By identifying coincidences in photon detection events at Alice and Bob, the protocol can effectively track the relative drift between their clocks.
When the two clocks operate at the same frequency, the cross-correlation of their recorded detection times reveals a time shift that aligns the photon detection events. The cross-correlation function is efficiently computed using the \ac{FFT}, ensuring high precision in determining the time offset even in noisy environments.
The time offset is determined by searching for the peak in the cross-correlation function, which corresponds to the maximum number of coincident photon events. The temporal resolution can be progressively refined by increasing the precision of detection time data. To ensure robust peak identification, a statistical significance test is also applied to distinguish true coincidences from background noise. 

If two clocks have slightly different frequencies, the photon-pair detection events remain correlated, but the time difference between detections varies linearly due to the frequency drift. The basic time-offset protocol can be extended to account for these frequency differences by modeling the detection-time relationship as a linear function of time.
Using a series of measurements at different times, the protocol can estimate both the time offset and the frequency difference between the two clocks. This process can handle clock drifts on the order of parts per million (ppm), thereby enabling the use of less stable crystal oscillators in the system.

Similarly, the \ac{QCS} protocol with correlated photons presented in \cite{STSKCCDRRS:23:PRAppl} synchronizes remote clocks by utilizing the time tags of entangled photon arrivals. This protocol first reduces the frequency difference between the clocks using a coarse frequency-skew compensation procedure. After this frequency-matching process, the time offset between the clocks is estimated by using the cross-correlation function of the time tags with the \ac{FFT}. We can further use a more computationally efficient method by calculating the cross-correlation via the start-stop method. This method represents the cross-correlation by measuring the time differences between the counters that: (i) start when the first photon arrives at Alice (or Bob) and (ii) stop when the second photon arrives at Bob (or Alice). However, this protocol requires a fairly good initial estimate of the timing offset. In a long communication-session scenario, the correlation peak is continuously traced, and a fast feedback loop is employed to mitigate clock frequency fluctuations.

The modified version of the time-offset QCS protocol \cite{HLK:09:NJP} was proposed in \cite{QDXLLZ:20:RSI} to improve the resolution and precision of peak coincidence detection. By exploiting the sparse nature of entangled photon states, the cross-correlation can be directly computed by counting time-sequence differences within a specified time window, thereby eliminating the need for \ac{FFT} and its inverse operations. The modified time-offset protocol first extracts a coarse estimate of the peak position. Then, using this value as a reference, a precise peak location is determined within a finer time window. This protocol enables sub-picosecond precision but depends highly on detector settings, such as timing jitter and detection rates.

Protocols in this class estimate timing offsets directly from correlated detection events. One approach embeds a low-power, low-frequency synchronization beacon into the quantum channel, which is detected with the same \acp{SPD} used for quantum signals. This method achieves timing jitter close to the system resolution limit with negligible channel impact, making it suitable for long-distance links subject to large clock drifts, including satellite scenarios \cite{LZM:23:ApplOpt}. Complementing this approach, hardware-efficient schemes such as iQSync enable precise clock-offset recovery on resource-constrained platforms, including field-programmable gate arrays and microcontrollers. Experimental demonstrations showed that iQSync maintains accurate synchronization within seconds (s), even under severe channel-loss conditions exceeding 70\;decibels (dB), extending the practicality of offset protocols to \ac{QKD} systems \cite{KWHF:25:PRAppl}.

\subsection{Two-Way Time-Transfer Clock Synchronization}

The nonlocal time-correlated characteristics of frequency-entangled photon pairs provide synchronization accuracy of two-way time-transfer protocols that surpass classical methods, achieving femtosecond (fs)-level precision. The quantum approach is inherently immune to variations in path length, as entanglement ensures precise correlation between detection events of the paired photons. When a signal photon is detected at a specific time, the corresponding idler photon can only be detected at a uniquely determined time due to the strict quantum correlations in the frequency-entangled state \cite{HDLZ:17:QIM, HQDXL:19:PRA}. The quantum-enhanced two-way time-transfer protocol is based on measurements of the fourth-order time-correlation function of the entangled photon pairs, and its synchronization accuracy depends critically on the bandwidth of the biphoton spectrum.

\begin{figure}[t!]	
\centering
\includegraphics{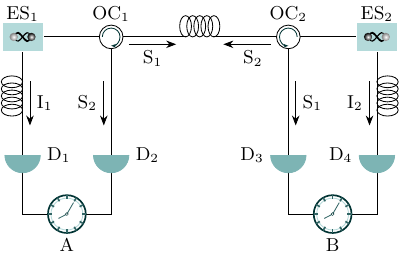}
\caption{
Illustration of the quantum-enhanced two-way time-transfer protocol. Entangled photon pairs are split by optical circulators, sending the signal photons to the opposite party while the idler photons remain local. Alice and Bob detect their respective signal and idler photons using SPDs and time-tagging modules synchronized to their local clocks. The relative clock offset is extracted by comparing the time correlations of the photon pairs.
}
\label{fig:8}
\end{figure}

The protocol requires Alice and Bob to have entangled photon pairs generated by \ac{SPDC}, which are frequency anti-correlated. Alice sends her signal photons to Bob while retaining the idler photons locally (see Fig.~\ref{fig:8}). Both Alice and Bob utilize \acp{SPD} and event timers synchronized to their respective local clocks to record the arrival times of the signal and idler photons. Alice determines the time difference between the registered signal and idler photons by finding the maximum coincidences in the photon arrival-time records. This measured time difference encodes the relative time offset between the local clocks of Alice and Bob, as well as the transit time of the photons. A similar procedure is followed by Bob, who sends his signal photons to Alice while retaining the idler photons. Bob also records the arrival times of the signal and idler photons to calculate the time difference. By comparing the time differences measured by both parties, the relative clock offset between the two parties can be accurately determined. Importantly, this approach inherently cancels out the effects of the photon transit time.


\subsection{Quantum Frequency Synchronization}

Quantum frequency synchronization is crucial for maintaining the stability of quantum clocks by aligning the oscillation frequency of their \acp{LO}. In centralized quantum networks, this can be achieved by stabilizing all \acp{LO}  using a shared reference clock \cite{KKBJSYL:14:NP}. However, this approach suffers from drawbacks such as vulnerability to network-wide failures caused by single-node disruptions and increased susceptibility to quantum noise. To overcome these challenges, distributed quantum networks have been proposed in \cite{UUS:20:QIP} for oscillator syntonization, allowing each node to adjust its \ac{LO} frequency independently using local quantum resources while maintaining overall network synchronization.
In this model, each node consists of clock qubits and syntonization particles. The clock qubits are prepared in cascaded \ac{GHZ} states, which undergo periodic interrogation to stabilize \acp{LO}. The syntonization particles serve two primary functions: (i) detecting and estimating systematic frequency shifts, and (ii) performing \ac{LO} resyntonization by cooperatively utilizing quantum resources shared among neighboring nodes. A key advantage of this architecture is that new clocks can be seamlessly integrated into the network without requiring a full restart of the synchronization process. This \emph{plug-and-syntonize} feature enables scalable and flexible quantum clock networks.

The quantum frequency synchronization process is carried out in two main steps: \emph{desyntonization detection} and \emph{\ac{LO} resyntonization}. The desyntonization detection step identifies nodes experiencing systematic frequency shifts due to external factors such as gravitational effects, Zeeman shifts, or Stark shifts. This task is accomplished by distributing singlet states between neighboring nodes and measuring their purity. If the local qubit noise is below a prescribed threshold, the node is classified as a branch node, capable of integrating new clocks and participating in network syntonization. If the noise exceeds a specified level, the node is designated as a leaf node, which can receive synchronization corrections but cannot perform syntonization independently.
Once a problematic node has been identified, the \ac{LO} resyntonization step is initiated. This task involves estimating the magnitude of the frequency shift and applying correction protocols to bring the affected \ac{LO} back into alignment with the network standard. The resyntonization process employs \ac{GHZ} states, which enhance sensitivity to frequency deviations by accumulating phase shifts proportional to the number of entangled qubits. However, increasing the size of \ac{GHZ} states also makes it more susceptible to quantum noise, requiring an optimal trade-off between precision and robustness. To mitigate noise effects, additional ancilla qubits can be introduced to improve error resilience and enhance the resyntonization stability.

The frequency synchronization protocol involving atomic qubits in a geographically separated network confined in optical-lattice resonators was proposed in \cite{NPSU:23:CN}. In this protocol, each node in the network contains a single qubit, initially oscillating at a distinct frequency due to local environmental differences or fabrication variability.
To synchronize these disparate oscillators, an external coherent optical field matched to the atomic transition properties is applied. This optical field propagates to each node through low-loss, polarization-maintaining optical fibers, enabling coherent coupling among the remote qubits. The driving field enforces frequency locking by aligning the phase oscillations of the qubits with its own frequency, effectively inducing a common oscillation mode across network nodes. The degree of synchronization is controlled by parameters such as the atom-field coupling strength, dissipation rate, and the photon number in each resonator. Optimal synchronization is achieved within specific coupling and dissipation regimes, where the spectral density exhibits a sharp peak at the external drive frequency, indicating coherent phase locking of the qubits.
Together, these advances position frequency-comb methods---classical and quantum---as a foundation for distributed clock networks.



\subsection{\ac{HQC} Clock Synchronization}

Hybrid synchronization protocols aim to integrate classical time-transfer methods with quantum resources, offering both improved precision and operational resilience. Recent reviews have proposed architectures where time-correlated entangled photons are combined with existing standards such as the precision-time protocol, enabling quantum-enhanced timing profiles that could deliver timing accuracies several orders of magnitude beyond classical limits \cite{B:25:SSRN}. Complementary simulation studies have further compared different quantum resources for network synchronization, notably time-correlated entangled photons and optical lattice clocks. The results indicated that while time-correlated entangled photons can deliver excellent accuracy in low-noise regimes, optical lattice clocks exhibit stable performance even under adverse noise conditions, highlighting the potential of \ac{HQC} systems that combine classical protocols with multiple quantum technologies \cite{NGBF:24:WCNC}.

\subsection{Space-to-Ground Clock Synchronization}

Satellite-based \ac{QCS} is a major frontier for enabling global timing distribution. One proposal envisions a constellation of \ac{LEO} satellites equipped with quantum resources that reinforce one another’s timing signals, collectively forming a master clock in space capable of distributing time worldwide with sub-nanosecond precision \cite{DATAH:25:PRAppl}. Other models consider satellite-carried qubits: for instance, three qubits placed on separate satellites can be synchronized via an applied optical field at 813.32\;nanometers (nm), achieving extremely high precision (on the order of $10^{15}$ signals per second) and enabling the dissemination of stable frequency standards across \acp{NTN} \cite{NLPNF:23:IEEE_CONF_AERO}. Simulation studies further showed that multi-qubit systems governed by realistic noise dynamics can reduce synchronization errors to below one meter in position-equivalent accuracy, demonstrating the scalability of quantum synchronization for satellite navigation and \ac{6G} networks \cite{NRHIBMPWBDCF:24:ComNet}. Complementary to these efforts, field trials have tested quantum-secured time transfer between geographically separated precise timing facilities. By encrypting synchronization data with \ac{QKD} and simulating satellite-based key distribution, these studies highlight how quantum communication can enhance the security of global timing infrastructures \cite{PVODZPAS:24:GPSSol}. Collectively, these results suggest that satellite-assisted \ac{QCS}, when combined with secure quantum links, could augment and strengthen future \ac{GNSS} and \acp{NTN}.

\subsection{Quantum Time-Transfer Coexistence}

For practical scalability, \ac{QCS} must operate alongside classical traffic within existing optical fiber infrastructure. Recent experiments have shown that quantum channels can coexist with classical synchronization signals in a single-mode fiber by carefully characterizing the optical noise spectrum between 1,500\;nm and 1,620\;nm. These results indicated the feasibility of supporting up to 100 quantum channels, each 100\;gigahertz (GHz) wide, in parallel with classical synchronization protocols such as white rabbit and pulsed-laser schemes \cite{BSHGRALSBP:23:OE}. The study also identified a practical coexistence distance limit of about 100\;km using commercial optical transceivers, with the potential for further extension with quantum-capable receivers. Such coexistence strategies are essential for integrating \ac{QCS} into wireless backbones without requiring dedicated fiber links, thereby reducing deployment costs.

\begin{table*}[t!]
\begin{center}
\caption{
\acs{QCS} protocols: Principles, Benchmarks, and limitations
}
\label{tab:QCS-Protocols}

\begin{tabular}{lllll}

\toprule
Protocol & Principle & Benchmark & Limitation & Reference \\
\midrule
\midrule

\makecell[l]{Ticking-qubit\\ handshake \acs{QCS}}
&
\makecell[l]{Round-trip exchange of a\\ phase-evolving qubit;\\ clock offset encoded in\\ accumulated phase and\\ extracted via interference}
&
\makecell[l]{Quantum phase estimation\\ determines $n$-bit offset\\ using $O(n)$ qubits; coherent\\ exchanges achieve\\ $O(\log_2 N_\mathrm{e}/N_\mathrm{e})$\\ uncertainty}
&
\makecell[l]{Naive sampling requires\\ $2^{2n}$ qubits; phase\\ estimation needs extra\\ processing qubits and\\ inverse \ac{QFT}; requires\\ stable tick frequency}
&
\makecell[l]{\cite{Chu:00:PRL,BB:05:PRA}\\ \cite{CEMM:98:PRSAMPE,HK:01:AQP}} \\

\midrule

\makecell[l]{Entanglement-\\ based \acs{QCS}}
&
\makecell[l]{Pre-shared singlet state\\ remains invariant under\\ identical evolution; local\\ measurements and classical\\ communication reveal offset}
&
\makecell[l]{Synchronization without\\ timing-signal exchange;\\ independent of transmission\\ medium after entanglement\\ distribution}
&
\makecell[l]{Requires reliable\\ entanglement distribution;\\ sensitive to spatial-frame\\ misalignment}
&
\makecell[l]{\cite{JADW:00:PRL,LLT:03:CTP,PSCLAMT:23:PRAppl}} \\

\midrule

\makecell[l]{Multiparty\\ entanglement\\ \acs{QCS}}
&
\makecell[l]{Multipartite entangled\\ states (W, \ac{GHZ}) shared\\ across nodes; measurements\\ encode and reveal local\\ clock offsets}
&
\makecell[l]{\ac{GHZ} protocols maintain\\ precision independent of\\ node number; multipartite\\ entanglement enables\\ Heisenberg-limit scaling}
&
\makecell[l]{W-state precision\\ decreases with participants;\\ requires global entanglement\\ distribution and joint\\ measurements}
&
\makecell[l]{\cite{KP:02:PRA,RH:12:PRA,RH:14:QCMC,BE:11:PRA,YZF:15:PRA,YYX:24:PRL}} \\

\midrule

\makecell[l]{Time-of-arrival\\ \acs{QCS}}
&
\makecell[l]{Clock offset inferred\\ from photon arrival-time\\ correlations; frequency\\ entanglement and number\\ squeezing reduce timing\\ uncertainty}
&
\makecell[l]{Classical scaling:\\ $\sqrt{M}$ pulses and $\sqrt{N}$\\ photons; entanglement and\\ squeezing yield $\sqrt{MN}$\\ improvement}
&
\makecell[l]{Highly sensitive to\\ photon loss; multi-pulse\\ entanglement difficult to\\ generate over long links}
&
\cite{GLM:01:N} \\

\midrule

\makecell[l]{Conveyor-belt\\ \acs{QCS}}
&
\makecell[l]{Bidirectional exchange\\ of modulated optical pulses;\\ offset derived from returned\\ interference pattern}
&
\makecell[l]{Insensitive to dispersion;\\ frequency-entangled pulses\\ allow quantum dispersion\\ cancellation}
&
\makecell[l]{Requires equal forward/\\ backward transit times and\\ stable clocks during signal\\ round-trip}
&
\makecell[l]{\cite{GLM:04:PRA,B:25:SSRN}} \\

\midrule

\makecell[l]{\acs{HOM}\\ synchronization}
&
\makecell[l]{Balances \ac{HOM}\\ interferometer using\\ reflected entangled photons;\\ correlated arrivals reveal\\ clock offset}
&
\makecell[l]{Distance-independent\\ synchronization; suitable\\ for optical and space links}
&
\makecell[l]{Requires interferometer\\ balancing and stationary\\ source configuration}
&
\cite{BG:04:AIP} \\

\midrule

\makecell[l]{Time-offset\\ \acs{QCS}}
&
\makecell[l]{Cross-correlation of\\ entangled-photon detection\\ timestamps estimates time\\ offset and frequency drift}
&
\makecell[l]{\ac{FFT} peak detection\\ enables sub-picosecond\\ precision and ppm-level\\ drift estimation}
&
\makecell[l]{Accuracy limited by\\ detector jitter, noise, and\\ coincidence statistics}
&

\makecell[l]{\cite{STSKCCDRRS:23:PRAppl,HLK:09:NJP} \\ \cite{QDXLLZ:20:RSI,LZM:23:ApplOpt}} \\

\midrule

\makecell[l]{Two-way\\ time-transfer\\ \acs{QCS}}
&
\makecell[l]{Signal photons exchanged\\ from frequency-entangled\\ pairs while idlers remain\\ local; transit delays cancel}
&
\makecell[l]{Quantum correlations\\ enable femtosecond-level\\ synchronization precision}
&
\makecell[l]{Requires \ac{SPDC} sources,\\ accurate time tagging, and\\ broadband biphoton spectra}
&
\makecell[l]{\cite{HDLZ:17:QIM,HQDXL:19:PRA}} \\

%

%

\midrule

\makecell[l]{Space-to-ground\\ \acs{QCS}}
&
\makecell[l]{Satellite-borne quantum\\ resources distribute timing\\ globally via optical links}
&
\makecell[l]{Sub-nanosecond global\\ timing distribution and\\ high-precision satellite\\ synchronization}
&
\makecell[l]{Requires space-qualified\\ quantum hardware and\\ robust channel models}
&
\makecell[l]{\cite{DATAH:25:PRAppl, NLPNF:23:IEEE_CONF_AERO,NRHIBMPWBDCF:24:ComNet}} \\

%

\bottomrule
\end{tabular}

\end{center}
\end{table*}

\section{\ac{QCS} Resources} \label{sec:4}
This section characterizes the principal quantum resources required for practical and scalable \ac{QCS} implementations. Table~\ref{tab:QCS-Resources-1} offers a structured comparison of quantum states employed as resources for \ac{QCS}, highlighting their performance benchmarks and limiting factors.

\subsection{Entangled and Squeezed Resources}

The \ac{QCS} protocol has been proposed in \cite{KKBJSYL:14:NP} for operating a network of geographically distributed optical atomic clocks, where quantum metrology, entanglement, and quantum communication techniques are employed to improve the accuracy and stability of timekeeping across the network---surpassing the limits of classical methods constrained by the \ac{SQL}. 

\subsubsection{Entangled-State Resources}

Utilizing entangled states, qubits across multiple clocks become quantum correlated, enabling the network to share timing information more precisely than individual clocks could achieve alone, thereby approaching Heisenberg-limited accuracy.
The use of maximally entangled \ac{GHZ} state has been shown to improve the accuracy and precision of clock synchronization while mitigating synchronization errors arising from asymmetry two-way delays \cite{SS:22:SR}. Simulations on quantum computers further demonstrated that the proposed \ac{GHZ}-based protocol can achieve the \ac{QCS} accuracy in ns.
Finding a more suitable entangled state can lead to significant performance enhancement of the \ac{QCS} protocol, as demonstrated in \cite{BE:11:PRA}. This work improves upon the multiparty \ac{QCS} protocol proposed in \cite{KP:02:PRA}, which utilizes W states but suffers from poor accuracy scaling. The proposed Z state generalizes the W state: instead of using the superposition of basis states with only a single qubit in the state $\ket{1}$, the Z state forms the superposition of basis states containing $k$ qubits in the state $\ket{1}$. Following the same protocol steps as in \cite{KP:02:PRA}, the probability of measuring the evolved collapsed state is given by
\begin{align}
P_{\ket{\pm}}^{(k)}
&=
\frac{1}{2}
\pm
A
\cos \left(\omega t\right)
\nonumber \\
&=
\frac{1}{2}
\pm
\frac{k\left(\nu-k\right)}
{\nu\left(\nu-1\right)}
\cos \left(\omega t\right),
\label{ZS}
\end{align}
which generalizes \eqref{WS}. The quantity $P_{\ket{\pm}}^{(k)}$ is optimized when $k=\lfloor \nu/2 \rfloor$, for which the amplitude $A$ approaches $1/4$ as $\nu \rightarrow \infty$, where $\lfloor x \rfloor$ is the greatest integer less than or equal to $x$. This generalization shows a significant improvement, allowing each party to save $O\left(\nu^2\right)$ qubits. Note that the unentangled state can achieve $A=1/4$, and the Z state can achieve a larger value of $A$ for any $\nu$ when $k$ is chosen optimally.

The protocol proposed in \cite{JADW:00:PRL} uses the singlet state due to its stability under time evolution. However, this robustness only holds when both parties experience identical time evolution, which can be achieved by slowly transporting the qubits to Alice and Bob. In more general scenarios---where transported qubits experience different proper times due to relativistic effects---the resulting proper-time difference is imprinted as a relative phase in the quantum state. This phase can be estimated by measuring the transported qubits, and it has been shown that entangled states offer better estimation accuracy than separable states in this setting~\cite{HAH:02:EPJ}. 
N00N states have also been proposed in the operation-triggered \ac{QCS} protocol to enhance the \ac{QCS} precision and achieve the Heisenberg-limited precision \cite{YZF:15:PRA}. Furthermore, N00N-type states obtain $\sqrt{\nu}$-level precision improvements in $\nu$-party clock synchronization compared to more restricted measurement-triggered protocols \cite{KP:02:PRA, RH:12:PRA,RH:14:QCMC}.
Entanglement has been shown to be a valuable resource for enhancing both the stability and precision of atomic clocks---an essential requirement for \ac{QCS} protocols, which typically assume highly stable clocks. In quantum clocks based on entangled atomic ensembles, entanglement suppresses quantum noise, thereby improving measurement accuracy and long-term stability.

\subsubsection{Squeezed-State Resources}

When spin-squeezed or entangled atomic ensembles are employed, the clock stability can improve with a scaling of $N^{1/6}$ relative to the shot-noise limit for $N$ atoms, surpassing the performance achievable with uncorrelated atomic states even under realistic noise conditions \cite{ASL:04:PRL}.
Spin-squeezing techniques have been used to synchronize networks of distant optical lattice clocks by generating collective spin-squeezed states through quantum nondemolition measurements \cite{PY:16:PRA}.
It enables correlated phase evolution among the clocks, allowing synchronized interrogation with precision surpassing the \ac{SQL}. The scheme supports Heisenberg-limited scaling with respect to both the number of atoms per clock and the number of clocks in the network, and remains robust even in the presence of significant optical losses through optimal cavity-enhanced design.


Another squeezing technique for improving atomic clock precision was demonstrated in \cite{LAROKP:10:NJP}, where a spin-squeezed state in cesium atoms is generated by means of quantum nondemolition measurements. This squeezing not only reduces the projection noise but also creates an entangled state that improves the \ac{SNR} by preserving the atomic coherence during clock interrogation. The experimental results reported a 1.1-dB reduction in quantum noise, resulting in an atomic clock with reduced measurement noise of the accumulated phase and leading to a more accurate estimate of the clock frequency. One of the main challenges in applying squeezing to atomic clocks is decoherence, which shortens the effective length of the collective spin vector. The experimental results showed that this decoherence-induced reduction can be modeled and partially compensated, enabling performance enhancement even under moderate decoherence.

The use of homodyne detection and squeezed light was proposed in \cite{LFT:08:PRL} to enhance the measurements of light-pulse arrival times beyond the \ac{SQL}. When combined with mode-locked fs lasers, homodyne detection allows for precise detection of phase and time-of-flight information within a pulse. Squeezed light further enhances the sensitivity of time-transfer measurements by reducing quantum noise in a selected quadrature below the \ac{SQL}, thereby directly increasing clock synchronization accuracy. Experimental realizations of squeezing have been achieved using optical parametric oscillators and Kerr media, where noise reductions of up to 10\;dB have been demonstrated \cite{VMCH:08:PRL}. However, the implementation of squeezed light faces challenges, primarily due to its sensitivity to optical losses and the need for precise stabilization. The performance of squeezed light degrades significantly with transmission and detection losses, making it more suitable for low-loss environments such as vacuum links between satellites.

The use of squeezed light for satellite-to-satellite clock synchronization was investigated in \cite{GMAGC:22:LATINCOM}, incorporating realistic imperfections of free-space optical inter-satellite channels. This study employed a detailed transmission-loss model that accounts for diffraction, pointing jitter, and detector inefficiencies, with the channel transmission characterized using a beam splitter introducing vacuum noise. Simulation results demonstrated that under high initial squeezing (e.g., 15\;dB), large aperture sizes (e.g., 0.6\;m), and low pointing jitter (e.g., 1\;macroradians), timing synchronization can achieve a two-fold improvement beyond the \ac{SQL} at inter-satellite distances of up to 300\;km. However, for small satellites such as CubeSats, where aperture dimensions are significantly constrained, the achievable quantum advantage is notably diminished. The analysis also revealed a saturation point for squeezing, where as the squeezing level increases, the incremental improvement in timing resolution becomes progressively marginal.

In \ac{LEO} satellite-to-satellite communication scenarios, the use of squeezed light was analyzed in \cite{GMAGB:23:GLOBE}, specifically employing \ac{TMSV} states under realistic lossy inter-satellite channels. The findings demonstrated that \ac{TMSV} states maintain a significant quantum advantage in synchronization precision when the effective transmissivity $\eta$ of inter-satellite channels remains moderately high ($\eta \geq 0.4$). This quantum advantage persists even under asymmetrical loss conditions, where the transmissivity differs between the forward and return paths. The analysis further indicated that increasing the squeezing level is beneficial primarily in high-transmissivity regimes ($\eta \geq 0.8$). Conversely, under highly lossy channel conditions ($\eta <0.3$), no measurable quantum advantage can be attained.

Entanglement is a central quantum resource for establishing non-classical timing correlations, and theoretical results show that genuine multipartite entanglement is required to obtain network-level quantum advantages in distributed synchronization. Moreover, a probabilistic protocol can---when successful---reach the Heisenberg limit for estimating a global parameter in distributed clock synchronization tasks \cite{YYX:24:PRL, GFDSYWGZ:24:IEEE_CONF_OGC, HQXX:22:IEEE_J_JLT}. 
Beyond entanglement, squeezed light offers noise reduction below the shot-noise limit. Recent experiments have demonstrated that asynchronous sampled and digitally reconstructed detection can distribute squeezed states over 10\;km of deployed fiber---without active phase-locking or clock-sharing---and support passive \ac{CV}-\ac{QKD} between laboratories, pointing to practical pathways for quantum-enhanced sensing and, prospectively, synchronization networks \cite{NDHE:25:NPJQI}. Together, entanglement (for provable network-level advantages) and squeezing (for measurement-noise suppression) delineate the two primary quantum resources for \ac{QCS}, while existing fiber demonstrations based on photon correlations establish scalable near-term building blocks.

\subsection{Quantum Frequency Combs}
Quantum frequency combs offer the potential to surpass the \ac{SQL} in \ac{QCS} by introducing non-classical properties into the comb structure, surpassing its classical counterpart and approaching the Heisenberg limit \cite{GAGLBM:24:APLP}. The non-classical properties include quadrature squeezing and quadrature entanglement, which are typically generated through nonlinear optical processes. Such quantum combs exhibit strong frequency-mode correlations, forming a high-dimensional entangled state that can be harnessed for enhanced metrology. The scalability of quantum advantage offered by quantum frequency combs has been studied in \cite{GAGLBM:24:APLP} under realistic constraints. While the Heisenberg limit offers the ideal scaling, achieving it in practice requires high state-generation efficiency, precise mode matching, and low-loss transmission and detection, which remain the deployment challenges of quantum frequency combs. Classical frequency combs have already demonstrated sub-femtosecond synchronization precision over hundreds of kilometers using advanced interferometric techniques. Quantum frequency combs offer comparable or superior achievable precision with fewer resources, making them particularly attractive for deployment in satellite constellations, where resource efficiency is critical. However, due to the complexity of generating and maintaining quantum states, a hybrid network architecture that combines classical and quantum combs is envisioned. It was demonstrated that the 1.5\;dB-quadrature-squeezed quantum comb can achieve a time-offset estimation precision of $7.5 \times 10^{-23}$\;s \cite{GAGLBM:24:APLP}.


\subsection{Single Photons}
\Ac{QCS} protocols leveraging single photons present significant advantages for quantum communication networks, primarily through hardware simplicity by eliminating additional complex hardware typically required in classical synchronization methods.
However, the continuous-wave single-photon sources suffer from random photon-emission times \cite{STRS:22:CLEO, CRHGS:24:CLEO}. In contrast, pulsed single-photon sources confine photon emission within fixed temporal envelopes dictated by the pump pulses, thereby greatly simplifying the synchronization procedure. The primary technical foundation of \ac{QCS} implementations using pulsed single photons involves a fast post-processing scheme tailored explicitly to the inherent structured arrival times of the photons. The protocol in \cite{SS:23:QST} employs a modulo-based cross-correlation technique, which uses the arrival times of photons within each pulse repetition interval as indicators of relative timing offsets. This approach requires only local processing of the photon arrival data at the receiver end, eliminating the need for continuous reference data exchanges or synchronization-specific signals. Experimental demonstrations were carried out using weak-coherent pulses, which achieved a synchronization jitter of 3\;ps over a 5-minute communication session \cite{SS:23:QST}.

In parallel, quantum-based clock recovery methods have been integrated into time-bin \ac{QKD} systems, eliminating the need for separate service channels while maintaining synchronization performance comparable to conventional methods \cite{ZRMRZGKOB:23:AVS_QSci}. Furthermore, photon indistinguishability has been established at telecom wavelengths using \ac{HOM} interferometry, with sub-picosecond synchronization jitter relative to an external clock, supporting scalable deployment in multi-node quantum networks \cite{LBBJKGSP:24:OE}. These results show that single-photon resources can act as both carriers of quantum information and precise synchronization signals.

\subsection{Superposition of Coherent States}

The \ac{SCS} has emerged as a significant quantum resource for enhancing the precision of \ac{QCS}. Coherent states exhibit properties closely resembling classical behavior. However, when coherent states are superposed in specific ways, they yield non-classical states with quantum interference patterns, providing enhanced capabilities for quantum measurements. In \ac{QCS} applications, \ac{SCS} can reduce phase variance, thereby improving the timing precision of synchronized clocks. 
The two types of \ac{SCS}---even and odd \acp{SCS}---have been examined in \cite{XPZ:12:IJTP} for enhancing clock synchronization. Even \ac{SCS} exhibits lower phase variance than odd \ac{SCS} within a specific range of average photon numbers, providing better \ac{QCS} accuracy. In particular, the even \ac{SCS} outperforms classical coherent states for $0.7\leq N \leq 3$, where $N$ denotes the average photon number. In contrast, the odd \ac{SCS} is less effective at achieving the same level of precision, particularly for small $N$. However, both the even and odd \acp{SCS} can only achieve the \ac{SQL} for large $N$.


\subsection{Quantum Memories for Synchronization}

Quantum memories provide the ability to buffer and release quantum states on demand, an essential capability for synchronizing \ac{QED} in networks subject to variable delays. Recent advances in materials have opened promising pathways toward telecom-band memories. 
For example, erbium-doped ceria (CeO$_2$) integrated with nanophotonic cavities has demonstrated enhanced emission and lifetime properties, paving the way for optically addressable spin qubits suitable for quantum networking \cite{PGMCMIBTM:25:arXiv}. Complementary experiments with atomic ensembles have achieved storage and retrieval of light pulses using photon-echo and electromagnetically induced transparency techniques, preserving coherence while enabling flexible operations such as time compression and multi-pulse recall \cite{HSH:09:N}.

In long-distance settings, quantum memories are indispensable for repeater-based architectures, where they allow heralded entanglement to be stored and swapped until successful distribution is confirmed \cite{SSRG:09:arXiv, FL:02:PRA}. These developments establish memory-assisted synchronization as a cornerstone for scaling \ac{QCS} and for building large-scale quantum networks.

%


\begin{table*}[t!]
\begin{center}
\caption{
Quantum-state resources for \acs{QCS}  
}
\label{tab:QCS-Resources-1}
\begin{tabular}{lllll}
\toprule

Resource
& State 
& Benchmark 
& Limitation 
& Reference \\

\midrule
\midrule

\multirow{ 4}{*}{\makecell[l]{Entanglement}} 
&\makecell[l]{Entangled\\ states} 
& \makecell[l]{Surpasses SQL; Heisenberg-limited\\ precision; network quantum advantage}
& \makecell[l]{Decoherence; generation and \\ distribution complexity}
& \makecell[l]{\cite{JADW:00:PRL,KP:02:PRA}\\ \cite{YZF:15:PRA,BE:11:PRA,YYX:24:PRL}}\\

\cmidrule{2-5}

&GHZ states
& \makecell[l]{Nanosecond accuracy; $1/N$ scaling}
& \makecell[l]{Requires global entanglement distribution}
& \cite{SS:22:SR} \\

\cmidrule{2-5}

&Singlet states
& \makecell[l]{Higher precision than separable states}
& \makecell[l]{Sensitive to relativistic phase shifts}
& \cite{JADW:00:PRL,HAH:02:EPJ} \\

\cmidrule{2-5}

&\makecell[l]{N00N states \\TMSV states}
& \makecell[l]{Nanosecond accuracy; $\sqrt{N}$ to Heisenberg\\ precision; robust for $\eta\ge0.4$}
& \makecell[l]{Loss sensitivity;\\ no advantage for $\eta < 0.3$}
& \cite{YZF:15:PRA,GMAGB:23:GLOBE} \\

\midrule

\multirow{ 3}{*}{\makecell[l]{Squeezing}} 
&\makecell[l]{Spin-squeezed \\ensembles}
& \makecell[l]{Stability $\propto N^{1/6}$;\\ 1.1\;dB noise reduction}
& \makecell[l]{Shortened collective spin\\ vector due to decoherence}
& \cite{ASL:04:PRL,PY:16:PRA,LAROKP:10:NJP} \\

\cmidrule{2-5}

&Squeezed light
& \makecell[l]{10\;dB suppression; $>2\times$ SQL improvement}
& \makecell[l]{Optical loss; phase stabilization}
& \cite{LFT:08:PRL,VMCH:08:PRL} \\

\cmidrule{2-5}

&\makecell[l]{Inter-satellite\\squeezed light}
& \makecell[l]{$\approx 2\times$ SQL improvement over 300\;km links}
& \makecell[l]{Aperture limits; pointing jitter;\\ diminishing returns at high squeezing}
& \cite{GMAGC:22:LATINCOM} \\

\midrule

\multirow{ 1}{*}{\makecell[l]{Hybrid}} 
&\makecell[l]{Entangled and\\ squeezing state}
& \makecell[l]{Heisenberg-like precision;\\ sub-picosecond stability in fiber}
& \makecell[l]{Integrated optical design complexity}
& \makecell[l]{\cite{GFDSYWGZ:24:IEEE_CONF_OGC,HQXX:22:IEEE_J_JLT, STSKCCDRRS:23:PRAppl}\\ \cite{NDHE:25:NPJQI}} \\

\bottomrule
\end{tabular}
\end{center}
\end{table*}

\section{\ac{QCS} Noise} \label{sec:5}

\ac{QCS} performance is fundamentally constrained by diverse noise sources---including optical, environmental, and relativistic effects---that degrade timing precision, alter photon statistics, and disrupt quantum interference. Table~\ref{tab:QCS-Noise} provides a consolidated overview of \ac{QCS} noise types, their physical sources, system-level impacts, mitigation strategies, experimental demonstration insights, and research directions.

\subsection{Optical Channel Dispersion}

One of the principal challenges limiting the high-accuracy realization of clock synchronization protocols is timing-signal dispersion due to the variations in the refractive index of the optical channel. This dispersion broadens the pulse width and shifts its mean temporal position \cite{GLMW:02:JOB}, thereby introducing arrival-time measurement errors that depend on the pulse width and leading to synchronization errors.
To overcome this limitation, several \ac{QCS} protocols intentionally avoid using timing signals, ensuring that channel dispersion does not affect their performance.
It is well established that, when using frequency-entangled photons, optical path-length differences can be measured with the \ac{HOM} interferometer with accuracy intrinsically immune to dispersion \cite{SKC:92:PRL, SKC:92:PRA}.
While perfect dispersion cancellation requires idealized and infinite-frequency entanglement, practical implementations inevitably involve finite-frequency entanglement. In this case, the optimal cancellation condition occurs when the group velocity dispersion experienced by the idler photon is proportional to that of the signal photon scaled by the spectral correlation coefficient \cite{XDLHQLZ:20:OE}. Unentangled photons can also mitigate dispersion effects, but only when both photons experience identical dispersion, which is not generally satisfied in frequency-entangled systems.

Frequency-entangled photons can be generated using the parametric down converter, while path-length differences can be introduced by moving mirrors or electro-optic modulators whose refractive index is tuned by varying electric fields \cite{GLMW:02:JOB}. The entangled photons are then directed to the beam splitter, where the coincidence rate at its output is measured by the photodetector. Based on the dip in the coincidence rate, Alice can infer whether her clock is synchronized with Bob's. Furthermore, using a feedback loop, the two parties can maintain synchronization provided that the relative clock drift is sufficiently slow.
The accuracy of time synchronization is determined by the width of the \ac{HOM} dip. This protocol fundamentally relies on dispersion cancellation, which remains valid under low-order dispersion effects.
However, attaining high synchronization accuracy requires accounting for high-order dispersion terms. Although such high-order contributions are often negligible, they become important to achieve high accuracy synchronization \cite{OOTINT:13:PRA}.

In \ac{QCS} protocols that utilize timing signals for synchronization (see, e.g., \cite{GLM:01:N}), channel dispersion directly affects synchronization accuracy by inducing pulse broadening. Although quantum resources such as entanglement and squeezing do not inherently eliminate dispersion effects, they can still improve synchronization accuracy.
In both fiber and free-space channels, chromatic dispersion and related effects introduce frequency-dependent group delays that degrade synchronization precision. The long-haul quantum networking experiments have identified path delay gradients, chromatic dispersion, polarization drift, and power fluctuations as major contributors to timing errors. For instance, metropolitan-scale fiber tests have demonstrated sub-picosecond time deviation with active stabilization, but uncompensated dispersion still limits achievable performance to the order of 10\;ps over extended distances \cite{MRPGAPBTHL:24:APL}. Techniques such as dispersion-shifted fibers, in-situ compensation, and digital post-processing have been investigated to counter these effects \cite{STSKCCDRRS:23:PRAppl}.  
These results show that while dispersion remains a dominant error source in fiber-based \ac{QCS}, advances in compensation and correlation-based processing continue to extend achievable precision.

\subsection{Phase Errors}
In \ac{QCS} protocols that rely on shared prior entanglement (see, e.g.,  \cite{JADW:00:PRL,KP:02:PRA}), the performance is highly sensitive to phase errors \cite{YD:02:PRA, Pre:00:AQP}. The phase offset introduced into the singlet state, for example:
\begin{align}
\ket{\M{\psi}}
=
\frac{1}{\sqrt{2}}
\left(
	\ket{0}_\mathrm{A} \ket{1}_\mathrm{B}
	-
	e^{\imath \omega \eta}
	\ket{1}_\mathrm{A} \ket{0}_\mathrm{B}
\right),
\end{align}
causes the protocol in \cite{JADW:00:PRL} to infer a time value of $t-\eta$, thereby introducing a systematic synchronization bias.
Furthermore, dephasing errors alter the probability of measurement outcomes, further degrading protocol accuracy. Although entanglement purification can yield a high-fidelity entangled state, this procedure also requires synchronous operations and the same basis convention. Hence, it is not feasible for the purpose of clock synchronization. 
Another approach is to apply quantum error-correction techniques by encoding entangled pairs into a code that is stable against phase errors. 
However, this code also suppresses the natural phase evolution that carries the timing information. Hence, when Alice measures her qubit, Bob's remaining qubit remains in a code space that is stable under time evolution and cannot serve as a ticking quantum clock, preventing the protocol from extracting relative timing information.

In channels experiencing complete dephasing---where random phase shifts disrupt the relative phases between quantum states---the exchanged timing information is lost entirely. This randomness fundamentally prevents Alice and Bob from establishing any common time reference, as neither separable nor entangled quantum strategies can counteract such complete phase randomization. In these channels, phase coherence is fundamentally essential for conveying timing information, and once it is irreversibly degraded, the encoded temporal data becomes inaccessible. Although postselection and partial-measurement techniques can mitigate dephasing in various quantum communications tasks, they offer no benefit for clock synchronization as the timing information remains inaccessible due to overwhelming noise from complete dephasing \cite{GLMS:02:PRA}.

To combat the phase offset in the preshared singlet state, a modified entanglement purification protocol was introduced in \cite{OTD:18:npjQI, ITDB:20:AIPCP}. 
Its key contribution is the introduction of a phase-delay operator that accounts for time delays caused by unsynchronized clocks. This operator ensures that the purification process can be executed asynchronously, enabling Alice and Bob to perform their quantum operations independently without requiring precise synchronization. 
Another notable feature of the protocol is its reliance on local basis conventions. Unlike traditional protocols that require a shared phase reference, this method allows Alice and Bob to operate exclusively in their respective local basis. Through modifications to the purification process, the protocol compensates for differences in basis conventions, ensuring that the purified state converges to a singlet state in the local basis, which eliminates systematic phase errors.

\ac{QCS} protocols are highly sensitive to phase fluctuations that accumulate during photon propagation, arising from sources such as fiber-length variations, thermal drift, and atmospheric turbulence. These fluctuations degrade interference visibility in protocols based on \ac{HOM} interference or time-bin encoding, which requires maintaining long-term phase stability  \cite{GFDSYWGZ:24:IEEE_CONF_OGC}. 
Looking ahead, space-based architectures employing quantum frequency combs emphasize that overcoming phase instabilities and related noise sources will be essential to push synchronization beyond the \ac{SQL} \cite{GAGLBM:24:APLP}. Accordingly, effectively suppressing phase errors thus remains a central challenge for scaling \ac{QCS} to global networks.

\subsection{Decoherence}

\Ac{QCS} protocols inherently face challenges related to quantum noise, which affects the accuracy of clock synchronization over quantum communication channels \cite{BCDS:06:LP,NA:24:APB}. Quantum noise degrades state coherence, thereby limiting the precision of time transfer between clocks. In the single-qubit transport scenario \cite{Chu:00:PRL}, the achievable accuracy remains relatively robust under bit-flip noise, scaling as $1/2$. 
Under phase-flip noise, the best accuracy scales as $1/2\left(1-2p\right)^2$ where $p$ is the noise parameter, indicating a substantial degradation in performance. The best achievable accuracy in the presence of amplitude damping scales as $1/2\left(1-p\right)$. In the entangled-state transport scenario, the best achievable accuracy in the bit-flip channel scales as $1/n\left(1-p\right)^n$, where $n$ is the number of entangled qubits, indicating a rapid decrease in performance as the number of qubits or noise increases. For phase-flip noise, the accuracy is given as $1/n\left|\left(1-2p\right)^n\right|$, and under amplitude damping it scales as $1/n\left(1-p\right)^{n/2}$.
It is evident that entangled states offer better accuracy in noise-free environments but are far more sensitive to quantum noise than single-qubit states. Therefore, single-qubit protocols may be preferable in practical deployments where noise is inevitable or difficult to mitigate. Conversely, entangled-state protocols remain attractive in controlled environments with low decoherence, where their superior accuracy can be fully leveraged \cite{OABY:13:PASIP, DJ:12:CTP}.

The decoherence effect has also been analyzed for coherent transport \ac{QCS} where a qubit---carrying the time information---is exchanged back and forth multiple times between the parties \cite{DH:16:JKPS}. Under bit-flip noise, the synchronization accuracy gradually degrades and eventually returns to the \ac{SQL} as the noise parameter $p$ approaches $0.5$, reflecting the complete randomization of bit values. Under phase-flip noise, the accuracy deteriorates exponentially with the noise parameter. As $p \rightarrow 0.5$, the synchronization precision diverges, indicating complete disruption of phase coherence and rendering the protocol ineffective. Under amplitude-damping noise, the synchronization accuracy decreases continuously with increasing $p$. As $p \rightarrow 1$, the accuracy diverges, corresponding to full loss of the excited-state component. Compared to single-qubit transport and entangled-state transport, the coherent transport protocol outperforms in low-noise regimes and remains closer to the Heisenberg limit. Its relative insensitivity to bit-flip errors highlights its robustness in environments where state flips dominate. However, the protocol performs poorly under phase-flip noise and amplitude damping. The repeated traversal of qubits through the noisy channel accumulates errors, thereby severely degrading synchronization accuracy.


\subsection{Thermal Noise}

Although \ac{QCS} protocols generally assume clocks are in a non-relativistic regime, analyzing their behavior under relativistic conditions is essential, as time dilation can potentially affect synchronization accuracy. Experimental studies have demonstrated that relativistic time dilation is detectable even at velocities as low as a few meters per second \cite{CHRW:10:Science}. For accelerated clocks, the Unruh effect predicts that an accelerated observer perceives vacuum fluctuations as a thermal bath. The temperature of this thermal bath is directly proportional to the observer’s proper acceleration \cite{CHM:08:RMP}. 

Multipartite \ac{QCS} has been considered in \cite{ZJ:18:AQP, ZJFW:19:AP} where each clock is modeled as a two-level atomic system, represented by an Unruh–DeWitt detector interacting with a massless scalar field. When one of the detectors is accelerated, its interaction with the field results in thermal noise that degrades the entanglement shared among the clocks and, in turn, reduces synchronization accuracy. Under this noise, quantum correlations decay, thereby reducing the distinguishability of measurement outcomes. Furthermore, the time evolution of an initially entangled state is affected when one clock undergoes acceleration. As the acceleration increases, the probability of measuring a particular clock state converges to $1/2$, making time synchronization indistinguishable from random guessing. These combined effects lead to a deterioration in synchronization accuracy. In this context, the choice of initial entangled states plays a crucial role in mitigating the thermal noise effects. The results show that Z-type entangled states outperform W-type states in the presence of Unruh thermal noise, retaining high entanglement fidelity under acceleration.

In deployed fiber networks, environmental fluctuations---e.g., path-delay gradients, chromatic dispersion, polarization drift, and optical power variations---introduce timing instabilities that limit long-term synchronization accuracy. Field tests on both underground and aerial fiber have shown that, while active stabilization techniques (e.g., electronically stabilized links) can achieve sub-picosecond stability and white-rabbit-precision time protocols can reach tens of picoseconds, residual variations remain challenges over long averaging times \cite{MRLBEADAJ:25:SPIE_PW}. These findings highlight the need for advanced compensation techniques to mitigate environmentally and temperature-induced fluctuations in future \ac{QCS} systems.

\subsection{Gravitational Effects}

In any realistic implementation of \ac{QCS} protocols, gravitational effects must be taken into account, as they can introduce measurable time differences between quantum clocks due to gravitational time dilation and gravitationally induced correlations \cite{WLJC:19:AQT}. Quantum clocks are typically modeled as two-level quantum systems oscillating between their energy eigenstates, and their ticking rate depends on the energy gap between these states. The gravitational field shifts these energy levels and, consequently, the clock frequency \cite{CGB:17:PNAS}. Furthermore, quantum clocks themselves induce gravitational effects due to their mass-energy equivalence. In networks of multiple quantum clocks, each clock influences the evolution of neighboring clocks, leading to time discrepancies that must be accounted for during synchronization.
If time is measured based on the evolution of quantum states, then clocks separated by a gravitational field inevitably become entangled \cite{CGB:17:PNAS}. This gravitationally induced entanglement directly impacts synchronization precision. Moreover, gravitational coupling alters the evolution of quantum clocks, leading to gradual desynchronization of initially synchronized quantum clocks if gravitational effects are not properly incorporated into the synchronization model. This gravitationally induced desynchronization depends on the energy gaps, separation distance, and the strength of the gravitational interaction between clocks.

To mitigate gravitationally induced desynchronization, \ac{QCS} protocols include corrections for gravitational time dilation and effects. One proposed protocol \cite{WLJC:19:AQT} is to prepare the clocks of Alice and Bob in superposition states and measure the clocks in dual bases at predetermined time intervals. If gravitational effects introduce a proper-time difference between the clocks, Bob’s measurement probabilities will systematically deviate from Alice’s. By evaluating the statistical deviation in these measurement probabilities, Bob estimates the gravity-induced time offset and adjusts his clock accordingly.
The relativistic effects introduced by Earth's gravitational field and the relative motion between ground and orbiting frames in satellite-based \ac{QCS} systems were examined in \cite{WTJF:16:PRD}. General relativistic phenomena, such as gravitational time dilation and spacetime curvature, significantly influence photon propagation and alter the coincidence detection statistics used in quantum interferometric synchronization protocols. As photons propagate between differing gravitational potentials, their spectral distributions undergo distortions, affecting wave-packet overlap and ultimately the coincidence rate. For an \ac{LEO} satellite at 400\;km altitude, the curvature-induced disturbance in the coincidence rate is on the order of $10^{-8}$, while for geostationary orbit at approximately 36,000\;km, this disturbance increases to $10^{-5}$.


\subsection{Background Light and Daylight Operation Noise}

One of the main challenges for free-space \ac{QCS} is noise arising from background sunlight and artificial illumination, as these photons can overwhelm \acp{SPD} and degrade synchronization fidelity. Despite its importance, this issue is not directly addressed in current large-scale field trials. For example, recent demonstrations of secure time transfer between precise timing facilities over 900\;km instead focused on integrating \ac{QKD} with satellite-assisted links to ensure the security of synchronization data \cite{PVODZPAS:24:GPSSol}. Although background-light effects were not explicitly part of that trial, they remain a critical obstacle for daylight operation in satellite-based \ac{QCS}, motivating ongoing research on advanced filtering and temporal gating to suppress noise in free-space quantum channels.

\subsection{Non-Reciprocal Path Effects}

Many \ac{QCS} protocols---particularly two-way schemes---rely on the assumption that channels are reciprocal. In practice, this assumption is often violated. Path-delay gradients, polarization drift, chromatic dispersion, and other asymmetries introduce systematic biases in the measured clock offsets. Experiments in metropolitan fiber networks have shown that, even with active stabilization, such non-reciprocal fluctuations remain a dominant source of error at the picosecond scale \cite{MRPGAPBTHL:24:APL}. While large-scale satellite demonstrations to date have primarily emphasized the secure distribution of timing information via \ac{QKD} \cite{PVODZPAS:24:GPSSol}, extending \ac{QCS} to free-space and orbital links requires explicit handling of asymmetries caused by atmospheric turbulence and relative motion. To address these challenges, advanced calibration methods, polarization tracking, and real-time reciprocity monitoring are being actively explored.

\begin{sidewaystable*}
\begin{center}
\caption{
QCS noise: types, sources, impacts, mitigation strategies, demonstration insights, and research directions
}
\label{tab:QCS-Noise}

\resizebox{\textwidth}{!}{
\begin{tabular}{lllllll}
\toprule

\makecell[l]{Noise Type} 
& \makecell[l]{Physical Source}  
& \makecell[l]{Synchronization\\ Impact} 
& \makecell[l]{Mitigation Strategy}  
& \makecell[l]{Demonstration Insight} 
& \makecell[l]{Research Direction}   
& Reference \\

\midrule
\midrule

\makecell[l]{Optical\\ dispersion}
& \makecell[l]{Variations in refractive\\ index and group-velocity; \\dispersion in optical or\\fiber channels}
& \makecell[l]{Timing errors due to\\ pulse spreading and\\ mean shift; \\ limited precision}
& \makecell[l]{Dispersion-shifted fibers;\\ digital or in-situ dispersion\\ compensation;\\ frequency-entangled photons} 
& \makecell[l]{Sub-picosecond stability;\\ dispersion domination\\ over long distances;\\ high bandwidth}
& \makecell[l]{Higher-order\\ dispersion corrections\\ for sub-picosecond \\global-scale QCS}
& \makecell[l]{\cite{GLMW:02:JOB,SKC:92:PRL, SKC:92:PRA,XDLHQLZ:20:OE,OOTINT:13:PRA,MRPGAPBTHL:24:APL} } \\

\midrule

\makecell[l]{Phase error}
& \makecell[l]{Phase fluctuation during\\ photon propagation;\\ dephasing in  entangled \\or interferometric states}
& \makecell[l]{Time offset and \\coherence loss\\ affecting interference\\ visibility and \\ measurement outcomes}
& \makecell[l]{Modified entanglement\\ purification using\\ phase-delay operators;\\ asynchronous local-basis\\ synchronization}
& \makecell[l]{Sub-picosecond stability\\ over 20-km fiber;\\ limited long-term\\ coherence}
& \makecell[l]{Robust phase stability\\ against turbulence and\\ thermal drift critical\\ for networked \acs{QCS}}
& \makecell[l]{\cite{GAGLBM:24:APLP} \\\cite{GFDSYWGZ:24:IEEE_CONF_OGC,OTD:18:npjQI}\\ \cite{JADW:00:PRL,KP:02:PRA}\\ \cite{ YD:02:PRA, Pre:00:AQP,GLMS:02:PRA,ITDB:20:AIPCP}} \\

\midrule

Decoherence
& \makecell[l]{Bit-flip, phase-flip,\\ amplitude-damping\\ noise in quantum\\ channels}
& \makecell[l]{Accuracy degradation\\ scaling with\\ noise parameter and\\ qubit number}
& \makecell[l]{Quantum error correction;\\ hybrid one-qubit transport\\ protocols}
& \makecell[l]{Entangled states\\outperform in low-noise\\ regimes but degrade\\ rapidly under noise}
& \makecell[l]{Low-overhead\\error correction; \\ decoherence-resistant\\ states for large-scale\\ \acs{QCS} networks}
& \makecell[l]{ \cite{Chu:00:PRL} \\ \cite{MRPGAPBTHL:24:APL,PVODZPAS:24:GPSSol} \\ \cite{BCDS:06:LP,NA:24:APB,OABY:13:PASIP, DJ:12:CTP,DH:16:JKPS}} \\

\midrule

\makecell[l]{Thermal noise\\ (Unruh effect)}
& \makecell[l]{Vacuum fluctuations\\ due to acceleration;\\increased thermal bath \\arising from\\ environmental heating} 
& \makecell[l]{Entanglement decay;\\ increased measurement\\ indistinguishability\\ due to acceleration}
& \makecell[l]{Z-type entangled states\\ resilient to thermal effects;\\ environmental and\\ temperature stabilization}
& \makecell[l]{Relevant for moving\\ or satellite-based clocks;\\ experimental detection\\ feasible at low\\ accelerations}
& \makecell[l]{Decoupling relativistic\\ and thermal effects;\\compensation for\\ accelerated or\\ orbital clocks}
& \makecell[l]{\cite{CHRW:10:Science} \\ \cite{CHM:08:RMP,ZJ:18:AQP, ZJFW:19:AP,MRLBEADAJ:25:SPIE_PW}} \\

\midrule

\makecell[l]{Gravitational\\ effect}
& \makecell[l]{Gravitational redshift \\and curvature\\ modifing quantum clock \\frequencies and\\ entanglement}
& \makecell[l]{Time-rate differences;\\ coincidence-rate\\ disturbances}
& \makecell[l]{Relativistic correction;\\ dual-basis comparison;\\ gravity-aware\\ synchronization models}
& \makecell[l]{Ground-to-satellite tests\\ confirming accuracy\\of $10^{-16}$ with\\ hydrogen clocks}
& \makecell[l]{Quantum-clock models\\ with consistent\\ gravitational redshift;\\relativistic QCS\\ for global networks}
& \makecell[l]{\cite{NRHIBMPWBDCF:24:ComNet,FSK:23:MeasTech} \\ \cite{ WTJF:16:PRD,WLJC:19:AQT,CGB:17:PNAS}} \\

\midrule

\makecell[l]{Background light\\ (Daylight noise)}
& \makecell[l]{Sunlight and artificial\\ photons overwhelming\\ detectors in \\free-space links}
& \makecell[l]{Detector saturation\\ and degraded\\ synchronization\\ fidelity for daylight}
& \makecell[l]{Temporal gating;\\ spectral filtering;\\ narrow-band detection}
& \makecell[l]{900-km secure time\\ transfer with QKD\\ integration; still limited\\ daylight operation}
& \makecell[l]{Noise-resilient QCS\\ with adaptive filtering\\ for bright-sky and\\ urban environments}
& \cite{PVODZPAS:24:GPSSol} \\

\midrule

\makecell[l]{Non-reciprocal\\ path effects}
& \makecell[l]{Channel asymmetry;\\ path-delay fluctuation;\\ chromatic dispersion;\\ polarization drift;\\ atmospheric variability} 
& \makecell[l]{Timing offset and\\ picosecond-scale\\ drift under active\\ stabilization}
& \makecell[l]{Real-time reciprocity\\ monitoring;\\ polarization tracking;\\ path calibration}
& \makecell[l]{Metropolitan fiber tests\\ for non-reciprocity\\ as dominant\\ picosecond-level\\ error source}
& \makecell[l]{Adaptive calibration\\ for dynamic atmospheric\\ and orbital conditions\\ in future QCS systems}
& \cite{PVODZPAS:24:GPSSol,MRPGAPBTHL:24:APL} \\

\bottomrule
\end{tabular}
}
\end{center}
\end{sidewaystable*}

\section{\ac{QCS} Security} \label{sec:6}

Distributed precision-time networks play a critical role in various modern systems, yet their security remains a critical vulnerability.
These networks rely on time signals to maintain synchronization, but the mechanisms for securing these signals are generally weak. As a result, they are vulnerable to security attacks such as spoofing, in which an adversary injects false signals to manipulate or corrupt the network’s timing data.
To mitigate these risks, military applications implement additional security measures designed to detect and deter spoofing. However, these countermeasures introduce considerable complexity, making their deployment more challenging. Furthermore, despite these enhanced protections, sophisticated adversaries may still be capable of circumventing security defenses, posing an ongoing threat to the integrity and reliability of precision timing networks.

Secure clock synchronization protocols must accurately measure and distribute information regarding the relative clock offset while ensuring that an adversary with access to the communication channels cannot manipulate the inferred offset undetected. This task falls within the broader domain of secure metrology, which combines principles from both secure communication and precise metrology. Such protections are critical for maintaining the integrity and reliability of time synchronization in distributed systems \cite{NH:18:IEEE_J_STSP}.

In clock synchronization protocols that rely on classical signals, security fundamentally depends on physical constraints governing an adversary’s capabilities by the laws of physics, as well as any known technical limitations. To ensure secure clock synchronization, Alice and Bob employ an authenticated encryption scheme to protect timing signals from being counterfeited by an adversary. This encryption ensures that only legitimate signals are used for synchronization and prevents unauthorized tampering. The signal propagation time between Alice and Bob cannot be reduced by more than a predefined fixed limit. This constraint determines the maximum achievable synchronization accuracy and prevents adversarial manipulation beyond a measurable threshold. Alice must possess prior knowledge of the expected round-trip time and be able to measure it with accuracy exceeding this bound. This capability allows her to reliably estimate the propagation time and to detect any discrepancies caused by an adversary attempting to manipulate the synchronization process \cite{NH:18:IEEE_J_STSP}. 

The purpose of these security requirements is to enable Alice to use her local clock to estimate the propagation time to Bob by measuring the round-trip time of the transmitted photons. By comparing her estimated propagation time with the expected value, Alice can detect the presence of an adversary if the deviation exceeds the predefined threshold. A critical challenge in achieving secure clock synchronization lies in Alice’s or Bob’s ability to determine the true channel propagation time with high precision and to establish a trustworthy lower bound on the extent to which this time can be reduced. The accuracy of this distance estimation directly impacts the secure limit of the clock synchronization protocol, making it a crucial factor in maintaining synchronization integrity against adversarial interference. Table~\ref{tab:QCS-Security} summarizes security threats in \ac{QCS} protocols, detailing attack vectors, adversary capabilities, impacts on synchronization, corresponding countermeasures, and their practical feasibility.

\subsection{Protocol Security}

Clock synchronization based on the second-order quantum correlations has been widely used in non-pulsed quantum cryptography to determine the relative time difference between Alice’s and Bob’s clocks. The first fundamental assumption in this protocol is that the signals received by each party originate from the same entangled photon pair. Quantum entanglement provides an intrinsic mechanism to verify this assumption. By performing a Bell-inequality test in the polarization degree of freedom, Alice and Bob can confirm that the correlated photon pairs are indeed linked. This verification process also ensures that an adversary has not accessed the polarization information, as any interference would alter the Bell-inequality results, revealing potential tampering \cite{LT:18:SPIE}.
However, verifying the Bell inequality in the polarization degree of freedom does not, by itself, guarantee that the timing information remains secure. An adversary could introduce a polarization-insensitive delay, leaving the Bell-test outcomes unchanged while still manipulating the timing synchronization. This vulnerability highlights why conventional one-way synchronization protocols are susceptible to delay attacks. Even classical two-way protocols must impose strict security conditions to prevent adversarial influence. In contrast, the quantum protocol mitigates these risks by employing symmetric photon sources and detection mechanisms. Under such symmetry, any adversarial time delay would have no impact on the calculated clock offset, making the synchronization process more secure.

Another critical assumption is the symmetry of propagation times through the communication channel. If an adversary introduces an arbitrary delay while Alice and Bob continue to assume channel symmetry, they accept a manipulated time offset. In this scenario, the estimated time difference deviates from the true value by half the magnitude of the introduced asymmetry. To counteract this risk, Alice and Bob must randomly sample their photon population to verify that the photons originating from the two parties remain truly indistinguishable. Any distinguishability creates an opportunity for an adversary to introduce asymmetric delays, compromising the accuracy and security of the synchronization process.
To compromise the security of the \ac{QCS} protocol, the adversary needs to introduce asymmetric propagation delays that depend on the direction of photon travel. This attack requires the adversary to reliably determine the photon’s propagation direction without disturbing the quantum state.
Hence, two key conditions must be met for this type of attacks to succeed. First, the adversary must identify the direction of travel with high probability in order to selectively introduce a propagation delay only when the direction is known. Second, this direction measurement must be nondestructive without absorbing the photon or altering any of its degrees of freedom, such as polarization or timing characteristics. If either condition fails, the adversary cannot introduce controlled asymmetries without producing detectable disturbances, preserving the security of the synchronization process.

If the adversary possesses the capability to perform a \ac{QND} measurement of photon presence with high success probability at two points along the channel, then both security conditions could, in principle, be satisfied, allowing the adversary to compromise \ac{QCS} protocols.  A \ac{QND} measurement allows the adversary to detect the presence of a photon without disturbing its quantum state, enabling precise control over the applied propagation delay. However, technical challenges associated with implementing \ac{QND} measurements or achieving controllable coherent single-photon nonreciprocity in practical settings remain significant obstacles to this attack strategy. Furthermore, at least one successful \ac{QND} measurement is an essential prerequisite for this attack. Since the adversary must reliably detect a photon at a specific location and determine its direction before introducing path-dependent delays, any inability to perform such measurements prevents a successful security breach. Consequently, the \ac{QCS} protocol remains secure as long as the adversary lacks the capability to perform \ac{QND} measurements of single-photon presence in the channel.


\subsection{Security Attacks}

Although quantum correlations were initially expected to protect clock synchronization from state tampering \cite{LT:18:SPIE} and to offer resilience against symmetric delay attacks \cite{LSCTLK:19:APL, LSUK:22:OE}, the \ac{QCS} protocols still remain vulnerable to asymmetric delay attacks. Such attacks manipulate propagation times in a bidirectional \ac{QCS} protocol without introducing detectable changes in quantum correlations \cite{LSCT:19:APL}. The vulnerability arises from the assumption in the second-order correlation \ac{QCS}: that the total round-trip time is evenly split between forward and reverse directions. An adversary who introduces a direction-dependent delay can violate this assumption and thereby force an incorrect synchronization offset.
This can be realized using optical circulators, which exploit the Faraday effect to break reciprocity in light propagation while leaving polarization and entanglement properties unchanged. As a result, an adversary can covertly introduce asymmetric propagation delays leading to a false clock offset estimate. The adversary can shift the midpoint of correlation peaks by selectively increasing the propagation time in only one direction without triggering conventional security checks. Experimental demonstrations have confirmed that an offset shift of 25\;ns can be introduced using this technique, all while preserving the observed entanglement characteristics.

A notable approach for securing quantum synchronization against time-delay and intercept-resend attacks is \ac{QSTT}, which leverages the security guarantees of \ac{QKD} to establish an authenticated and tamper-proof time-transfer protocol. In these attacks, 
an adversary attempts to intercept synchronization signals and resend their modified versions with manipulated timing information. Such tampering can introduce timing biases that cause desynchronization between remote clocks, potentially leading to erroneous timekeeping and creating security breaches in quantum communication protocols.
\ac{QSTT} mitigates this risk by encoding timing signals within polarization-encoded single photons, which adhere to the quantum no-cloning theorem. Any attempt to intercept and resend these photons inevitably disturbs their quantum states, producing detectable anomalies in the \ac{QBER}. By monitoring the \ac{QBER}, Alice and Bob can verify the authenticity of the received timing signals and discard any data blocks suspected of being compromised. 

In the case of time-delay attacks, \ac{QSTT} measures the propagation time of timing signals to monitor anomalous delays. 
If a discrepancy is detected between the measured and expected round-trip propagation times, the affected timing data is discarded, effectively nullifying the attack. While quantum timing signals benefit from inherent protection provided by quantum mechanics, classical timing data exchanged between the parties must also be protected from eavesdropping and tampering.
In \ac{QSTT}, the classical timing data is encrypted using keys generated through \ac{QKD}, ensuring that any intercepted data cannot be deciphered or modified without detection. This dual-layer security mechanism guarantees that both the physical timing signals and their corresponding classical timing data remain secure throughout the synchronization process.
Satellite-based \ac{QSTT} has been experimentally demonstrated in \cite{DSWLLC:20:NP}, where the protocol was tested using single-photon-level timing signals exchanged between the Micius satellite and a ground station, achieving a time-transfer precision of approximately 30\;ps and maintaining \ac{QBER} below 1\,$\%$. These results surpass the accuracy of classical synchronization methods such as the T2L2 time-transfer system aboard the Jason-2 satellite, which relies on laser pulses and achieves lower timing precision. Additionally, the \ac{QSTT} system supports higher data rates (e.g., 9\;kHz) compared to traditional optical time-transfer methods, further enhancing its applicability in quantum networking and high-precision timekeeping.

The time-delay attack can also be mitigated by other mechanisms, such as decoy-state analysis and Franson interferometry \cite{WSZQKF:21:QCCS}.  Decoy states can be randomly inserted into the transmitted quantum pulses to detect photon-number-splitting attacks, ensuring that an adversary cannot extract timing information without inducing detectable anomalies. Complementarily, Franson interferometry can be used to verify the time-energy entanglement of transmitted photon pairs, providing a further layer of security by detecting any unauthorized measurements performed by an eavesdropper. When synchronization signals are generated using weak coherent pulses rather than true single-photon sources, photon-number splitting attack becomes relevant \cite{WSZQKF:21:QCCS}. In this case, multi-photon pulses may occasionally be emitted, allowing an adversary to split off one photon while allowing the remaining photons to continue toward the receiver. The attacker can then perform delayed measurements on the retained photon to extract partial timing information without fully collapsing the transmitted quantum state. Although the attack does not completely disrupt synchronization, it introduces information leakage and statistical bias in the timing distribution, which may gradually degrade synchronization accuracy. Mitigation strategies include decoy-state insertion and photon-number statistical analysis, which allow legitimate nodes to detect deviations from the expected photon-number distribution.

Security risks also arise through time-stamp spoofing or replay attacks. In this scenario, an adversary injects or replays previously valid timing messages in the classical communication channel used to exchange synchronization data \cite{NH:18:IEEE_J_STSP}. Because the manipulated messages may appear authentic, the attack can introduce false synchronization events or incorrect clock offsets without directly interacting with the quantum channel. Such threats are mitigated through authenticated classical communication, encrypted timestamp exchange, time-window validation, and replay protection mechanisms, which ensure that only valid timing messages are accepted by the synchronization protocol. Hardware-level vulnerabilities may also enable detector side-channel attacks, which exploit imperfections in single-photon detection systems. An attacker may manipulate detector nonlinearities, saturation behavior, or timing jitter by injecting carefully controlled optical signals \cite{NH:18:IEEE_J_STSP}. This can lead to biased coincidence statistics and distorted timestamp measurements, ultimately affecting the accuracy of the synchronization process. To mitigate these risks, practical \ac{QCS} implementations incorporate optical power limiting, detector monitoring circuits, and regular calibration procedures to detect abnormal detector responses.

In addition to active attacks on quantum states themselves, an adversary may also manipulate the physical propagation channel to induce systematic synchronization errors. Such channel delay manipulation attacks can arise from environmental control of optical fiber length, temperature-induced refractive-index variations, or deliberate path-length modifications \cite{NH:18:IEEE_J_STSP}. Over time, these disturbances may accumulate and generate systematic timing offsets or long-term synchronization drift between distributed nodes. Continuous delay calibration, two-way synchronization verification, and authenticated communication channels are therefore required to ensure that propagation delays remain consistent with expected physical models. Furthermore, denial-of-service attacks target the availability of \ac{QCS} rather than the integrity of the synchronization data. By introducing strong attenuation, blocking the optical channel, or injecting intense classical noise, an adversary can prevent successful photon detection or disrupt synchronization exchanges \cite{NH:18:IEEE_J_STSP}. Although such attacks are typically detectable, they may still cause temporary loss of synchronization or link outages, which can degrade the performance of distributed quantum networks. Robust system design therefore requires link monitoring, redundant communication paths, and path-diversity mechanisms to maintain synchronization availability even under adverse conditions.

\begin{sidewaystable*}
\begin{center}
\caption{QCS security threats: types, attack vectors, countermeasures, and feasibility}
\label{tab:QCS-Security}

\resizebox{\textwidth}{!}{
\begin{tabular}{lllllll}

\toprule

Threat Type&
Attack Vector &
Adversary Capability &
Synchronization Impact  &
Countermeasure &
Feasibility &
Reference \\

\midrule 
\midrule

\makecell[l]{Entanglement\\ hijacking} &
\makecell[l]{Tampering with correlated\\ photon pairs or entanglement\\ channels; access to polarization\\ information without disturbance} &
\makecell[l]{Ability to intercept or\\ interact with entangled\\ photons while \\preserving apparent\\ polarization  correlations} 
&\makecell[l]{False correlations;\\ degraded Bell visibility;\\ corrupted timing inference} &
\makecell[l]{Bell-inequality\\ verification; polarization\\ correlation checks} &
\makecell[l]{Very low\\ (any intrusion disturbing\\ entanglement)} &
\cite{LT:18:SPIE} \\

\midrule

\makecell[l]{Symmetric\\ delay} &
\makecell[l]{Injection of polarization-\\insensitive and equal\\ delays in both directions} &
\makecell[l]{Control of bidirectional\\ optical paths while\\  preserving reciprocity
\\ and polarization} &
\makecell[l]{Hidden bias in offset\\ estimation, particularly\\in one-way protocols} &
\makecell[l]{Random sampling\\ of photon population;\\ indistinguishability\\ verification} &
\makecell[l]{Low} &
\makecell[l]{\cite{LSCTLK:19:APL}\\ \cite{LSUK:22:OE}} \\

\midrule

\makecell[l]{Asymmetric\\ delay} &
\makecell[l]{Direction-dependent delay\\ insertion using optical\\ circulators (Faraday effect);\\ shift of correlation midpoints} &
\makecell[l]{QND-meaurement\\ sensing and controlled \\ path-dependent delay\\ insertion} &
\makecell[l]{False clock offset without\\ altering polarization or\\ entanglement observables} &
\makecell[l]{Decoy-state analysis;\\ Franson interferometry;\\ asymmetry monitoring} &
\makecell[l]{Extremely difficult\\ (scalable QND\\ impractical)} &
\makecell[l]{\cite{LSCT:19:APL}\\ \cite{WSZQKF:21:QCCS}} \\

\midrule

\makecell[l]{Intercept-\\and-resend} &
\makecell[l]{Capture and retransmission\\ of timing photons with\\ manipulated delay} &
\makecell[l]{Perfect quantum-state\\ replication (violating\\ no-cloning theorem
)} &
\makecell[l]{Timing bias;\\ desynchronization\\ between nodes} &
\makecell[l]{Quantum secure time\\ transfer with QKD\\ authentication;\\ QBER monitoring} &
\makecell[l]{Physically impossible\\ (ideal case); detectable\\ when attempted} &
\cite{DSWLLC:20:NP} \\

\midrule

\makecell[l]{Photon-number\\ splitting} &
\makecell[l]{Splitting multi-photon pulses\\ to extract partial timing\\ information without fully\\ destroying the state} &
\makecell[l]{Non-destructive\\ photon separation\\ and path control} &
\makecell[l]{Partial information\\ leakage; bias in\\ timing statistics} &
\makecell[l]{Decoy-state insertion;\\ QBER statistical\\ analysis} &
\makecell[l]{Partial but detectable\\ by decoy analysis} &
\cite{WSZQKF:21:QCCS} \\

\midrule

\makecell[l]{Time-stamp\\ spoofing\\ (replay)} &
\makecell[l]{Injection or replay of\\ valid-looking timing\\ events} &
\makecell[l]{Access to or emulation\\ of authenticated classical\\ timing channels with\\ reliable replay capability} &
\makecell[l]{False events;\\ offset errors;\\ log inconsistencies} &
\makecell[l]{Authenticated\\ encryption;\\ time windowing;\\ replay protection} &
\makecell[l]{Moderate\\ (classical risk)} &
\cite{NH:18:IEEE_J_STSP} \\

\midrule

\makecell[l]{Detector\\ side-channel} &
\makecell[l]{Exploitation of\\ detector nonlinearity,\\ saturation response,\\ or timing jitter} &
\makecell[l]{Control of incident\\ optical power or\\ timing at receivers} &
\makecell[l]{Skewed time-stamp\\ statistics; biased\\ coincidence detection} &
\makecell[l]{Power limiting;\\ detector monitoring;\\ calibration procedures} &
\makecell[l]{Moderate} &
\cite{NH:18:IEEE_J_STSP} \\

\midrule

\makecell[l]{Channel delay\\ manipulation} &
\makecell[l]{Biasing one-way delays\\ via environmental or\\ path-length manipulation} &
\makecell[l]{Optical path access or\\ control; environmental\\ manipulation} &
\makecell[l]{Systematic timing offset;\\ long-term drift} &
\makecell[l]{Two-way transfer;\\ delay calibration;\\ authenticated channels} &
\makecell[l]{Moderate\\ (detectable with\\ calibration)} &
\cite{NH:18:IEEE_J_STSP} \\

\midrule

\makecell[l]{Denial-of-\\service (DoS)} &
\makecell[l]{High attenuation;\\ channel blocking;\\ deliberate noise injection} &
\makecell[l]{Physical access or\\ intense classical\\ co-channel noise} &
\makecell[l]{Synchronization loss;\\ link outages} &
\makecell[l]{Link monitoring;\\ redundancy;\\ path diversity} &
\makecell[l]{High but\\ obviously detectable} &
\cite{NH:18:IEEE_J_STSP} \\

\bottomrule
\end{tabular}
}
\end{center}
\end{sidewaystable*}

\section{\ac{QCS} Applications} \label{sec:7}

This section reviews the major application domains enabled by \ac{QCS}, highlighting its relevance to secure communication, distributed quantum processing, precision geodesy, and time distribution in quantum Internet architectures. Table~\ref{tab:QCS-Apps} summarizes representative \ac{QCS} application domains, highlighting the role in each domain, experimental demonstration setups, and achieved synchronization accuracy.

Beyond enabling precise time distribution, \ac{QCS} directly impacts the performance of several quantum-network services, including \ac{QKD}, quantum teleportation, entanglement distribution, and distributed quantum computing and sensing. Many quantum protocols rely on temporal indistinguishability of photons, accurate coincidence detection, and stable phase relationships between remote nodes. Imperfect synchronization introduces timing jitter and clock offsets that degrade key performance metrics such as \ac{QBER}, interference visibility, entanglement fidelity, and secret-key generation rate. Consequently, \ac{QCS} serves as a foundational network-layer capability that supports reliable operation of higher-layer quantum communication, computing, and sensing protocols.

\subsection{\ac{QKD}}
Cryptography fundamentally depends on secure and efficient key distribution, ensuring that only authorized parties can access the shared secret keys necessary for encrypting and decrypting messages. \Ac{QKD} provides a quantum-mechanically secure mechanism for key exchange, offering inherent resistance to eavesdropping attempts \cite{ZCPWZ:20:PRL}. While \ac{QKD} is widely recognized for its provable security, practical deployment also requires high efficiency, including high key-generation rates and scalability suitable for modern communication networks. Meeting these requirements involves addressing technical challenges, such as achieving precise timing synchronization, optimizing detector sensitivity and timing resolution, and mitigating noise arising from environmental disturbances \cite{RR:17:ICMIP}.

\subsubsection{Synchronization-Jitter Impacts}

Clock synchronization plays a pivotal role in the integrity and performance of \ac{QKD} systems. The precision of time alignment between sender and receiver directly influences data sampling, parameter estimation, and ultimately the security of secret-key generation \cite{XGLZ:18:PLA}. Among synchronization imperfections, clock synchronization jitter has a pronounced impact on \ac{QKD} performance, which is inherently tied to the physical limitations of timing electronics. Jitter can cause deviation in data acquisition through homodyne or heterodyne detection and introduces sampling noise that deteriorates the measured signal amplitude, with the error magnitude increasing exponentially with the jitter amplitude and inversely with the square of the pulse width. The impact of jitter cascades through the \ac{QKD} protocol and leads to an underestimation of channel parameters due to signal amplitude attenuation. This underestimated excess noise can pose a security risk, as it conceals the underlying vulnerability introduced by timing inaccuracies in the synchronization process. An adversary exploits this discrepancy by injecting tailored timing distortions or mimicking jitter effects to obscure eavesdropping attempts.

\subsubsection{Detector-Timing Issues}

In \ac{QKD} systems, synchronization is especially challenging due to the use of gated-mode \acp{SPD}, which require tight timing alignment with the arrival of quantum signals. If the timing window of \acp{SPD} is misaligned, quantum states may not be detected efficiently, leading to low key-generation rates and potential security loopholes such as time-shift attacks and detector-efficiency-mismatch attacks. A major issue in \ac{QKD} synchronization arises from detector-efficiency-mismatch attacks, where an adversary manipulates synchronization pulses to cause different detectors to have varying efficiencies. If Alice and Bob independently calibrate the timing for each \ac{SPD}, the adversary can selectively shift the arrival times of synchronization pulses, creating a timing bias that the adversary can later exploit through time-shift attacks. This allows the adversary to gain partial information about the key without introducing detectable errors, thereby compromising the \ac{QKD} security. To mitigate these risks, a secure and efficient synchronization scheme for \ac{QKD} has been developed, significantly improving synchronization accuracy while eliminating vulnerabilities to timing-based attacks. The scheme is built on two key principles: (i) fixing the relative delay of the signal time window among all \acp{SPD} and (ii) performing parallel synchronization using multiple \acp{SPD} \cite{LY:19:OSAC}.

\subsubsection{High-Speed Clock Recovery}

The work in \cite{BGMHN:04:OE} achieves a sifted key rate of 1\;Mbits/s (Mbps) over a 730-m free-space link by using a 1.25-Gbps clock-recovery technique. This key rate is significantly improved over earlier asynchronous \ac{QKD} systems that need to send timing signals before each quantum pulse \cite{HNDP:02:NJP}. By using synchronous timing recovery, significantly higher continuous transmission rates are feasible, leading to increased key rates and extended operational distances.
In satellite-to-ground \ac{QKD} over distances up to 1200\;km, precise timing and noise mitigation are achieved using a beacon laser \cite{LCLZL:17:N}.
Due to the large physical satellite-to-ground separation and their independent reference clocks, precise-time synchronization is crucial to accurately timestamp photon arrivals and distinguish \ac{QKD} signals from background noise. A 532-nm beacon laser operating at 10\;kHz with a pulse width of approximately 0.88\;ns, is employed for this purpose. The beacon laser achieves synchronization jitter of 0.5\;ns, enabling accurate time tagging of signal photons. The received \ac{QKD} photons exhibit a typical temporal distribution with a standard deviation of approximately 500\;ps. These photons are registered within a narrow time window of 2\;ns, effectively excluding events caused by noise outside this interval.

\subsubsection{Qubit4Sync Family}

The work in \cite{WSZQKF:21:QCCS,AACSFZ:20:Optica, CSAADVV:20:PRAppl, CG:21:Entropy, SAACFS:22:AQT, M:24:JOPCS} has proposed clock synchronization in \ac{QKD} by extracting timing information directly from qubits used for key distribution. The Qubit4Sync protocol introduced in \cite{AACSFZ:20:Optica} can synchronize the clocks without requiring separate synchronization signals in polarization-based \ac{QKD} systems. This approach demonstrated that synchronization precision remains stable even under high channel losses up to 40\;dB---corresponding to fiber distances of 26\;km---while maintaining an intrinsically low \ac{QBER} of 0.05\,$\%$. The protocol \cite{CSAADVV:20:PRAppl} was tested in a polarization-based \ac{QKD} system, achieving a synchronization precision of 2\;ns, which is sufficient to support GHz-rate \ac{QKD}. The protocol remained effective at channel losses up to 40\;dB, making it suitable for long-distance \ac{QKD} in both free-space and fiber-optic channels. Notably, synchronization could still be recovered even when detecting only a small fraction of transmitted qubits.

The Qubit4Sync protocol was also employed in a cross-encoded \ac{QKD} system exploiting both time-bin and polarization degrees of freedom \cite{SAACFS:22:AQT}. The relative delay between two parties is resolved by transmitting an initial sequence of known qubit states---encoded in the first $10^6$ quantum states of the transmission---which is used to align the timing reference. The performance of this synchronization mechanism was validated over a 12-hour continuous \ac{QKD} session through a 50-km fiber spool. The Qubit4Sync synchronization maintained stable temporal alignment, achieving a \ac{QBER} of 0.765\,$\%$ and a secret key rate of approximately 16\;kbps.
The Qubit4Sync-type protocol was further demonstrated in experiments involving the Micius satellite and a Russian optical ground station \cite{M:24:JOPCS}. The protocol compared the time intervals between consecutive synchronization pulses and photon detections, enabling correction of the relative timing without requiring absolute position or velocity data of the satellite. Synchronization precision achieved during satellite \ac{QKD} sessions with Micius has ranged from 467\;ps to 497\;ps.

\subsubsection{Time-Bin Synchronization}

The use of time-bin encoded states for clock synchronization in \ac{QKD} systems was implemented in \cite{WSZQKF:21:QCCS}, where Alice transmits a structured sequence of quantum pulses containing synchronization pulses, dead-time intervals, and encoded key bits. The synchronization pulses serve as reference markers, allowing Bob to determine when each time-bin sequence begins. By analyzing the timing of these reference pulses, Bob reconstructs Alice’s clock ticks, thereby ensuring precise alignment between their local oscillators. Experimental results have demonstrated the robustness of this implementation under high-loss conditions, achieving a raw synchronization accuracy of 20\;ps even at 55\;dB of channel loss, which is comparable to long-distance fiber-optic transmission. The quantum synchronization protocol also operates at a high data rate, with a measured raw bit rate of 5\;Mbps.
The Bayesian clock-offset estimation protocol was proposed in \cite{CG:21:Entropy}. This approach exploits all public information exchanged in \ac{QKD} to optimize synchronization precision while preserving the secret key rate. Simulations of the three-state BB84 prepare-and-measure protocol with decoy states confirm that the Bayesian synchronization method achieves 95-$\%$ synchronization confidence within 4,140 communication bin widths, even under channel conditions with high loss and dark-count probabilities.

\subsubsection{Rubidium Synchronization}

Synchronization of rubidium clocks was considered in \cite{STSKSR:21:CLEO} for quantum network applications such as \ac{QKD}.
The experimental setup utilizes free-running rubidium oscillators at both Alice’s and Bob’s locations, operating as independent time references. The frequency offset between the two rubidium clocks introduces a linear drift, which can be pre-compensated using computational corrections. However, residual time-dependent frequency instabilities pose a greater challenge, as they exhibit non-deterministic variations that require continuous post-processing to ensure high synchronization accuracy. Experimental findings show that in a 50-m free-space \ac{QKD} link, the residual nonlinear drift amounted to 7\;ns over a 30-minute interval, corresponding to additional synchronization jitter of 0.39\;ps per 100-ms data integration window.

\subsubsection{Doppler Compensation}

Explicitly accounting for time drift caused by Doppler effects arising from the high relative velocity between the satellite and the Earth,
clock synchronization in satellite-to-ground \ac{QKD} systems was demonstrated in \cite{WLCLLP:21:OE}. The proposed \ac{QCS} method mitigates this issue by incorporating frequency recovery and Doppler compensation mechanisms. Specifically, a target frequency scanning technique is employed to rapidly estimate the correct operating frequency by searching within a predefined range. Subsequently, polynomial fitting is applied to model the variation of the received frequency over time, effectively counteracting Doppler-induced fluctuations. Experimental validation was performed using satellite-to-ground \ac{QKD} data obtained from the Micius satellite, achieving a best synchronization precision of 711\;ps.

\subsubsection{Twin-Field and Device-Independent Classes}

An analytical model investigating the impact of time synchronization errors on twin-field \ac{QKD} was presented in \cite{XWTFG:23:PRA}. The study introduced a detailed analytical framework for describing the degradation in interference visibility caused by temporal misalignment between twin-field pulses. This visibility directly affects the error probability in distinguishing the correct detector click during measurement. The study found a critical threshold at which synchronization error begins to drastically impact system performance. Specifically, when the timing jitter exceeds a threshold determined by the coherence time of optical pulses, both the key rate and the maximum achievable communication distance sharply decline.
Moreover, \ac{HOM}–based synchronization algorithms have been integrated into \ac{MDI}-\ac{QKD}, eliminating additional synchronization hardware and simplifying large-scale deployment \cite{HLLWWYHCGH:25:IEEE_J_JLT}. Collectively, these developments highlight the growing convergence between \ac{QCS} and \ac{QKD}, positioning quantum-enhanced synchronization as a core enabler for secure and scalable quantum communication networks.


\subsection{Quantum Teleportation}

\begin{figure}[t!]	
\centering
\includegraphics{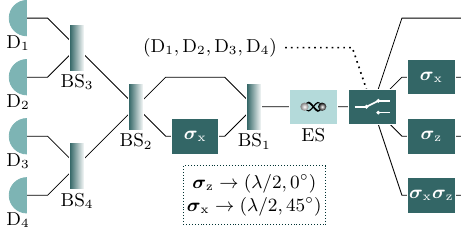}
\caption{
Illustration of the quantum teleportation setup. An entangled photon pair is generated at the entanglement source and routed to Alice and Bob. Alice mixes her unknown input state with her half of the entangled pair and performs a complete Bell-state measurement (BSM) using beam splitters (BS) and detectors (D). The measurement outcomes determine which Pauli correction Bob should apply---implemented using half-wave plates---to reconstruct the quantum state prepared by Alice.
}
\label{fig:9}
\end{figure}

Quantum teleportation is a fundamental quantum communication protocol that allows the transfer of unknown quantum states from one location to another without physically moving the particle itself.
By exploiting shared entanglement and classical communication between two parties (say, Alice and Bob), quantum teleportation enables secure and efficient state transfer without violating the no-cloning theorem (see Fig.~\ref{fig:9}). 
As such, it constitutes a key primitive for quantum communication, quantum computing, and quantum networks. Quantum teleportation protocols require strict temporal synchronization among distributed nodes because the \ac{BSM} relies on high-visible two-photon interference between independently generated photons \cite{MHSW:12:N}. In photonic teleportation architectures, the interfering photons must overlap in all degrees of freedom---particularly temporal mode, spectral profile, and polarization---such that their arrival-time difference is much smaller than the photon coherence time. Any clock desynchronization between nodes introduces a relative timing offset between photon-wave packets, reducing their temporal overlap and consequently degrading the \ac{HOM} interference visibility that underpins the \ac{BSM} process \cite{RXY:17:N}. A reduction in interference visibility directly impacts teleportation performance because the fidelity of the teleported state is determined by the quality of the Bell-state projection. Timing offsets broaden the coincidence window and increase the probability of accidental or distinguishable detection events, thereby lowering the success probability of the \ac{BSM} and degrading the average teleportation fidelity. In long-distance or multi-photon teleportation experiments, where coincidence rates are already limited by channel loss and detector noise, even small synchronization errors can significantly reduce the \ac{SNR} and overall teleportation throughput \cite{RRLMCI:25:OC}.


\subsubsection{Free-Space Long-Distance Teleportation}

The work in \cite{MHSW:12:N} demonstrated a long-distance quantum teleportation using active feed-forward over 143-km free-space channels. This experiment was enabled by advanced technologies such as low-noise photon detection, frequency-uncorrelated entangled photon sources, and entanglement-assisted \ac{QCS}. One of the key experimental aspects is the clock synchronization between Alice and Bob, which is crucial for maintaining temporal coherence during the teleportation process. In this work, two-photon coincidence counts were analyzed to synchronize Alice's and Bob's local clocks with a precision of approximately 1\;ns, allowing them to maintain a small coincidence window and reduce noise during the teleportation protocol. The work successfully demonstrated real-time teleportation, achieving an average teleportation fidelity well above the classical limit of $2/3$ and confirming genuine quantum teleportation over long distances.

\subsubsection{Satellite-Based Teleportation}

Quantum teleportation of single-photon polarization qubits over 1,400\;km was experimentally demonstrated in \cite{RXY:17:N} using a free-space link between a ground station and the Micius satellite. The Bell state was prepared using \ac{SPDC} with one photon retained at the ground station and its entangled counterpart transmitted to the satellite. A \ac{BSM} was performed at the ground station between the input qubit and the locally retained qubit of the entangled pair, thereby projecting the satellite photon onto the input qubit conditioned on the \ac{BSM} outcome. Clock synchronization has been achieved with a timing accuracy of 0.7\;ns, allowing accurate photon-coincidence counting within a 3-ns window and an average teleportation fidelity of $0.8 \pm 0.01$.

\subsubsection{\ac{HOM} Interference}

The \ac{HOM} interference---a key component of entanglement swapping and quantum teleportation---between independently generated photonic sources over 4.3-km optical fiber was demonstrated in \cite{RRLMCI:25:OC}. The setup employed a heralded single-photon source, which was synchronized through a shared optical clock signal. The experiment achieved a maximum \ac{HOM} interference visibility of $0.58 \pm 0.04$, exceeding the nonclassical threshold of 0.5 and meeting the requirement for performing \acp{BSM} in quantum teleportation protocols. This nonclassical interference was preserved even in the presence of timing jitter on the order of 20\;ps.

\subsubsection{High-Dimensional \ac{QED}}

Long-distance distribution of high-dimensional energy-time entangled photon pairs over a 10.2-km free-space quantum link was demonstrated in \cite{BPHKLE:23:PRX}. The distributed entangled states were realized through \ac{SPDC} in an interferometer, generating hyperentanglement in both polarization and energy-time degrees of freedom. The preservation of entanglement over a turbulent and lossy free-space channel was verified using a nonlocal Franson-type interferometric setup, which facilitates postselection-free measurements in the time-superposition basis. Entanglement was certified by a witness, where the witness expectation value exceeds 1.5, which is realized by a clock synchronization accuracy of about 20\;ps. Notably, the experiment demonstrated that the entanglement remains robust for more than 1.5 hours after sunrise, during which background photon counts exceeded the signal count by more than a factor of twenty. Such stable, high-dimensional \ac{QED} is crucial for the implementation of quantum teleportation protocols.

\subsubsection{Satellite-to-Ground \ac{QED}}

Distribution of polarization-entangled photon pairs from the Micius satellite to two ground stations separated by 1,203\;km was demonstrated in \cite{YCL:17:Sci}. By employing polarization compensation and clock synchronization with a timing jitter of 0.77\;ns, the entanglement fidelity exceeding 0.87 was achieved. The Bell test was conducted using rapidly and independently chosen measurement settings at both ground stations, yielding a Bell parameter of $2.73 \pm 0.09$. This result confirmed the preservation of nonlocal entanglement after long-distance satellite-to-ground transmission. 
To extend such precision to metropolitan and satellite-scale networks, frequency-comb-based synchronization has been investigated both theoretically and experimentally. These approaches exploit temporal-mode correlations and quantum-enhanced frequency combs to reduce timing deviations by up to an order of magnitude and to approach \ac{SQL} performance \cite{GM:25:QST, GAGLBM:24:APLP}. Collectively, these advances position \ac{QCS} as an enabling technology for high-fidelity, long-distance quantum teleportation and future repeater-based quantum networks.

\subsection{Quantum Networks}

Timing precision and clock synchronization are critically important in quantum networks due to the inherent sensitivity and fragility of quantum information encoded in quantum states \cite{WEH:18:S}. Accurate synchronization ensures that the quantum information---such as photons generated and detected at different network nodes---maintains precise temporal alignment, enabling coherent quantum-state transfer and reliable implementation of quantum communication protocols. Even minor deviations in synchronization can lead to photon misidentification, reduced quantum interference visibility, and substantial degradation in state fidelity, thus undermining the effectiveness and reliability of the entire quantum communication system.

\subsubsection{Fiber-Optic \ac{QCS}}

Fiber-optic-based \ac{QCS} that achieves picosecond-scale precision suitable for practical quantum networks was introduced in \cite{VNDL:22:JLT}. This synchronization method leveraged optical clock pulses operating at telecom wavelengths to align the timing of photon emission and detection across network nodes. Utilizing a central node equipped with a photon-pair source and synchronization transmitters, clock pulses were generated and combined with photon pairs into single-mode fibers for distribution to distant nodes. At the receiving nodes, the photon pairs were detected, where the recorded detection events were further processed to ensure alignment with the local oscillators. The synchronization system exhibited the 2-ps timing jitter over one minute, enabling quantum operations at high clock rates on the order of hundreds of megahertz.

\subsubsection{Satellite-Assisted Global \ac{QCS}}

Global \ac{QCS} by forming a globally distributed quantum-assisted master clock using a constellation of satellites was proposed in \cite{ DATAH:25:PRAppl, HABL:23:PRA, HAT:23:PRA}. This network exploits quantum resources, particularly entangled photon pairs that possess extremely tight time-of-birth correlations in the femtosecond regime. These photons are exchanged between satellites and ground stations, effectively linking clocks separated by large distances. The synchronization protocol operates by exchanging pairs of entangled photons and analyzing cross-correlation functions derived from photon detection events at different stations. Critical technical challenges are the relative motion of satellites, which degrades synchronization precision. To mitigate these effects, the protocol employs collective synchronization across a strategically designed satellite constellation. Simulation results showed that a constellation of approximately 50 satellites distributed in multiple orbits at an altitude of around 500\;km can synchronize clocks across the globe at a sustained sub-nanosecond precision.
The influence of satellite velocities on the \ac{QCS} precision was further examined in \cite{HAT:23:PRA}, where an \ac{LEO} satellite scenario incorporating imperfections---such as transmission losses, atmospheric absorption, and background noise---was analyzed. The simulations demonstrated that even at realistic relative satellite-ground velocities, sub-nanosecond to picosecond synchronization remains achievable over distances up to 4,000\;km by optimizing data acquisition windows. The impact of motion-induced timing shifts can be effectively mitigated by adjusting the photon-detection duration according to relative velocities between the satellite and ground stations.

\subsubsection{Multi-User \ac{QCS}}

A multi-user \ac{QCS} scheme in \ac{QED} networks was demonstrated in \cite{TTCH:23:EPJQT}, employing a single entangled photon source at a central server. The entangled photons were distributed to multiple remote users using wavelength-division multiplexing. Each user partially measured the received entangled signal photons and then returned their portion to the server. This round-trip scheme exploits the intrinsic time correlations of entangled photons to determine the clock time offsets. In experimental realizations, synchronization between the server and a single user over a 75-km fiber link achieved the clock-difference uncertainty of 4.45\;ps.
A more scalable multi-user \ac{QCS} network based on a silicon-chip dual-pumped entangled photon source was proposed in \cite{LHHT:24:arXiv}. This source produces one signal photon and three idler photons, supporting simultaneous synchronization of multiple users through the second-order correlation function. A proof-of-principle experiment was conducted for 11.1 hours involving the server and three users separated by 10\;km and 25\;km of standard optical fibers. This system achieved the minimum clock-difference standard deviation of 6.28\;ps, with the lowest observed time deviation of 0.82\;ps for averaging times up to 8,000\;s.

\subsubsection{Security-Aware \ac{QCS}}

Two-way clock synchronization was employed to detect system-level interception attempts by analyzing timestamps of timing messages exchanged among parties in multiparty quantum communication \cite{AHRZ:23:QIP}. Communicating nodes exchange synchronization and delay-response messages containing timestamps, which are then used to compute the path delay and clock offset. If the measured round-trip delay exceeded a predefined threshold---indicating a potential interception---the communication session would be immediately terminated. This mechanism enables the detection of channel-level attacks such as man-in-the-middle or intercept-resend attacks, because any additional optical interaction introduced by an attacker produces measurable timing deviations in the synchronization process. Additionally, the study proposed a quantum authentication scheme based on encoding sequences of keys into the polarization angles of optical quantum states to address the data-level threats. The sender encodes information using secret polarization angles and distributes authentication information through rotated quantum states that are subsequently decoded by legitimate receivers. The use of secret rotations and Pauli bit-flip operations makes it extremely difficult for an adversary to reconstruct the encoded qubits without knowledge of the secret angles and authentication keys. Consequently, attacks such as beam-splitting, photon-number-splitting, and collective attacks introduce detectable disturbances or bit errors in the measured quantum states.

\subsubsection{Quantum Access Networks}

\Ac{QCS} has also been utilized in \aclp{qan}, where multiple users share a central quantum receiver through time-division multiplexing. Conventional synchronization approaches in \aclp{qan} typically rely on dedicated classical synchronization hardware, which introduces considerable cost and design complexity. The integration of qubit-based synchronization was proposed in \cite{HCLH:24:SCPMA}, which enables \ac{QCS} directly through the quantum states used for communication, thereby simplifying the network infrastructure. In this approach, synchronization qubits were inserted at regular intervals within a larger qubit frame, upon which the Fourier transform and the least-squares optimization were employed to get a fine estimate of relative clock offsets. Simulations of \ac{QKD} networks under realistic scenarios demonstrated that the system can support up to 64 users, with each achieving secure key rates up to 1,070\;bps. Experimental results showed the feasibility of this synchronization scheme over a 56-km fiber using two users with distinct wavelengths. Each user achieved average secure key rates of 53.84\;kbps and 71.90\;kbps, with \acp{QBER} below 1\,$\%$.

\subsubsection{Foundational Network Services}

In emerging quantum networks, distributed nodes must maintain precise phase and frequency alignment to enable secure communication, entanglement distribution, and coordinated quantum processing. A recent field deployment demonstrated that clock synchronization can be directly integrated into entanglement-based \ac{QKD} links without requiring any dedicated timing channel \cite{PSCLAMT:23:PRAppl}. In this metropolitan-scale network connecting multiple nodes over 50\;km of commercial fiber, energy–time–entangled photon pairs provided both secure key generation and clock alignment through photon-counting techniques alone. This experiment illustrated that synchronization can be achieved natively within quantum communication channels, suggesting that \ac{QCS} evolves into a foundational service for future quantum networks, supporting both secure timing dissemination and high-performance quantum communication.

\subsection{Distributed Quantum Computing and Sensing}

Precise clock synchronization is essential for distributed quantum computing and sensing, where entangled processors or sensors must share a stable phase and frequency reference. Recent studies have proposed and simulated multiparty \ac{QCS} protocols based on \ac{GHZ} entanglement, demonstrating that networks of quantum clocks can be synchronized simultaneously across multiple nodes and large distances \cite{YBGKR:24:IEEE_CONF_QCNC}. Theoretical analyses further reveal that genuine multipartite entanglement is a necessary resource for achieving quantum advantages in distributed sensing and parameter estimation, thereby defining the ultimate precision limits attainable in quantum networks \cite{YYX:24:PRL}. In parallel, advances in quantum frequency combs have highlighted their potential to push synchronization and interferometric sensing toward the Heisenberg limit, paving the way for coherent quantum-enhanced telescopes and navigation systems \cite{GAGLBM:24:APLP}. Collectively, these developments indicate that \ac{QCS} is not only foundational for communication networks but also a key enabler for scalable distributed quantum computing and sensing infrastructures.

\subsubsection{Teleportation-Based Distributed Quantum Computation}
Teleportation-based computation enables spatially separated multiple quantum processors to collaboratively execute quantum algorithms by sharing entanglement and exchanging quantum states through teleportation. Remote quantum gates are implemented by teleporting qubits between nodes or by teleporting gate operations through entangled channels. Successful execution of these protocols requires precise temporal coordination because entanglement generation, \ac{BSM}, and classical feed-forward corrections must occur within the coherence time of the distributed qubits \cite{CBPL:26:IEEE_M_IC}. \Ac{QCS} provides a shared temporal reference that aligns photon emission, interference, and measurement events across network nodes. Accurate synchronization minimizes timing jitter and phase drift between processors, thereby preserving interference visibility and entanglement fidelity required for reliable teleportation-based remote gate execution.

\subsubsection{Entangled Sensor Network}
Distributed quantum sensing can be realized through entangled or correlated sensor networks in which spatially separated probes cooperatively estimate multiple physical parameters. Within the framework of multiparameter quantum metrology, correlations between parameters determine the achievable measurement variance. In cascaded sensing architectures---where multiple phase parameters are encoded sequentially along a single optical path and distinguished through temporally resolved detection---the sensing process relies critically on the precise temporal alignment of the probe evolution and measurement stages~\cite{KCG:25:PRA}. In such systems, the phase accumulation of the probe state follows a unitary evolution, implying that the extracted phase information is directly proportional to the interrogation time. Consequently, any clock offset between spatially separated sensing nodes introduces discrepancies in the effective interrogation interval, leading to relative phase errors that increase the estimator covariance.

\begin{table*}[t!]
\begin{center}
\caption{QCS applications: types, roles, experimental setups, and synchronization accuracy}
\label{tab:QCS-Apps}
\small
\begin{tabular}{lllll}
\toprule
\makecell[l]{Type}
& QCS Role
& Experimental Setup
& \makecell[l]{Synchronization \\ Accuracy}
& Reference \\
\midrule
\midrule

\multirow{11}{*}{QKD}
& \multirow{11}{*}{\makecell[l]{Enabling precise temporal \\ alignment between communicating\\ parties, ensuring high-fidelity\\ photon arrival correlations, \\coincidence-window optimization,\\ high SNR, and accurate sifting\\ of detection events; minimizing\\ QBER; supporting high-rate key\\ generation; enhancing resistance\\ to timing-based security attacks;\\ improving detector gating accuracy; \\ reducing dark-count contamination;\\ providing reliable timestamping\\ for long-distance QKD links;\\ ensuring clock stability and\\ coincidence accuracy}
}
& 730-m free space & N/A & \cite{BGMHN:04:OE} \\
\cmidrule{3-5}
&  & 1,200-km satellite-ground & 0.5\;ns & \cite{LCLZL:17:N} \\ 
\cmidrule{3-5}
& & 26-km fiber & N/A & \cite{AACSFZ:20:Optica} \\
\cmidrule{3-5}
& & N/A & 2\;ns & \cite{CSAADVV:20:PRAppl} \\
\cmidrule{3-5}
& & 50-km fiber & N/A & \cite{SAACFS:22:AQT} \\
\cmidrule{3-5}
& & Satellite-ground & 467\;ps & \cite{M:24:JOPCS} \\
\cmidrule{3-5}
& & 20-km fiber & 20\;ps & \cite{WSZQKF:21:QCCS} \\
\cmidrule{3-5}
& & N/A & N/A & \cite{CG:21:Entropy} \\
\cmidrule{3-5}
& & 50-m free space & 0.39\;ps & \cite{STSKSR:21:CLEO} \\
\cmidrule{3-5}
& & Satellite-ground & 711\;ps & \cite{WLCLLP:21:OE} \\
\cmidrule{3-5}
& & Fiber & N/A & \cite{XWTFG:23:PRA} \\

\midrule

\multirow{5}{*}{Teleportation}
& \multirow{5}{*}{\makecell[l]{Ensuring synchronous joint\\  measurements and time-aligned\\ detection of entangled photons\\ across distant nodes---essential for\\ validating measurement outcomes,\\preserving teleportation fidelity, \\and reliablely teleporting states}
}
& 143-km free space & 1\;ns & \cite{MHSW:12:N} \\
\cmidrule{3-5}
& & 1,400-km free space & 0.7\;ns & \cite{RXY:17:N} \\
\cmidrule{3-5}
& & 4.3-km deployed fiber & 20\;ps & \cite{RRLMCI:25:OC} \\
\cmidrule{3-5}
& & 10.2-km free space & 20\;ps & \cite{BPHKLE:23:PRX} \\
\cmidrule{3-5}
& & 1,203-km satellite-ground & 0.77\;ns & \cite{YCL:17:Sci} \\

\midrule

\multirow{7}{*}{\makecell[l]{Quantum\\ networks}}
& \multirow{7}{*}{\makecell[l]{Providing network-wide time\\ synchronization required for\\ coherent quantum-state transfer,\\ entanglement swapping, and\\ multipartite quantum repeating;\\ ensuring temporal stability across\\ distributed quantum memories;\\ enabling synchronized photon\\ generation or detection cycles;\\ supporting time-bin encoding}
}
& 11-km fiber & 2\;ps & \cite{VNDL:22:JLT} \\
\cmidrule{3-5}
& & 500-km altitude satellite-ground & 1\;ns & \cite{DATAH:25:PRAppl} \\
\cmidrule{3-5}
& & 500-km altitude satellite-ground & 1\;ns & \cite{HAT:23:PRA} \\
\cmidrule{3-5}
& & 75-km fiber & 4.45\;ps & \cite{TTCH:23:EPJQT} \\
\cmidrule{3-5}
& & N/A & N/A & \cite{AHRZ:23:QIP} \\
\cmidrule{3-5}
& & 50-km fiber & N/A & \cite{HCLH:24:SCPMA} \\
\cmidrule{3-5}
& & 25-km fiber & 6.28\;ps & \cite{LHHT:24:arXiv} \\

\midrule

\multirow{3}{*}{\makecell[l]{Distributed\\ quantum\\ computing\\ and sensing}}
& \multirow{3}{*}[0.55em]{\makecell[l]{Maintaining stable frequency and\\phase  coherence across distributed\\quantum processors and sensors\\ for multiparty quantum operations\\ and high-precision sensing;\\ Ensuring temporal referencing for\\ distributed processing tasks}
}
& \makecell[l]{Multi-party networks\\ using GHZ entanglement} & N/A & \cite{YBGKR:24:IEEE_CONF_QCNC} \\
\cmidrule{3-5}
& & \makecell[l]{Quantum clock networks with\\ multipartite entanglement} & \makecell[l]{Heisenberg-limited\\ precision} & \cite{YYX:24:PRL} \\
\cmidrule{3-5}
& & \makecell[l]{Quantum frequency-comb\\ synchronization} & \makecell[l]{Sub-picosecond\\ (Heisenberg regime)} & \cite{GAGLBM:24:APLP} \\

\bottomrule
\end{tabular}
\end{center}
\end{table*}

\section{\ac{QCS} Implementations} \label{sec:8}

This section details practical \ac{QCS} implementations across diverse experimental platforms, highlighting their operational scenarios, achieved timing accuracies, and deployment constraints. Table~\ref{tab:QCS-Implementations} summarizes \ac{QCS} implementations, including the employed protocols, system platforms, deployment scenarios, and achieved synchronization performance.

\subsection{\ac{TQH-QCS}}

The implementation of the \ac{TQH-QCS} protocol on a three-qubit \ac{NMR} system was reported in \cite{ZLD:04:PRA}. The \ac{NMR} platform enables reliable realization of quantum algorithms and precise control of quantum states. Experimental results were validated through quantum state tomography, reconstructing the density matrix of the final state to verify the accuracy of the \ac{TQH} implementation. The reported fidelities range from 75\,$\%$ to 91\,$\%$, demonstrating successful implementation of the \ac{TQH} algorithm. The primary sources of error were attributed to pulse imperfections, magnetic-field inhomogeneities, and decoherence.
The work in \cite{QLL:06:CPL} implemented the \ac{TQH-QCS} protocol using linear-optical setups where single photons act as qubits. Optical systems are particularly attractive for such implementations due to their low decoherence rates and high controllability of single-photon states. Using linear optics, two distinct schemes for implementing the \ac{TQH-QCS} protocol were proposed: location-plus-polarization and all-location qubit schemes. In the location-plus-polarization scheme, qubits are encoded in both the spatial path and the polarization degree of photons. In the all-location qubit scheme, both ticking and ancillary qubits are represented solely in spatial modes. This work implemented three- and four-qubit \ac{TQH-QCS} protocols, demonstrating its feasibility in photonic systems.


\subsection{\ac{EB-QCS}}

The implementation of \ac{EB-QCS} was first experimentally demonstrated using a four-qubit \ac{NMR} system operating at room temperature \cite{KXWWWLL:18:QIP}. This experiment validated the multiparty synchronization protocols proposed in \cite{KP:02:PRA} and \cite{ BE:11:PRA} where one qubit serves as the master clock while the remaining three qubits serve as clocks to be synchronized. The system was initialized from a thermal equilibrium state into a pseudo-pure $\ket{0000}$ state with a fidelity exceeding 99.02\,$\%$. Subsequently, four-qubit entangled W and Z states were prepared to realize the protocols in \cite{KP:02:PRA} and \cite{BE:11:PRA}, respectively. After the first qubit, serving as the master clock, was measured in the Hadamard basis---thereby encoding timing information into the shared state, the remaining qubits were measured in the same basis. Experimental results showed strong agreement with theoretical predictions for both protocols. Notably, the protocol proposed in \cite{BE:11:PRA} utilizing the Z state outperformed the W-state-based protocol \cite{KP:02:PRA} in terms of synchronization accuracy. These results also demonstrated that higher ticking rates $\omega$ lead to improved synchronization accuracy.


\subsection{\ac{HOM-QCS}}

\subsubsection{Dispersion Cancellation}

The \ac{QCS} protocol exploiting the dispersion-cancellation effect of frequency-entangled photons in \ac{HOM} interferometers was experimentally implemented in \cite{HDQZ:12:ASR} using fiber-optic links. In this setup, Alice generates frequency-entangled photon pairs and sends one photon to Bob via the optical fiber, while retaining the other photon locally and routing it through the fiber coiling with adjustable length. Both photons are then directed into the \ac{HOM} interferometer. Alice adjusts the length of her fiber coiling to balance the \ac{HOM} interferometer and records the photon creation times, while Bob records the arrival times of the transmitted photons. The clock offset between Alice and Bob is subsequently estimated from these time records. The experimental results demonstrated synchronization accuracy below 1\;ps over a fiber distance of 10\;km. This accuracy was shown to depend largely on the spectral bandwidth of the entangled-photon source and on fluctuations in the fiber links due to temperature variations.

\subsubsection{Second-Order Quantum Coherence}

\ac{HOM-QCS} exploiting the second-order quantum coherence of entangled photons was demonstrated in \cite{QDZWW:15:CLEO, QZWHW:16:SR, QDZLZ:18:CLEO, QDZH:19:OL}, achieving sub-picosecond timing stability over a 4-km fiber link. The entangled photon pairs generated via \ac{SPDC} were labeled as signal and idler photons, and distributed to two distant clocks through independent optical fibers. The \ac{HOM} interferometer served as the synchronization hub, enabling precise correlation measurements of photon arrival times. Each photon was split by a polarizing beam splitter, and adjustable optical delay lines were used to balance optical paths within the interferometer. At the receiver end, fiber couplers directed photons toward detectors, and the relative clock offset was extracted from the measured second-order correlation function. A synchronization accuracy of 73.2\;ps was reported in \cite{QZWHW:16:SR}. The system performance was primarily limited by the photon generation rate, detection efficiency, and path-length stability. 

\subsubsection{Dispersion-Free Quantum Feedback}

Enhanced \ac{HOM-QCS} incorporating dispersion-free quantum feedback was reported in \cite{XZLL:21:OE}, employing twin-beam states. The twin-beam state is a frequency-entangled photon-pair state generated through \ac{SPDC}, where the sum of signal and idler photon frequencies remains constant. This property preserves strong time correlations between the photon pairs, enabling synchronization precision to be maintained even over long distances. Experimental results showed that with dispersion compensation, the system achieves a synchronization accuracy of 4\;ps and a time offset precision of 1.8\;ps over a 10-s acquisition interval. Furthermore, with extended averaging over 5,500\;s, the time deviation is reduced to 0.15\;ps, highlighting the long-term stability of the synchronization process.

\subsubsection{Segmented Fiber Architectures}

One of the primary challenges in \ac{HOM-QCS} is maintaining stable \ac{HOM} interference fringes over extended distances. The degradation of \ac{HOM} interference due to fiber dispersion, temperature fluctuations, and mechanical instabilities introduces phase drifts, which can significantly impair synchronization precision. To address this limitation, an approach utilizing multiple segmented fiber links instead of a single long-length fiber has been demonstrated to reduce path-drift accumulation and enhance synchronization stability \cite{LQXH:21:APL}. By carefully balancing the fiber lengths and integrating real-time feedback control mechanisms, it is possible to maintain interference visibility, ensuring robust synchronization over extended fiber-optic networks. The experimental implementation was conducted over a 20-km fiber network composed of three segmented fibers ($5+4+1$\;km). This setup was designed to mitigate path-delay fluctuations, which are a major limitation in long-haul quantum synchronization. Experimental results showed a maintained visibility of 60\,$\%$ after 20\;km of fiber transmission. The implementation achieved a minimum timing stability of 74\;fs at 48,000\;s. Additionally, an absolute synchronization accuracy of $50.56 \pm 7.2$\;ps was obtained, in close agreement with the theoretical prediction of 44.14\;ps.

\subsection{Remote-Detection \ac{QCS}}

The implementation of clock synchronization by coincidence detection of entangled photons was first experimentally demonstrated in \cite{VSS:04:APL} where signal and idler photons were transmitted to two detectors placed 3\;km apart, simulating a space-to-ground synchronization scenario. The protocol was executed in two measurement rounds. In the first round, the signal photon was sent to one detector and the idler photon to the other. In the second round, the photon paths were swapped. By analyzing the difference in detection times across the two rounds, the time offset between clocks was estimated with high precision. The experiment demonstrates that the time correlations of entangled photon pairs are preserved over long fiber transmission, confirming the feasibility for one-way clock synchronization with picosecond-level accuracy.

In subsequent experiments \cite{QDXLLZ:20:RSI}, a synchronization accuracy of 0.72\;ps was reported with a 4.5-s acquisition time and a coincidence rate of 1,140\;counts/s (cps). This can be further enhanced to the femtosecond regime by increasing the photon-pair rate and acquisition time.
The 10-km implementation over optical fiber was demonstrated in \cite{LSUK:22:OE}, achieving an offset fluctuation uncertainty of 88\;ps over a 100-s measurement window with a coincidence rate of 160\;cps. Further extension to a 50-km deployed fiber link was reported in \cite{PSATM:23:EQEC}, which comprises two key stages---namely, time-offset determination and drift compensation. This adaptive estimation achieved a 8-ps synchronization accuracy. More broadly, remote-detection \ac{QCS} schemes utilize temporal correlations of photon arrival times to synchronize distant clocks without requiring optical interference at a central node \cite{STSKCCDRRS:23:PRAppl}. These methods have been validated under realistic, emulated high-loss conditions---including atmospheric turbulence---demonstrating that quantum temporal correlations can replace conventional optical or electronic timing channels. Since independent stations can synchronize through coincidence analysis without ultra-stable clocks or dedicated reference lasers, remote-detection \ac{QCS} is especially promising for scalable network architectures and potential satellite-to-ground implementations where centralized interference-based synchronization is impractical.

\subsection{Two-Way Time-Transfer \ac{QCS}}

The two-way \ac{QCS} protocols estimate relative clock offsets through reciprocal photon exchange and correlation analysis, providing inherent robustness against asymmetric delays and channel fluctuations. Recent experimental demonstrations have confirmed their feasibility across fiber, free-space, and hybrid transmission links.

\subsubsection{Proof-of-Principle Implementations}

The experiment in \cite{FRRXTS:18:CLEO} demonstrated a proof-of-principle implementation of two-way \ac{QCS} using frequency-entangled photon pairs transmitted over a 20-km fiber link. The protocol relies on the fourth-order time correlation measurements to achieve precise time synchronization between two clocks. To mitigate dispersion effects, a 2.6-km dispersion compensation fiber was incorporated into the setup. Photon arrival times were recorded using high-precision event timers synchronized with the respective clocks. Cross-correlation peak detection was used to determine the time difference between the photon arrival times at the detectors.
In measuring the time stability without fiber coiling, the \ac{RMS} fluctuation of the time offset was measured to be 7.48\;ps, with the time deviation falling below 1\;ps at an averaging time of 160\;s. When the averaging time was extended to 1,280\;s, the time deviation improved to 0.54\;ps.
By adding the 20-km fiber and 2.6-km dispersion-compensation coiling, the system achieved an \ac{RMS} fluctuation of the 41.4-ps time offset and a minimum time deviation of 0.85\;ps at an averaging time of 5,120\;s, demonstrating the feasibility of long-distance two-way \ac{QCS}.

\subsubsection{Deployed Fiber Implementations}

Two-way \ac{QCS} utilizing time-energy entangled photon pairs over a 7-km deployed fiber was implemented in \cite{QHXQZ:22:OE}. Experimental results demonstrated a synchronization time stability of approximately 1.9\;ps at an averaging time of 30\;s, closely matching theoretical predictions of 1.7\;ps under idealized conditions. After thoroughly evaluating and compensating for various systematic biases, the synchronization accuracy was calibrated and verified to 7.6\;ps.
A 50-km fiber-optic two-way \ac{QCS} implementation was demonstrated in \cite{HQXX:22:IEEE_J_JLT}, where dispersion cancellation was achieved using fiber Bragg grating modules with dispersion values of 825\;ps/nm inserted into the idler photon paths. With a common reference clock, the synchronization accuracy reached the picosecond regime after calibration. The measured synchronization stability achieved approximately 2.6\;ps at an extended averaging interval of 7\;s, further improving to 54.6\;fs at extended 57,300\;s.
When operating with independent reference clocks, the synchronization stability remained comparable, achieving 2.7-ps stability at 7\;s and 89.5\; fs at 57,300\;s and highlighting the robustness and practical feasibility under realistic operational conditions.

In \cite{XXHQCD:24:OE}, the protocol was implemented over varying fiber lengths, with dispersion-compensation fibers employed to mitigate chromatic dispersion effects. The measured standard deviations of the time offsets were 0.46\;ps, 1.14\;ps, and 3.98\;ps over 11.3-km, 22.4-km, and 55.6-km fibers, respectively. Synchronization stability was measured over an averaging time of 12,680\;s, resulting in time deviations of 0.49\;ps, 0.59\;ps, and 1.01\;ps for the same distances. In \cite{SXHLZ:24:APL}, the protocol was implemented over a 50-km single-mode fiber link at both parties. Nonlocal dispersion introduced during fiber propagation was mitigated using fiber Bragg grating modules. The experiment achieved a 3.61-ps standard deviation of the time offset with a synchronization stability of 0.69\;ps over an averaging time of $10^4$\;s employing predictive algorithms.

\subsubsection{Hybrid and Urban Implementations}

A hybrid two-way \ac{QCS} configuration combining a 2-km turbulent free-space link with a 7-km deployed optical fiber was demonstrated in \cite{XSQLX:23:QST}. Despite significant photon loss and atmospheric turbulence, the two-way \ac{QCS} maintained sub-picosecond time stability, achieving a minimum time deviation of 1.71\;ps at a 50-s averaging time and improving to approximately 144\;fs at 6,400\;s.
Two-way \ac{QCS} over a 103-km urban fiber link was demonstrated in \cite{HQXLLC:24:IEEE_J_JLT}, where the fiber link was subject to thermal drift, polarization changes, and signal crosstalk from adjacent telecom services. The transmitted photons were routed through dispersion-compensating fiber modules to enable nonlocal dispersion cancellation. With a total link attenuation of 38\;dB and low correlated photon detection rates of 40\;cps, the system demonstrated a time deviation of 0.28\;ps at 40,000\;s. The total uncertainty in the extracted time offset was estimated to be 13.6\;ps, accounting for the measurement error, calibration drift, non-reciprocal delay due to the spectral asymmetry and polarization-mode dispersion. 

An experimental demonstration of two-way \ac{QCS} under conditions emulating an \ac{LEO} satellite link in the free-space quantum communication testbed was reported in \cite{LENGS:24:PRAppl}. The experiment was conducted over a 1.6-km free-space channel, where atmospheric turbulence and background sky radiance were emulated to mimic a 700-km \ac{LEO} channel. The standard deviation of the synchronized clock offset was measured to be 27.1\;ps during nighttime operation and 39.7\;ps during daytime.

\subsection{Satellite-Based \ac{QCS}}

Extending \ac{QCS} beyond terrestrial networks, recent studies have explored satellite-based architectures for global-scale time distribution. Simulation results indicated that multi-qubit quantum synchronization can reduce timing errors to below 1\;m, enhancing the precision and scalability of satellite-based navigation and communication systems \cite{NRHIBMPWBDCF:24:ComNet}. Complementary work proposes a network of satellites equipped with quantum resources that mutually reinforce timing stability, effectively forming a distributed master clock capable of sub-nanosecond synchronization on a global scale \cite{DATAH:25:PRAppl}. Together, these developments delineate a clear pathway toward quantum-enhanced global navigation satellite systems and space-based quantum networks, bridging communication, sensing, and timekeeping in a unified framework.

\subsection{Chip-Integrated \ac{QCS} Platforms}

Recent advances in compact quantum synchronization techniques and quantum frequency-comb technology are paving the way toward scalable and potentially chip-integrated \ac{QCS} systems. Quantum frequency combs, exhibiting nonclassical features such as squeezing and entanglement, have demonstrated the potential to surpass the \ac{SQL}, providing precise timing for optical and space-based networks \cite{GAGLBM:24:APLP}. Complementary developments, such as synchronization in \ac{MDI}-\ac{QKD} \cite{HLLWWYHCGH:25:IEEE_J_JLT}, quantum-assisted clock recovery without service channels \cite{KWHF:25:PRAppl}, and photon-pair-based synchronization achieving sub-picosecond stability in fiber links \cite{GFDSYWGZ:24:IEEE_CONF_OGC}, illustrate a broader trend toward simplified and hardware-efficient synchronization architectures. As these technologies continue to mature, their integration with nanophotonic and on-chip frequency-comb platforms is expected to enable portable and low-power \ac{QCS} devices suitable for field and satellite deployment.


\begin{table*}[t!]
\begin{center}
\caption{QCS implementations: protocols, system types, experimental setups, and synchronization performance}
\label{tab:QCS-Implementations}
\small
\begin{tabular}{lllll}
\toprule

Protocol 
& \makecell[l]{System\\ Type} 
& Experimental Setup 
& \makecell[l]{Synchronization\\ Performance} 
& Reference \\

\midrule
\midrule

\multirow{2}{*}{TQH-QCS}
& NMR 
& Three-qubit system 
& \makecell[l]{$>$\,100\;ps\\ fidelity 75--91\,$\%$} 
& \cite{ZLD:04:PRA} \\

\cmidrule{2-5}

& Photonic 
& Four-qubit linear-optical system 
& $>$\,100\;ps 
& \cite{QLL:06:CPL} \\

\midrule

Multiparty EB-QCS
& NMR
& Four-qubit W- or Z-state system
& 29-$\mu$s synchronization delay
& \cite{KXWWWLL:18:QIP} \\

\midrule

\multirow{4}{*}{HOM-QCS}
& Photonic 
& 10-km fiber  
& 0.54-ps accuracy 
& \cite{HDQZ:12:ASR} \\

\cmidrule{2-5}

& Photonic 
& 4-km fiber  
& 73.2-ps accuracy 
& \cite{QZWHW:16:SR} \\

\cmidrule{2-5}

& Photonic 
& \makecell[l]{22-km fiber \\ (dispersion-free feedback)} 
& \makecell[l]{4-ps accuracy\\ 0.15-ps deviation\\ (5,500-s avgeraging)} 
& \cite{XZLL:21:OE} \\

\cmidrule{2-5}

& Photonic 
& \makecell[l]{20-km segmented fiber} 
& \makecell[l]{$50.6\pm7.2$-ps accuracy\\ 74-fs stability} 
& \cite{LQXH:21:APL} \\

\midrule

\multirow{4}{*}{Time-offset \acs{QCS}}
& Photonic 
& 3-km space-to-ground emulation 
& 1-ps accuracy 
& \cite{VSS:04:APL} \\

\cmidrule{2-5}

& Photonic 
& \makecell[l]{Fiber\\ (4.5-s acquisition)} 
& 0.72-ps accuracy 
& \cite{QDXLLZ:20:RSI} \\

\cmidrule{2-5}

& Photonic 
& 10-km fiber  
& \makecell[l]{88-ps fluctuation\\ (100-s averaging)} 
& \cite{LSUK:22:OE} \\

\cmidrule{2-5}

& Photonic 
& \makecell[l]{50-km deployed fiber\\ (drift compensation)}
& \makecell[l]{8-ps accuracy} 
& \cite{PSATM:23:EQEC} \\

\midrule

\multirow{8}{*}{Two-way QCS}
& Photonic 
& 20-km fiber  
& \makecell[l]{41.4-ps RMS\\ $<$\,1-ps deviation\\ (160-s averaging)} 
& \cite{FRRXTS:18:CLEO} \\

\cmidrule{2-5}

& Photonic 
& 7-km deployed fiber link 
& \makecell[l]{7.6-ps accuracy\\ 1.9-ps stability} 
& \cite{QHXQZ:22:OE} \\

\cmidrule{2-5}

& Photonic 
& \makecell[l]{50-km fiber\\ (dispersion compensation)} 
& \makecell[l]{36.6-ps accuracy\\ 54.6-fs stability} 
& \cite{HQXX:22:IEEE_J_JLT} \\

\cmidrule{2-5}

& Photonic 
& \makecell[l]{11.3--55.6-km fiber} 
& 0.46--3.98-ps accuracy 
& \cite{XXHQCD:24:OE} \\

\cmidrule{2-5}

& Photonic 
& 50-km fiber 
& \makecell[l]{3.61-ps accuracy\\ 0.69-ps stability} 
& \cite{SXHLZ:24:APL} \\

\cmidrule{2-5}

& Photonic 
& \makecell[l]{$2+7$-km free-space and fiber\\ (hybrid)} 
& \makecell[l]{1.6-ps accuracy\\ 0.14-ps deviation} 
& \cite{XSQLX:23:QST} \\

\cmidrule{2-5}

& Photonic 
& 103-km urban fiber
& \makecell[l]{13.9-ps accuracy\\ 0.28-ps deviation\\ (40,000-s averaging)} 
& \cite{HQXLLC:24:IEEE_J_JLT} \\

\cmidrule{2-5}

& Photonic 
& \makecell[l]{1.6-km free-space\\ (emulated 700\;km)} 
& \makecell[l]{27.1-ps accuracy (night)\\ 39.7-ps accuracy (day)} 
& \cite{LENGS:24:PRAppl} \\

\midrule

\multirow{2}{*}{Space-based QCS}
& Photonic 
& Multi-satellite quantum network 
& $<$\,1-ns global synchronization 
& \cite{DATAH:25:PRAppl} \\

\cmidrule{2-5}

& Photonic 
& Satellite architecture (simulation) 
& $<$\,1-m timing error 
& \cite{NRHIBMPWBDCF:24:ComNet} \\

\midrule

\multirow{3}{*}{\makecell[l]{Chip-integrated QCS}}
& Photonic 
& Integrated quantum frequency comb 
& $<$\,1-ps timing precision 
& \cite{GAGLBM:24:APLP} \\

\cmidrule{2-5}

& Photonic 
& On-chip photon-pair source 
& $<$\,1-ps stability 
& \cite{GFDSYWGZ:24:IEEE_CONF_OGC} \\

\cmidrule{2-5}

& Photonic 
& \makecell[l]{MDI-QKD/qubit-based QCS\\ (no service channel)} 
& $<$\,1-ps stability 
& \makecell[l]{\cite{KWHF:25:PRAppl} \\ \cite{HLLWWYHCGH:25:IEEE_J_JLT}} \\

\bottomrule

\end{tabular}
\end{center}
\end{table*}

%
\section{Open Challenges and Research Directions}
\label{sec:9}
This section outlines the major open challenges and future research directions for scalable and deployable \ac{QCS} systems.

\subsubsection{Large-Scale Quantum Networks}

Although \ac{QCS} protocols have been successfully demonstrated in point-to-point and small-scale setups, extending them to large, distributed quantum networks remains a major open challenge. Recent theoretical work shows that achieving a genuine quantum advantage in large-scale synchronization requires multipartite entanglement shared across the entire network \cite{YYX:24:PRL}. Simulations using \ac{GHZ}-based protocols on fiber-connected quantum networks suggest that simultaneous synchronization of multiple atomic clocks is feasible, providing a blueprint for future distributed architectures \cite{YBGKR:24:IEEE_CONF_QCNC}. However, scaling to hundreds of nodes while preserving entanglement fidelity and sub-picosecond precision requires advances in entanglement generation, routing, and error mitigation under heterogeneous channel conditions.

\subsubsection{Integration with Classical Time Infrastructures}

A realistic pathway toward large-scale deployment of \ac{QCS} lies in hybrid integration with established classical timing systems such as \ac{GNSS}, the precision-time protocol, the network-time protocol, and two-way satellite time and frequency transfer. Field demonstrations already show that quantum communication and synchronization channels can coexist with classical traffic in shared fiber infrastructures, with up to 100 quantum channels operating alongside classical synchronization signals when properly spectrally isolated \cite{BSHGRALSBP:23:OE}. Similarly, secure quantum-assisted time transfer has been tested between distant precision-time facilities, where \ac{QKD} has been used to protect classical synchronization data over satellite and terrestrial links \cite{PVODZPAS:24:GPSSol}. At a larger scale, theoretical models envision satellite constellations equipped with quantum resources acting as a global master clock---potentially complementing or enhancing \ac{GNSS} through sub-nanosecond synchronization across the globe \cite{DATAH:25:PRAppl}.
Despite these promising advances, full interoperability between quantum and classical time infrastructures has not yet been standardized. Future efforts must focus on defining interface protocols, failover mechanisms, and metrological alignment with the International Bureau of Weights and Measures and the Consultative Committee for Time and Frequency Standards to ensure compatibility and cost-effectiveness relative to enhanced classical systems.

\subsubsection{Standardization and Interoperability}

Standardization of \ac{QCS} still remains in its early stages. Unlike \ac{QKD}, which is progressing under established bodies such as the European Telecommunications Standards Institute and the International Telecommunication Union, \ac{QCS} currently lacks a unified framework or cross-platform benchmarking standard. Existing demonstrations employ diverse encoding approaches---such as time-bin, polarization, and quantum frequency-comb methods---making it difficult to compare performance or interoperability across implementations \cite{GM:25:QST, GAGLBM:24:APLP}.
Recent studies have begun quantifying timing precision limits using quantum estimation theory, revealing how parameters such as timing centroid and relative intensity affect achievable synchronization accuracy \cite{GM:25:QST}. Similarly, satellite-based proposals envision networks of optically linked quantum clocks forming a globally synchronized reference, potentially serving as the foundation for next-generation quantum navigation and timing infrastructures \cite{DATAH:25:PRAppl}. Establishing standardized metrics---such as covering precision, stability, and compatibility with existing timing systems---is essential to ensure interoperability and to integrate \ac{QCS} into future metrological and communication standards.

\subsubsection{Deployment in Adverse Environmental Conditions}

Deployment of \ac{QCS} faces substantial environmental challenges across both free-space and fiber-based links. In satellite and ground-based systems, issues such as background light, atmospheric turbulence, and high channel loss can severely limit synchronization fidelity and secure time transfer, especially under daylight or long-distance conditions \cite{PVODZPAS:24:GPSSol}. Similarly, fiber-based metropolitan testbeds experience instability from temperature fluctuations, path delay gradients, polarization drift, and optical power variations, which degrade long-term precision unless compensated by active stabilization \cite{MRPGAPBTHL:24:APL}. Recent coexistence studies also show that optical noise and link-length limitations constrain the performance of \ac{HQC} networks \cite{MW:24:ApplOpt}. Addressing these challenges requires adaptive control schemes, environment-aware calibration, and learning-based prediction models to ensure reliable synchronization performance in adverse conditions, particularly for global-scale space-ground quantum networks.

\subsubsection{\ac{QCS} for Wireless Network Systems}

Clock synchronization is a fundamental requirement in communication networks, including \acl{IIoT}, \aclp{CPS}, and digital-twin platforms. Recent studies in networking areas have explored intelligent distributed synchronization mechanisms, such as data-driven and digital-twin frameworks, to improve synchronization accuracy and scalability in large-scale systems while reducing communication overhead \cite{PXX:21:IEEE_J_IOT,PXK:23:IEEE_J_IINF}. These works highlight the growing importance of precise and robust timing infrastructures for coordinated sensing, control, and data processing in distributed network environments.
Building on these developments, an important research direction is the integration of quantum-enhanced synchronization and control with communication infrastructures, including wireless networks, satellite systems, and integrated terrestrial and non-terrestrial networks \cite{RKC:25:IEEE_J_NSE,PXX:23:IEEE_J_IOT}. By potentially enabling ultra-precise and secure time transfer, \ac{QCS} can complement classical synchronization mechanisms in scenarios where network latency, propagation asymmetry, and/or environmental noise limit conventional approaches.

\subsubsection{Demonstrating Quantum Advantage}

A central open question in \ac{QCS} research is how to rigorously define and verify quantum advantage, i.e., a measurable and reproducible improvement over the best classical synchronization methods. Theoretical analyses indicate that quantum frequency combs, which exploit squeezing and entanglement, can surpass the \ac{SQL} and potentially reach Heisenberg-limited precision in clock synchronization \cite{GAGLBM:24:APLP}. However, existing implementations typically achieve precision levels comparable to advanced classical techniques such as two-way optical time transfer or the white rabbit protocol.
To demonstrate a genuine quantum advantage, future experiments must benchmark \ac{QCS} against classical baselines in realistic conditions. Promising directions include satellite-based synchronization, where entanglement-assisted protocols can outperform \ac{GNSS} and non-terrestrial network timing by achieving sub-meter accuracy \cite{NRHIBMPWBDCF:24:ComNet}, and fiber-based experiments demonstrating femtosecond-level stability over tens of kilometers using correlated photon pairs \cite{GFDSYWGZ:24:IEEE_CONF_OGC, HQXX:22:IEEE_J_JLT}. Establishing standardized benchmarking criteria---covering precision, scalability, and security---is crucial for determining whether \ac{QCS} can evolve from laboratory demonstrations into a foundational technology for global timekeeping.

\section{Conclusion} \label{sec:10}

\ac{QCS} has emerged as a rapidly advancing paradigm at the intersection of quantum communication, precision timekeeping, and distributed network coordination. This survey highlights that \ac{QCS} has demonstrated strong theoretical advantages---such as surpassing classical timing precision limits and enabling intrinsically secure time transfer. However, transitioning these advantages from laboratory-scale experiments to deployable large-scale infrastructure remains a central challenge. Current \ac{QCS} performance is constrained by photon loss over long-distance transmission, phase instability in optical channels, environmental decoherence, limited quantum memory lifetimes, and hardware imperfections such as detector timing jitter. In addition, the coexistence of \ac{QCS} with classical optical communication networks raises new security considerations, including vulnerability to timing-manipulation and man-in-the-middle attacks if quantum and classical layers are not co-designed. Despite these challenges, recent advances in satellite quantum communication, integrated photonics, quantum frequency combs, and entanglement distribution architectures strongly indicate the feasibility of continental-scale---and ultimately global---quantum-synchronized networks. Future developments are expected to involve \ac{HQC} timing architectures, error-corrected \ac{QED}, and standardization for synchronization benchmarks and security models. \ac{QCS} engineering is poised to play a foundational role in the emerging quantum Internet, distributed quantum computing and sensing, 6G \acp{NTN}, secure navigation systems, and relativistic science applications. Continued advances in scalable photonic integration, quantum repeater networks, and satellite constellation architectures will be critical to realizing robust and high-precision \ac{QCS} systems capable of redefining how global timing is established, maintained, and secured.


\balance



\end{document}